\documentclass[fleqn,12pt]{wlscirep}
\usepackage[normalem]{ulem}
\usepackage[many]{tcolorbox}
\tcbset{before upper={\parindent0em}}
\usepackage{bbold}
\usepackage{multicol}

\usepackage{scrextend}
\deffootnote[2em]{2em}{1em}{\textsuperscript{\thefootnotemark}\,}

\def\aj{AJ}%
\def\araa{ARA{\&}A}%
\def\apj{ApJ}%
\def\apjl{ApJ}%
\def\apjs{ApJS}%
\def\aap{A{\&}A}%
\def\jcap{J. Cosmology Astropart. Phys.}%
\def\mnras{MNRAS}%
\def\na{New A}%
\def\prd{Phys. Rev. D}%
\def\prl{Phys. Rev. Lett.}%
\def\pasj{PASJ}%
\def\nat{Nat.}%
\def\physrep{Phys.~Rep.}%
\def\physrep{Physics Reports}

\title{Cosmological Simulations of Galaxy Formation}

\author[1]{Mark Vogelsberger}
\author[2]{Federico Marinacci}
\author[3]{Paul Torrey}
\author[4]{Ewald Puchwein}
\affil[1]{Department of Physics, Kavli Institute for Astrophysics and Space Research, Massachusetts Institute of Technology, Cambridge, MA 02139, USA}
\affil[2]{Department of Physics and Astronomy, University of Bologna, via Gobetti 93/2, 40129 Bologna, Italy}
\affil[3]{Department of Astronomy, University of Florida, 316 Bryant Space Sciences Center, Gainesville, FL 32611 USA}
\affil[4]{Leibniz-Institut f\"ur Astrophysik Potsdam (AIP), An der Sternwarte 16, 14482 Potsdam, Germany}

\newcommand{\gsim}{\,\lower.7ex\hbox{$\;\stackrel{\textstyle>}{\sim}\;$}}
\newcommand{\lsim}{\,\lower.7ex\hbox{$\;\stackrel{\textstyle<}{\sim}\;$}}
\newcommand{\ion}[2]{#1{\sc #2}}

\begin{abstract}
{\bf 
Over the last decades, cosmological simulations of galaxy formation have been instrumental for advancing our understanding of structure and galaxy formation in the Universe. 
These simulations follow the non-linear evolution of galaxies modeling a variety of physical processes over an enormous range of scales. A better understanding of the physics relevant for shaping galaxies, improved numerical methods, and increased computing power have led to simulations
that can reproduce a large number of observed galaxy properties. Modern simulations model dark matter, dark energy, and ordinary matter in an expanding space-time starting from well-defined initial conditions. The modeling of ordinary matter is most challenging due to the large array of physical processes affecting this matter component.  Cosmological simulations have also proven useful to study alternative cosmological models and their impact on the galaxy population. This review presents a concise overview of the methodology of cosmological simulations of galaxy formation and their different applications.}
\end{abstract}

\begin{document}

\flushbottom
\maketitle
\thispagestyle{empty}

\section{Introduction}

\begin{figure}[ht!]
\centering
\includegraphics[width=1.00\linewidth]{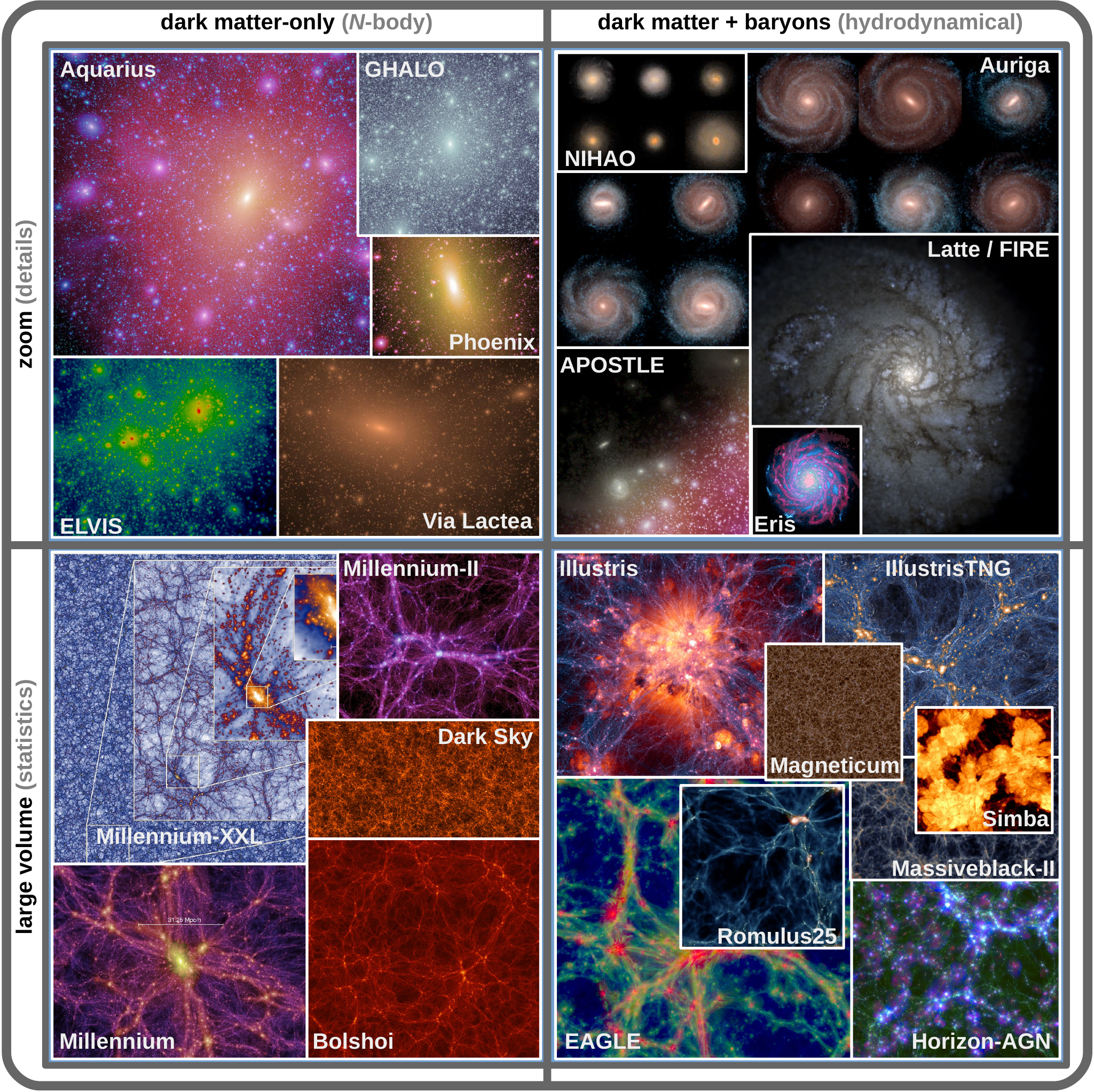}
\caption{\footnotesize Visual representations of some selected recent structure and galaxy formation simulations. The simulations are divided in large volume simulations providing statistical samples of galaxies, and zoom simulations resolving smaller scales in more detail. Furthermore, they are also divided in dark matter-only, i.e. $N$-body, and dark matter plus baryons, i.e. hydrodynamical simulations. Dark  matter-only simulations have now converged on a wide range of predictions for the large-scale clustering of dark matter and the dark matter distribution within gravitationally bound dark matter halos. Recent hydrodynamical simulations reproduce galaxy populations that agree remarkably well with observational data. However, many detailed predictions of these simulations are still sensitive to the underlying implementation of baryonic physics.}
\label{figure_1}
\end{figure}

\begin{figure}[ht!]
\centering
\includegraphics[width=\linewidth]{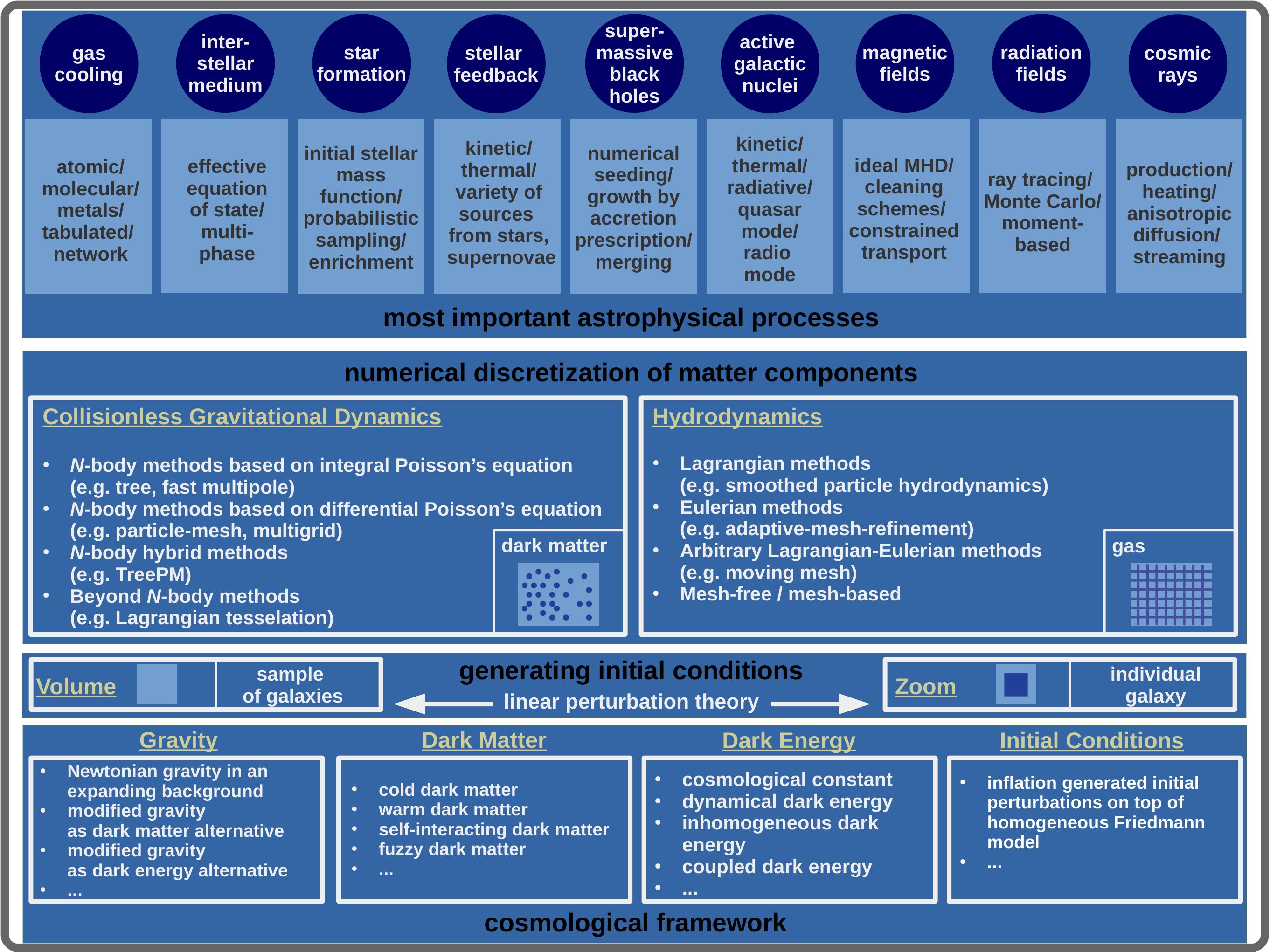}
\caption{\footnotesize Overview of the key ingredients of cosmological simulations. These simulations are performed within a given cosmological framework, and start from specific initial conditions. This framework includes physical models for gravity, dark matter, dark energy, and the type of initial conditions. Two types of simulations are typically performed: either large volume simulations or zoom simulations. The evolution equations of the main matter components, dark matter and gas, are discretized using different techniques and evolved forward in time. The dark matter component follows the equations of collisionless gravitational dynamics that are in most cases solved through the $N$-body method using different techniques to calculate the gravitational forces. The gas component of baryons is described through the equations of hydrodynamics that are solved, for example, with Lagrangian or Eulerian methods. Various astrophysical processes must also be considered to achieve a realistic galaxy population. Many of these are implemented through effective sub-resolution models. }
\label{figure_2}
\end{figure}

Modern astronomical surveys provide enormous amounts of observational data confronting our theories of structure and
galaxy formation. Interpreting these observations demands accurate theoretical predictions. However, galaxy formation is a challenging problem due to its intrinsic multi-scale and multi-physics character. Cosmological computer simulations are,
hence, the method of choice for tackling these complexities when studying the properties, growth and evolution of
galaxies. These simulations are important to understand the detailed workings of structure and galaxy formation. Dark matter builds the backbone for structure formation and is therefore a key ingredient of these simulations. In addition, dark energy is responsible for the accelerated expansion of the Universe and must also be considered. Despite the fact that the nature of dark  matter and dark energy are not known, simulations can make detailed and reliable predictions for these dark components based on their general characteristics. Ordinary matter, e.g. stars and gas, contribute only about five percent to the energy budget of the Universe. Nevertheless, simulating this matter component is essential to study galaxies, but, unfortunately, it is also the most challenging aspect of galaxy formation. Recent simulations follow the formation of individual galaxies and galaxy populations from well-defined initial
conditions and yield realistic galaxy properties \cite{Somerville2015}. Visual representations of the predictions of some of these simulations are shown in Figure~\ref{figure_1}. At the heart of these simulations are detailed galaxy formation models. Among others, these models describe the cooling of gas, the formation of stars, and the energy and momentum injection caused by supermassive black holes and massive stars\cite{Naab2017}. More recent simulations also model the impact of radiation fields, relativistic particles, and magnetic fields  leading to a more and more complex description of the galactic ecosystem and the detailed evolution of galaxies in the cosmological context. Galaxy formation simulations have also become important for cosmological studies since they can, for example, explore the impact of alternative cosmological models on the galaxy population. Cosmological simulations of galaxy formation therefore provide important insights into a wide range of problems in astrophysics and cosmology. The most important components of cosmological galaxy formation simulations are discussed in this review. 
A schematic overview of the different ingredients of cosmological simulations is presented in Figure~\ref{figure_2}.

\section{Cosmological Framework}
Cosmological simulations of galaxy formation are performed within a cosmological model and start from specific initial conditions. Both of these ingredients are now believed to be known to high precision.

\subsection{Cosmological Model}
Various observations revealed that our Universe is geometrically flat and dominated
by dark matter and dark energy  accounting for about $\sim 95\%$ of the energy density. Standard model particles make up for the remaining $\sim 5\%$ and are collectively referred to as baryons. The leading model for structure formation assumes that dark matter is cold, with negligible
random motions when decoupled from other matter, and collisionless, so-called cold dark matter, and dark energy is represented by a cosmological constant $\Lambda$, which drives the accelerated expansion of the Universe. This leads to the concordance $\Lambda$CDM model, which builds the framework for galaxy formation. Measurements of the cosmic microwave background combined with other observations such as the distance-redshift relation
from Type Ia supernovae, abundances of galaxy clusters, and galaxy clustering
constrain the fundamental parameters of the $\Lambda$CDM model\cite{Planck2016}. 

\subsection{Initial Conditions}
Initial conditions for cosmological simulations specify 
the perturbations imposed on top of a homogeneous 
expanding background. The background model is generally taken to be a spatially flat Friedmann-Lema\^{i}tre-Robertson-Walker space-time with a defined composition of dark matter, dark energy and baryons. Inflation predicts Gaussian perturbations, where the joint probability distribution of density fluctuations is a multidimensional Gaussian completely specified by its matter power spectrum $P(|k|)$. 
The post-recombination density field is the linear convolution of the primordial fluctuation field as predicted by inflation with a transfer function $T(k)$\cite{Seljak1996,Peacock1997,Eisenstein1998,Eisenstein1999}. Therefore, the power spectrum used to initialize simulations
generally takes the form ${P(k) = A k^n |T(k)|^2}$ with $n\approx 1$.
Once the linear density fluctuation field has been specified at some initial
time, typically at redshift $z \sim 100$, dark matter particle
positions and velocities are assigned along with baryon density, velocity,
and temperature fields. The standard approach for dark matter is to displace simulation particles from a uniform Cartesian lattice or glass-like\cite{Baugh1995,White1996} particle configuration using a linear theory approximation\cite{Zeldovich1970} or low-order perturbation theory\cite{Bertschinger2001,Jenkins2010,Hahn2011,Garrison2016}. A gravitational glass is made
by advancing particles from random positions using the opposite sign of gravity
until they freeze in comoving coordinates. Baryon positions and velocities are set in a similar way, and the baryon
temperature is often roughly initialized to the redshift-dependent microwave background temperature. 
Two types of initial conditions are commonly employed: uniformly sampled periodic large volumes or zoom initial conditions, where a low resolution background realization of the density fields surrounds a high resolution region of interest. The computational cost of these zoom simulations increases with the mass of the object that is studied for a given mass resolution. Zoom simulations of dwarf galaxies are therefore computationally less expensive than zoom simulations of large galaxy clusters given the larger number of resolution elements. Some simulations also employ constrained initial conditions to mimic, for example, the local Universe, e.g. nearby dark matter overdensities\cite{Hoffman1991,Salmon1996}. \\

\footnotesize
\begin{tcolorbox}[enhanced,drop fuzzy shadow,colback=yellow!10!white,colframe=yellow!50!black,title={\bf Generating initial conditions}]
\vspace{-15pt}
\begin{flalign}
&\text{\bf initial positions: } {\bf x} = {\bf q} + D(t) {\boldsymbol \Psi} ({\bf q}) 
& & \text{\bf initial velocities: } a(t) \,\dot{\bf x} = a(t) \,\frac{{\rm d}D(t)}{{\rm d}t} \, {\boldsymbol \Psi} ({\bf q}) =  a(t)\, H(t)\, \frac{{\rm d} \ln D}{{\rm d} \ln a} \, D(t) {\boldsymbol \Psi} ({\bf q}) \nonumber
\end{flalign}
\noindent\rule{\textwidth}{1pt}
\noindent Comoving initial positions, ${\bf x}$, are assigned based on the unperturbed particle position, ${\bf q}$, the linear growth factor, $D(t)$,  and the scale factor, $a$, which is related to the initial redshift, $z=1/a - 1$. The curl-free displacement field $\boldsymbol \Psi$ is computed by solving the linearized continuity equation ${{\boldsymbol \nabla} \cdot {\boldsymbol \Psi} = -\delta / D(t)}$, where $\delta$ is the relative density fluctuation. 
\end{tcolorbox}
\normalsize

\section{Simulating Dark Matter}

Dark matter builds the backbone for the formation of galaxies, which are expected to form at the centers
of dark matter overdensities, so-called halos. The continuum limit of non-interacting dark matter particles is described by the collisionless Boltzmann equation coupled to Poisson's equation. This pair of equations has to be solved in an expanding background Universe dictated by the Friedmann equations, which are derived from the field equations of general relativity. Most cosmological simulations employ Newtonian rather than relativistic gravity, which provides a good approximation since linear structure growth is identical in
the matter dominated regime in the two theories, and non-linear large-scale structure induces velocities far below the
speed of light. Cosmological simulations are also typically performed with periodic boundary conditions to mimic the large-scale homogeneity and isotropy of the matter distribution of the Universe, i.e. the cosmological principle.\\

\footnotesize
\begin{tcolorbox}[enhanced,drop fuzzy shadow,colback=yellow!10!white,colframe=yellow!50!black,title={\bf Modeling dark matter}]
\vspace{-15pt}
\begin{flalign}
& \text{\bf collisionless Boltzmann equation: } \frac{\mathrm{d}f}{\mathrm{d}t} = \frac{\partial f}{\partial t} + {\bf v} \, \frac{\partial f}{\partial{\bf r}} -\frac{\partial\Phi}{\partial{\bf r}} \, \frac{\partial f}{\partial{\bf v}} = 0 
& &\text{\bf Poisson's equation: } {\bf \nabla}^2 \Phi = 4\pi G \int  f {\rm d}{\bf v}\nonumber
\end{flalign} 
\noindent\rule{\textwidth}{1pt}
\noindent The collisionless Boltzmann equation describes the evolution of the  phase-space density or distribution function of dark matter, $f=f({\bf r}, {\bf v}, t)$, under the influence of the collective gravitational potential, $\Phi$, given by Poisson's equation. The collisionless Boltzmann equation states the conservation of the local phase-space density; i.e. Liouville's theorem.
\end{tcolorbox}
\normalsize

\subsection{Numerical Techniques}

The high dimensionality of the collisionless Boltzmann equation prohibits efficient numerical solution methods based on standard discretization techniques for partial differential equations. Therefore, over the past decades, other numerical techniques have been developed to solve this problem more efficiently. An overview of some selected simulation codes and the employed dark matter simulation techniques is presented in Table 1.\\[-0.3cm]

\noindent {\bf The $\mathbf{N}$-Body method:} $N$-body methods are often employed to follow
the collisionless dynamics of dark matter, where the phase-space density is sampled by an ensemble of $N$ phase-space points ${\mathbf{r}_i
, \dot{\mathbf{r}}_i}$, $i = 1 \ldots N$ with masses $m_i$. The conservation of $f$ along the flow implies that
the masses $m_i$ remain unchanged along each trajectory. $N$-body methods therefore solve the collisionless Boltzmann equation by the method of characteristics. Alternatively, this method can also be interpreted as a Monte Carlo technique since any initial sample of $N$ phase-space points drawn from the same phase-space density at $t = 0$ results in an $N$-body model for the time evolution of $f(\mathbf{r}, \dot{\mathbf{r}}, t)$. The ensemble of all $N$ particles together represents the coarse grained phase-space density $\langle f \rangle \approx \sum_i m_i\,f({\bf r}_i(t),\dot{{\bf r}}_i(t))$. The latter represents a typical Monte Carlo estimate that can be applied also to other quantities, like the configuration space density. This sampling is subject to Poisson noise, and high particle numbers are therefore desirable to reduce noise in these estimates. To avoid unphysical two-body scatterings between nearby particles, gravitational interactions are
softened on small scales so that the particle collection represents a smoothed density field. A variety of kernel-based smoothing techniques are implemented, and some simulations also implement adaptive softening schemes to reduce the softening length in high density regions to reach higher spatial force resolution\cite{Price2007}. The main challenge of $N$-body simulations is to efficiently calculate the gravitational force that governs the motion of the dark matter sample particles. Once the forces
have been calculated, the particles are advanced based on symplectic integration schemes commonly implemented through a Leapfrog integrator. Symplectic integrators exactly solve an approximate
Hamiltonian such that the numerical time evolution is a canonical map and preserves certain conserved
quantities, such as the total angular momentum, and the phase-space volume. Cosmological simulations are further confronted with a large
dynamic range in timescales; i.e. in high-density regions orders of magnitude smaller
timesteps are required than in low-density regions. Integration schemes with individual timesteps
are therefore typically employed. The time integration is no longer
symplectic in a formal sense when individual short-range
timesteps are chosen for different particles. 
Methods to calculate gravitational forces of the $N$-body system can roughly be divided in two groups: approaches to accelerate the direct summation problem through approximations, or mesh-based methods to calculate the forces. The former
approaches aim for efficient numerical solutions of the integral form of Poisson's equation. The latter methods aim for efficient techniques to solve the differential form of Poisson's equation.  \\[-0.4cm]

\noindent {\bf \textit{Solving the integral form of Poisson's equation:}} 
The integral form of Poisson's equation, $\Phi({\bf r}) = -G\int\mathrm{d}{\bf r}^\prime\, {\rho({\bf r}^\prime)/|{\bf r}-{\bf r}^\prime|}$, can be translated to a discrete direct summation problem with complexity $\mathcal{O}(N^2)$. Solving this problem directly results in the so-called particle-particle scheme, and the earliest simulations employed this brute-force approach. 
The most common method to accelerate the direct summation through approximations is the
so-called tree approach\cite{Barnes1986}. Here, contributions to the gravitational potential from distant particles are
approximated by the lowest order terms in a multipole expansion of the mass distribution at a
coarse level of the tree reducing the computational cost to $\mathcal{O}(N\log N)$. The approximation used in the tree method is formally
obtained by Taylor expanding the force around some expansion center of the particle group. Often an octree is implemented in cosmological simulations, where each cubic
cell is split into up to eight child cells resulting in a
tree-like hierarchy of cubic nodes with the root node containing all particles at its bottom. The particles within each
of the tree nodes constitute a well-defined and localized group that build the basis for the tree force calculation. A further improvement to $\mathcal{O}(N)$ complexity is possible through the use of the fast multipole method, where forces are calculated between two tree nodes rather than between individual particles and nodes. This method is best implemented using a tree structure\cite{Dehnen2000}, although the original proposed method was based on a fixed mesh\cite{Greengard1987}.
Implementing periodic boundary conditions for these direct summation-based schemes typically requires Ewald summation techniques\cite{Hernquist1991} originally developed for solid-state physics\cite{Ewald1921}.\\[-0.4cm]

\noindent {\bf \textit{Solving the differential form of Poisson's equation:}}
Mesh-based methods aim to solve the differential form of Poisson's equation, ${\bf \nabla}^2\Phi({\bf r}) = 4\pi G\rho({\bf r})$.
This equation can be solved efficiently through fast Fourier transform-based methods, with Poisson's equation in Fourier space $k^2 \widetilde\Phi({\bf k}) = -4 \pi G \widetilde\rho({\bf k})$, leading to the so-called particle-mesh method\cite{Hockney1981}. 
To obtain forces, the
potential is then differentiated using a finite difference approximation and the forces are interpolated to the particle
positions.
The calculation of the gravitational forces via a fast Fourier transform  has only a $\mathcal{O}(N\log N)$ complexity, where $N$ is the number of mesh cells. The computational cost does not depend on
the details of the particle distribution, and no explicit force softening is necessary for this scheme since the force is automatically softened on the grid scale.
Combining the particle-mesh method with a set of nested grids of
increasing resolution enables an efficient force solver for inhomogeneous systems resulting in adaptive-mesh-refinement schemes. Multigrid or multilevel methods, which solve the discretized form of Poisson's equation using relaxation methods, such as Gauss-Seidel iterations, are also commonly employed\cite{Brandt1977}. The advantage of this technique over the fast Fourier transform approach is that the grid does not need to be equidistant, but can be locally adapted according to the particle density. The structure of such an adaptively refined mesh is identical to that of a shallow octree. \\[0.2cm]
\noindent {\bf \textit{Hybrid schemes:}} A variety of schemes combine direct summation-based techniques, for short range forces, with Fourier transform-based methods, for long range forces. The most basic example of this is the particle-particle plus particle-mesh method\cite{Efstathiou1985}. A common hybrid scheme is the tree-particle-mesh method \cite{Bode2003} where the direct summation for short range interactions is approximated by a tree-like method. Combinations of the multigrid method with the fast Fourier transform are also employed, where the Fourier transform is used as a force solver on the
coarsest grid \cite{Kravtsov1997}. Most modern simulations implement these hybrid solvers to achieve high efficiency.\\[-0.3cm]

\noindent {\bf Beyond $\mathbf{N}$-Body method:} Conceptually different methods to solve the collisionless Boltzmann equation have also been developed. However, none of these alternatives have so far been widely used for general structure formation simulations. These different methods are motivated, among others, by the desire to resolve the fine-grained structure of the phase-space density and to avoid numerical inaccuracies of the $N$-body approach like the artificial clumping of simulation particles for dark matter models with a cut-off in the initial power spectrum\cite{Wang2007}. Among the methodological alternatives to the $N$-body method are, for example, a reformulation of the Boltzmann-Poisson system as a Schr\"odinger equation \cite{Widrow1993,Schaller2014,Uhlemann2014}, the waterbag method \cite{Colombi2014,Colombi2015}, geodesic deviation equation-based methods\cite{Vogelsberger2008,Vogelsberger2011}, Lagrangian tessellation
techniques~\cite{Hahn2013}, and direct integration schemes using finite volume approaches based on positive flux conservation methods of plasma physics\cite{Yoshikawa2013}. 

\scriptsize
\begin{tcolorbox}[left=-4pt, title={\bf \hspace{0.1cm}Table 1: Major galaxy formation simulation codes},width=\textwidth]
\centering
\begin{tabular}{llllll}
\hline \\[-2ex]
\multicolumn{1}{l}{\bf code} &
\multicolumn{1}{l}{\bf gravity} &
\multicolumn{1}{l}{\bf hydrodynamics} &
\multicolumn{1}{l}{\bf parallelization} & {\bf code} &
\multicolumn{1}{l}{\bf primary} \\
\multicolumn{1}{l}{\bf name} &
\multicolumn{1}{l}{\bf treatment\footnote{\scriptsize PM: particle-mesh; TreePM: tree + PM, FM: fast multipole, P$^3$M: particle-particle-particle-mesh; ML: multilevel; MG: multigrid}} &
\multicolumn{1}{l}{\bf treatment\footnote{\scriptsize SPH: smoothed particle hydrodynamics, CRK-SPH: conservative reproducing kernel smoothed particle hydrodynamics , AMR: adaptive-mesh-refinement, MMFV: moving-mesh finite volume, MLFM/MLFV: mesh-free finite mass / finite volume}} &
\multicolumn{1}{l}{\bf technique\footnote{\scriptsize data-based: data parallelism focuses on distributing data across different nodes, which operate on the data in parallel; task-based:  task parallelism focuses on distributing tasks concurrently performed}} &
\multicolumn{1}{l}{\bf availability\footnote{\scriptsize private: private code; public: publicly available code (in some cases with limited functionality)}} &
\multicolumn{1}{l}{\bf reference} \\ [0.5ex]
\hline \\[-2ex]
ART \hspace{1.5cm} & PM/ML \hspace{1.6cm} & AMR \hspace{1.6cm} & data-based \hspace{1.6cm} & public \hspace{1.5cm} & Kravtsov~(1997) \cite{Kravtsov1997}\\
RAMSES & PM/ML & AMR &  data-based  & public & Teyssier~(2002) \cite{Teyssier2002} \\
GADGET-2/3 & TreePM & SPH & data-based  & public & Springel~(2005) \cite{Springel2005b} \\
Arepo  & TreePM & MMFV & data-based  & public & Springel~(2010) \cite{Springel2010} \\
Enzo & PM/MG & AMR & data-based  & public & Bryan et al.~(2014) \cite{Bryan2014} \\
ChaNGa\footnote{\scriptsize gravity solver is based on PKDGRAV3 \label{pkdgrav}} & Tree/FM & SPH & task-based & public & Menon et al.~(2015) \cite{Jetley2008,Gioachin2010,Menon2015} \\
GIZMO\footnote{\scriptsize based on the GADGET-3 code} & TreePM & MLFM/MLFV & data-based & public & Hopkins et al.~(2015) \cite{Hopkins2015} \\ 
HACC & TreePM/P$^3$M & CRK-SPH &  data-based & private & Habib et al. (2016)\cite{Habib2016}\\
PKDGRAV3 & Tree/FM & $-$ &  data-based & public & Potter et al. (2017)\cite{Potter2017}\\
Gasoline2 & Tree & SPH & task-based  & public & Wadsley et al.~(2017) \cite{Wadsley2017} \\
SWIFT & TreePM/FM & SPH &  task-based  & public & Schaller et al.~(2018) \cite{Schaller2018}\\ [0.5ex]
\hline
\end{tabular}
\end{tcolorbox}

\normalsize

\subsection{Some Key Results of $N$-body Simulations}
The earliest dark matter simulations studied halo population models\cite{Press1974},
the assembly of massive clusters\cite{White1976}, and the growth
of large-scale structure\cite{Aarseth1979, Efstathiou1979}.
Since then the resolution of these simulations has grown exponentially starting from a few thousand to multi-trillion particle simulations today \cite{Skillman2014}. Table 2 presents some selected recent structure and galaxy formation simulations. The findings of these simulations can roughly be divided in two categories: the large-scale distribution of dark matter and the structure of dark matter halos. The interaction between baryons and dark matter does affect the structure of dark matter on smaller scales, which is especially important for the internal structure of dark matter halos \cite{Gnedin2004,DiCintio2014b,Zhu2016,Benitez-Llambay2018,Bose2018,Katz2018,Read2019}. Studying these phenomena requires, however, simulations that model both dark matter and baryons.\\[-0.3cm]

\noindent {\bf The large-scale distribution of dark matter:} Cold dark matter simulations predict that the large-scale distribution of dark matter is not completely homogeneous, but instead exhibits a web-like structure consisting of  voids, walls, filaments, and halos quantified through, among others, the halo mass and matter correlation functions. \\[-0.4cm]

\noindent {\bf \textit{Halo mass function:}} The halo mass function quantifies the comoving number density of dark matter halos as a function of their virial mass, $M_{\rm vir}$, typically defined as the mass, $M_{200}$, enclosed within a radius
$r_{200}$ containing a mean density $200$ times the critical density of the Universe. Recently, also other halo boundary definitions like the splashback radius\cite{Diemer2014,More2015}, which corresponds to the outermost caustic originally discussed in symmetric analytic halo formation models\cite{Fillmore1984,Bertschinger1985}, have been proposed to avoid, for example, the pseudo-evolution of the halo mass and radius\cite{Diemer2013}. In simulations, dark matter halos are identified through cluster finding methods like the friend-of-friends algorithm\cite{Davis1985} and extensions of this based on gravitational unbinding\cite{Springel2001} or phase-space structure finding taking into account also velocity space information\cite{Behroozi2013}. 
In the cold dark matter cosmogony, structure forms through the hierarchical merging of dark matter halos\cite{Davis1985}, and the corresponding evolution of the halo mass function has been
studied extensively\cite{Press1974,Bond1991,Jenkins2001,Sheth2002,White2002,Reed2003,Warren2006,Reed2007,Tinker2008}.
Most importantly these studies revealed that the low-mass end of the halo mass function has a power law slope close to $-2$. Furthermore, the high mass end of the halo mass function is exponentially suppressed. 
The halo mass function is also an important probe of the nature of dark matter since many particle candidates predict strong, scale-dependent deviations from the expectations of the cold dark matter model\cite{Angulo2013,Schneider2015}. The high mass shape
and evolution of the halo mass functions also constrains cosmological parameters \cite{Eke1996}. 
Simulation-based empirical halo mass functions are often expressed as $M{\rm d n}/{\rm d}M = \rho_0 \, {\rm d}\ln \sigma^{-1}/{\rm d}M f(\sigma(M))$, where $\rho_0$ is the mean mass density of the Universe, $\sigma(M)$ is
the variance of the linear density field within a top-hat filter containing mass $M$, and $f(\sigma)$ is a function that is determined empirically by fitting the simulation results.
This functional form of the halo mass function is motivated and also predicted by the analytic Press-Schechter
model\cite{Press1974}. However, the shape of $f(\sigma)$ found in simulations differs significantly from the analytic model originally proposed, agreeing much better with a version motivated
by ellipsoidal rather than spherical collapse\cite{Sheth2001}.
The detailed form of $f(\sigma)$ depends, among others, on simulation details and halo mass definitions, and a variety of empirical fitting functions have been published\cite{Tinker2008, Angulo2012,Watson2013, Heitmann2015, Bocquet2016}. \\[-0.4cm]

\noindent {\bf \textit{Dark matter distribution:}} A major success of cold dark matter simulations is their ability to predict the matter distribution on large scales\cite{Springel2006}, which is described through the two-point correlation function $\xi(r)$. For a set of points this function is defined as $\xi(r) = \langle N_p\rangle / N_m - 1$, where $\langle N_p\rangle$ indicates the average number of pairs in a thin shell of radius $r$ centered on one point of the set and $N_m$ the expected number of pairs in the same shell given a uniform distribution of points. Although this function can be estimated analytically in the linear regime, dark matter simulations are needed to probe its evolution into the non-linear regime. The dark matter correlation function signal grows with time and develops a characteristic shoulder at small scales\cite{Springel2018}. This effect can be explained by the relative contribution of the one-halo term, i.e. pairs composed of particles within the same halo, and the two-halo term, i.e. pairs formed by particles in different halos, to the clustering signal\cite{Cooray2002}. Finally, the dark matter correlation function has a markedly different shape than the galaxy correlation function. The latter has a power law shape over a significant range of scales and an amplitude nearly constant at all redshifts\cite{Springel2006}. Indeed, galaxies trace the highest peaks of the dark matter distribution, and their clustering does not change significantly with time, as more and more dark matter structures grow. This bias needs to be taken properly into account when estimating the large-scale total matter distribution using galaxy tracers\cite{Benson2000}.\\[-0.3cm]

\noindent{\bf Structure of dark matter halos:} Cold dark matter simulations have also established multiple characteristics of the dark matter distribution within collapsed and virialized dark matter halos. This has most importantly led to the discovery of a nearly universal radial density profile of dark matter halos.  \\[-0.4cm]

\noindent {\bf \textit{Internal halo structure:}} The dark matter mass distribution within halos is well described by a near-universal spherically averaged density profile, the so-called Navarro-Frenk-White profile\cite{Navarro1996,Navarro1997}: $\rho(r)=\rho_s/[(r/r_{\rm s})(1+(r/r_{\rm s}))^2]$ with a characteristic density $\rho_s$, and a transition radius $r_s$.
This form of density profiles 
has been shown to arise also in the absence of hierarchical growth like, for example, in hot dark matter models\cite{Wang2007}, or models with truncated initial power spectra\cite{Moore1999}. The central slope of dark matter halos has been debated for a long time and is also affected by baryonic physics effects that require hydrodynamical simulations. 
More recent higher resolution simulations found a central slope shallower than $-1$, indicating that the density profile is better described by a functional form with a gradually changing slope profile\cite{Navarro2004}: $\ln(\rho(r)/\rho_{-2})=(-2/\alpha)[(r/r_{-2})^\alpha-1]$ with slope $\alpha$ and transition radius $r_{-2}$.
This profile had previously been used to fit
star counts in the Milky Way\cite{Einasto1965}, and is known as 
the Einasto profile.
The adjustable shape parameter, $\alpha$, shows considerable scatter but increases systematically with halo mass at $z = 0$.
The ratio of the virial radius, $r_{\rm vir}$, and the transition radius, $r_{\rm s}$, is called the concentration parameter, $c$, that correlates with the mass of the halo leading to the mass-concentration relation\cite{Ludlow2014,Ludlow2016,Klypin2016} ($c \propto M^{-\delta}, \delta \approx 0.1$). 
Simulations demonstrated that the dependence of halo concentration on mass, initial fluctuation spectrum and cosmological parameters all reflect a dependence of concentration on
the actual halo formation time\cite{Navarro1997}. Specifically, lower mass halos typically assemble earlier, and thus have higher concentration, due to the higher density of the Universe at the time of their formation.
The shapes of halos have also been studied, and those depart from sphericity, with halos typically being prolate
and increasingly so towards their centers. Major-to-minor axis ratios of two or greater are
not uncommon, and more massive halos tend to be less spherical than lower mass halos\cite{Jing2002,Allgood2006,Bett2007}. The exact shapes of dark matter halos also depend on the dark matter particle physics model. Simulations also provide information on the velocity structure of halos. The averaged velocity anisotropy profile, $\beta(r)=1-0.5\sigma_{\rm t}^2/\sigma_{\rm r}^2$, grows from zero, i.e. isotropic, to about 0.5, i.e. mild radial anisotropy, towards the outer regions\cite{Diemand2007,Navarro2010}. Here, $\sigma_{\rm t}$ denotes the tangential and $\sigma_{\rm r}$ the radial velocity dispersion, with the total velocity dispersion being $\sigma^2 = \sigma_{\rm t}^2 + \sigma_{\rm r}^2$.
A $\beta$ value of $1$ and $\beta \rightarrow -\infty$ correspond to
systems where dark matter particles have purely radial and
purely circular orbits, respectively. 
Simulated dark matter halos therefore turn out to be almost
isotropic in their inner regions and to be somewhat radially biased at larger radii. 
Although both $\rho(r)$ and $\sigma(r)$ are not close to a
power law, the combination $f(r) = \rho(r)/\sigma^3(r)$ ,
also called pseudo-phase-space density, is remarkably close to a power law, with slope  $\approx -1.875$\cite{Navarro2010}. This power law index is identical to that of solutions for self-similar infall onto a point mass from an otherwise uniform
Einstein-de Sitter Universe\cite{Bertschinger1985}. \\[-0.4cm]

\noindent {\bf \textit{Halo substructure:}} As the resolution of dark matter simulations increased, halos within halos, so-called subhalos, could be resolved\cite{Ghigna1998,Moore1999}. Subhalos have cuspy, Navarro-Frenk-White-like density profiles but they tend to be less extended than comparable halos in the field due to tidal stripping\cite{Diemand2007,Springel2008}. Bound subhalos with $\mu = M_{\rm sub}/M_{\rm vir} > 10^{-7}$ contain
about $10\%$ of the halo mass within the virial radius \cite{Springel2008}. Lower mass halos tend to have fewer subhalos and lower subhalo mass fractions at a given $\mu$.  This shift is due to the difference in the relative dynamical age of halos; e.g. substructure is more effectively destroyed by tides
in older, galactic halos compared to more massive galaxy cluster halos. The cumulative subhalo mass function is a power law $N(> \mu) \propto \mu^{-s}$ for $\mu \ll 1$, with $s$ close to the critical value of unity\cite{Springel2008,Gao2012}. For $s=1$
each logarithmic mass bin contributes equally to the total
mass in substructure. This is logarithmically divergent as $\mu$
approaches zero, and implies that a significant fraction of
the mass could, in principle, be locked in halos too small
to be resolved by the simulations. This can, for example, have important implications for the prediction of dark matter annihilation signals since these small unresolved halos can boost the overall resolved annihilation emission\cite{Springel2008b}. The abundance of subhalos also varies systematically
with other properties of the parent halo, like, for example, the concentration leading to a lower amount of substructure with increasing halo concentration\cite{Gao2004}. The radial distribution of subhalos varies only little with the mass or
concentration of the parent halo. It is much less centrally
concentrated than the overall dark matter profile\cite{Springel2008}. 

\section{Simulating Baryons}
Dark matter and dark energy dominate the energy budget of the Universe, but the visible components of galaxies consist of baryons. Simulating baryons is therefore crucial to make predictions for the visible Universe. Initially,
the baryon component is solely comprised of gas, mostly hydrogen and helium. Some of this gas eventually turns into stars during structure formation.  
Astrophysical gases in cosmological simulations are typically described as inviscid ideal gases following the Euler equations, which can be expressed in different forms leading to different numerical discretization schemes. Hydrodynamics in cosmological simulations is numerically demanding due to the large dynamic range, highly supersonic flows, and large Reynolds numbers.\\

\footnotesize
\begin{tcolorbox}[enhanced,drop fuzzy shadow,colback=yellow!10!white,colframe=yellow!50!black,title={\bf Modeling cosmic gas}]
\vspace{-15pt}
\begin{flalign}
&\text{\bf Eulerian formulation:} 
& &\text{\bf Lagrangian formulation:}
& &\text{\bf Arbitrary Lagrangian-Eulerian formulation:} \nonumber \\
&\frac{\partial\rho}{\partial t} + {\bf \nabla} \cdot (\rho {\bf v}) = 0   
& &\frac{\mathrm{D}\rho}{\mathrm{D} t}  = -\rho{\bf \nabla}\cdot{\bf v} 
& & \frac{{\rm d}}{{\rm d}t} \int_V \rho {\rm d}V = - \int_S\rho ({\bf v}-{\bf w}) \cdot {\bf n} {\rm d}S\nonumber \\
&\frac{\partial \rho {\bf v}}{\partial t} + {\bf \nabla} \cdot (\rho {\bf v} \otimes {\bf v} + P\mathbb{1}) = 0 
& &\frac{\mathrm{D}{\bf v}}{\mathrm{D} t} = - \frac{1}{\rho}{\bf \nabla} P  
& & \frac{{\rm d}}{{\rm d}t} \int_V \rho {\bf v} {\rm d}V = -\int_S \rho {\bf v} ({\bf v} - {\bf w}) \cdot {\bf n} {\rm d}S - \int_S P {\bf n} {\rm d}S\nonumber \\
&\frac{\partial \rho e}{\partial t} + {\bf \nabla} \cdot (\rho e+P) {\bf v} = 0 
& &\frac{\mathrm{D} e}{\mathrm{D} t} = - \frac{1}{\rho}\nabla \cdot P {\bf v} 
& & \frac{{\rm d}}{{\rm d}t} \int_V  \rho e {\rm d}V = -\int_S \rho e ({\bf v} - {\bf w}) \cdot {\bf n} {\rm d}S - \int_S P {\bf v}\cdot {\bf n}{\rm d}S\nonumber
\end{flalign}
\noindent\rule{\textwidth}{1pt}
\noindent Different forms of the hydrodynamical equations. ${\rm D}/{\rm d}t\equiv\partial /\partial t + {\bf v}\cdot \nabla$ denotes the Lagrangian derivative and ${e=u + {\bf v}^2/2}$ the total energy per unit mass. The equations are closed through $P = (\gamma - 1)\rho u$ with $\gamma=5/3$.  For the arbitrary Lagrangian-Eulerian formulation the grid moves with velocity ${\bf w}$ and cell volumes evolve as ${{\rm d} V/{\rm d}t =\int_V (\nabla \cdot {\bf w}) \, {\rm d}V}$.
\end{tcolorbox}
\normalsize

\subsection{Numerical Techniques}
The hydrodynamical equations can be discretized in different ways employing methods that roughly fall into three classes: Lagrangian, Eulerian or arbitrary Lagrange-Eulerian techniques. 
The Lagrangian specification of the field assumes an observer that follows an individual fluid parcel, with its own properties like density, as it moves through space and time. The Eulerian specification, on the other hand, focuses on specific locations in space through which the fluid flows as time passes. In addition, numerical approaches can also be distinguished between mesh-free and mesh-based algorithms. Mesh-free methods do not require connections between nodes, but are rather based on interactions of each node with its neighbors. An overview of some selected simulation codes and the employed hydrodynamical simulation techniques is shown in Table 1.\\[-0.3cm]

\noindent {\bf Eulerian Methods:} Eulerian methods are the traditional method to solve the system of
hyperbolic partial differential equations that constitute ideal hydrodynamics. 
The most common approaches include finite volume, finite difference, finite element, spectral or wavelet methods. For example, accurate Godunov finite volume schemes offer high-order spatial accuracy, have negligible post-shock oscillations and low numerical diffusivity. For these methods a Riemann problem is solved across cell faces, which yields the required fluxes at each cell face to  update the conserved quantities. 
If the cell is assumed to have uniform
properties, this is called a first-order Godunov solver. Modern implementations employ
parabolic interpolation, known as the piecewise parabolic method \cite{Colella1984,Woodward1986}. The large dynamic range of cosmological simulations requires adaptive meshes, where the mesh size can be reduced based on some refinement criterion. This leads to the class of adaptive-mesh-refinement schemes\cite{Berger1984,Berger1989,Klein1994,Bryan1995}, which were first developed for solving general problems involving hyperbolic partial differential equations, and then later were also applied to cosmological simulations.
Recently also discontinuous Galerkin methods\cite{Cockburn1989,Mocz2014b,Guillet2019} became more popular in computational astrophysics since
they offer a framework for discretizing hyperbolic problems at any order of spatial accuracy, together with attractive data locality by combining  features of spectral element
and finite volume methods.  \\[-0.3cm]

\noindent {\bf Lagrangian Methods:} Smoothed particle hydrodynamics is a widely used mesh-free Lagrangian technique for approximating the continuum dynamics of fluids through the use of sampling particles, which may also be viewed as interpolation points, following the equations of motion derived from the hydrodynamical equations\cite{Lucy1977,Gingold1977,SpringelSPH,Price2012}. Energy, linear momentum, angular momentum, mass, and entropy, assuming no artificial viscosity operates, are all simultaneously conserved.  The local resolution follows the mass flow, which is convenient to represent large density contrasts. Over the last years various improved formulations of the smoothed particle hydrodynamics method have been developed and applied to cosmological simulations \cite{Read2012,Hopkins2013,Keller2014,Schaye2015,Wang2015,Barnes2017}.
A few cosmological simulations have also employed Lagrangian mesh-based hydrodynamics schemes, which are based on grid deformation techniques\cite{Gnedin1995,Pen1998}. However, mesh-tangling effects are a major problem of such multi-dimensional mesh-based Lagrangian hydrodynamics methods.\\[-0.3cm]

\noindent {\bf Arbitrary Lagrangian-Eulerian Methods:} 
For arbitrary Lagrangian-Eulerian methods, the grid velocity can be freely chosen. For astrophysical applications, such a scheme has recently been realized through a Voronoi tessellation of a set of discrete mesh-generating points, which are allowed to move freely\cite{Springel2010}. A finite volume hydrodynamic scheme
with the Voronoi cells as control volumes can then be
consistently defined. Most importantly, due to the mathematical properties of the Voronoi tessellation, the mesh
continuously deforms and changes its topology as a result
of the point motion, without ever leading to problematic
mesh-tangling effects. Similar methods have over the last years also been implemented in other simulation codes\cite{Duffell2011,Vandenbroucke2016}. Most recently new types of arbitrary Lagrangian-Eulerian, mesh-free, finite mass and finite volume
methods have been successfully applied to astrophysical and galaxy formation problems\cite{Hopkins2015}.

\subsection{Baryonic Physics}\label{sec:baryons}
The hydrodynamical equations have to be complemented by various astrophysical processes that shape the galaxy population. Most of these processes are implemented through effective, so-called sub-resolution models, which are necessary due to the limited numerical resolution of simulations.\\[-0.3cm]

\noindent{\bf Gas cooling:} Gas dissipates its internal energy through cooling processes, like 
collisional excitation and ionization, inverse Compton, recombination and free-free emission. Cooling processes are coupled to the energy equation using cooling functions that are either tabulated or extracted from chemical networks.  Cosmological simulations often assume
that the gas is optically thin and in ionization equilibrium and neglect three-body processes that are typically unimportant. In addition to primordial cooling also cooling due to heavy elements, so-called metals, is important. Metal line cooling dominates for temperatures $10^5 \lesssim T \lesssim 10^7\,{\rm K}$. Early simulations typically employed
cooling rates assuming collisional ionization equilibrium\cite{Sutherland1993}, but
most later galaxy formation models account for the photo-ionization of
metals by the metagalactic radiation field\cite{Wiersma2009}. For most post-reionization simulations this metagalactic radiation field is assumed to be spatially uniform but time-dependent\cite{Haardt2012}. Simulations that resolve the cold phase of the interstellar medium also include gas cooling below $10^4\,{\rm K}$ via fine-structure and molecular cooling. In neutral atomic gas, the efficiency of cooling is sensitive to the residual ionization degree. In molecular gas ($n \gsim 100\,{\rm cm^{-3}}, T \lsim 50\,{\rm K}$), the CO molecule dominates the cooling at low densities while at higher densities \ion{C}{i}, ${\rm O_2}$ and ${\rm H_2 O}$ start to contribute\cite{Tielens2010}. Gas cooling is a direct physical process that is not implemented through a sub-resolution model. However, following all cooling processes in detail requires sufficient numerical resolution to resolve the different gas phases. \\[-0.3cm]

\noindent{\bf Interstellar medium:} Carefully modeling the interstellar medium is important since its properties directly impact star formation. However, simulating the interstellar medium is challenging due to its complex multi-phase structure including magnetic fields and relativistic particles. Especially modeling the cold phase is technically difficult because of the short timescales associated with the dense gas. These timescales require very small time-steps to reliably follow the cold gas evolution. Moreover, the implementation of additional physical processes is needed to accurately model such a phase. To circumvent this problem, this dense gas phase is often not directly modeled but rather described by an effective polytropic equation of state\cite{Springel2003, Agertz2011, DallaVecchia2012}; i.e. $T \propto \rho^{\gamma(\rho)}$ , which naturally emerges from an equilibrium two-phase interstellar medium where a hot, supernova-heated and volume-filling phase co-exists with a colder phase containing the bulk of the mass\cite{Springel2003}. More recent modeling efforts started to abandon the effective equation of state approach and instead aimed towards resolving the multi-phase structure directly. Such simulations are starting to be able to
resolve the Jeans mass of gas, corresponding to the scale of molecular cloud complexes. Therefore, a more direct modeling of the multi-phase interstellar medium is possible\cite{Hopkins2012}.
In such simulations, the gas density and temperature distributions follow a multi-modal distribution\cite{Hopkins2012,Agertz2013,Rosdahl2015}. Generally, the cold gas phase dominates ($\sim 90\%$) the gas mass budget, but occupies a very small volume fraction ($\sim 1\%$), which is mostly comprised of hotter gas \cite{Emerick2019}.
Simulating the molecular phase of the interstellar medium is challenging because it requires detailed modeling of the interaction between gas, dust, and radiation, which tends to destroy molecules unless gas is able to effectively self-shield from ionizing radiation\cite{Krumholz2011}. Detailed models of the interstellar medium have also to take into account the various feedback sources that ultimately shape the structure of the interstellar medium. Thus, future simulations have to consider how this complex interplay of such a wide range of physical processes affects the properties of the interstellar medium. \\[-0.3cm]

\noindent{\bf Star formation:} Cold and dense gas eventually forms stars, and simulations therefore transform a portion of this gas into  collisionless star particles, representing co-eval, single-metallicity stellar populations described by an underlying initial stellar mass function. 
Observations support a nearly universal star formation efficiency in molecular gas, where
about $1\%$ of the gas is converted into stars per free fall time\cite{Bigiel2011, Krumholz2012}.
Based on a calculated star formation rate, the gas is converted into star particles typically using a probabilistic sampling scheme. The star formation rate is usually computed based on a Kennicutt-Schmidt type  relation as $d M_{\star} / dt = \epsilon M_{\rm g} / t_{\rm ff}$, where $M_{\rm g}$ is the gas cell/particle mass, $t_{\rm ff}$ is the gravitational free fall time and $\epsilon$ is a conversion efficiency typically in the range $0.01-1$\cite{Springel2003, Hopkins2014, Hopkins2018}. However, not all the gas elements are eligible for star formation. Commonly adopted criteria are based on: a density threshold \cite{Springel2003,Stinson2006,Teyssier2013,Hopkins2014,Vogelsberger2014,Schaye2015,Wang2015,Hopkins2018,Pillepich2018}, restricting star formation to gravitationally bound regions identified via the virial parameter  -- that quantifies the degree of pressure support against gravitational collapse\cite{Semenov2017,Hopkins2018}, Jeans length-based criteria -- that is gas must be prone to gravitational instability \cite{Teyssier2013,Trebitsch2017,Hopkins2018,Rosdahl2018}, restricting star formation to the molecular gas phase\cite{Gnedin2011,Christensen2012,Feldmann2012,Kuhlen2012,Monaco2012,Hopkins2014,Hopkins2018} or converging flows ($\nabla \cdot {\bf v} < 0$ \cite{Stinson2006}).
Alternative to the probabilistic sampling scheme, and to better model the clustered nature of star formation, a few simulations also consider star clusters as the unit of star formation by allowing the growth of star particles through accretion from the ambient medium\cite{Li2017}.
Once stellar particles have been formed, modern galaxy formation models also track the stellar evolution and mass return of these stars to the gas component. This leads to an enrichment of the gas with metals. Early models tracked only Type II supernova enrichment, but recent models also follow asymptotic giant branch stars\cite{Wiersma2009a}, Type Ia supernovae, which are important for iron enrichment \cite{Vogelsberger2013}, and neutron star mergers for r-process element enrichment\cite{Naiman2018}. The actual enrichment is based on metal yield models derived from detailed stellar evolution calculations. These yields are however still rather uncertain, at least by a factor of two, particularly at low metallicities and for more massive stars. This uncertainty then propagates into predictions for metal abundances in simulations. Future cosmological simulations will still have to implement star formation as sub-resolution models with individual stars as their building blocks.    \\[-0.3cm]

\noindent{\bf Stellar feedback:} Stars interact with their surrounding gas through the injection of energy and momentum leading to a feedback loop regulating star formation. To regulate star formation, stellar feedback must be efficient in launching galactic-scale outflows to eject gas from galaxies, and a plethora of sub-resolution schemes exists to achieve an efficient generation of galactic winds. Those differ in the way energy and momentum, most notably in the form of supernova explosions, are coupled to the surrounding gas. Essentially the energy can be deposited thermally or kinetically. In the first case, excessive radiative gas cooling must be avoided. While cooling in dense and cold gas is physically expected, at the comparatively low resolution of cosmological simulations it cannot be modeled reliably. The result is then an artificial excessive cooling of the gas, which leads to the unphysical loss of the supernova feedback energy via radiation and greatly reduces its effectiveness. Some approaches therefore disable the radiative cooling of the affected gas for a prescribed amount of time ($\sim 10^7\,{\rm yr}$)\cite{Stinson2006}, or heat the gas probabilistically to reach high enough temperatures ($T \sim 10^6\,{\rm K}$) for radiative cooling to become ineffective on time scales of $\sim 10^7\,{\rm yr}$\cite{DallaVecchia2012}. In the second case, kinetic energy cannot be radiated away until it thermalizes. However, the use of hydrodynamically-decoupled wind particles, to realize a non-local injection of momentum in the gas surrounding active star forming regions, can still be necessary to obtain large-scale galactic outflows \cite{Springel2003,Oppenheimer2010,Vogelsberger2013, Pillepich2018}. Recently, more explicit models for stellar feedback have been developed. In addition to supernova feedback they also take into account other feedback channels, such as energy and momentum injection by stellar winds and photoionization and radiation pressure due to radiation emitted by young, massive stars \cite{Stinson2013a, Agertz2013,Hopkins2014,Hopkins2018,Smith2018}. The combination of these processes then leads to a regulation of star formation to the observed low gas to star conversion efficiency of 1\% per free-fall time\cite{Bigiel2011, Krumholz2012}.
Stellar feedback must be efficient in launching galactic-scale outflows to eject gas from galaxies, thereby also explaining the low baryon retention fraction in galaxies\cite{Behroozi2010,Moster2013}. Recent explicit feedback models can make direct predictions for the outflow rates of these outflows\cite{Muratov2015}, whereas older models typically prescribe the mass loading of these galactic-scale outflows close to the galaxies. Sub-resolution models of stellar feedback vary widely among different galaxy formation models. More work is required to understand in detail which stellar feedback channels are most important for shaping the different types of galaxies. \\[-0.3cm]

\noindent{\bf Supermassive black holes:} 
Supermassive black holes are observed in
massive galaxies \cite{Gehren1984,Kormendy1995}, in small, bulge-less disc
galaxies \cite{Filippenko2003,Shields2008} as well as
in dwarf galaxies \cite{Reines2011,Moran2014}.  
Simulations therefore include models for supermassive black holes, and numerically seed them typically
in dark matter haloes with masses $\gtrsim 10^{10}-10^{11}\,{\rm M}_\odot$ since the true seeds cannot be resolved, and their origin is not yet fully understood. They then accrete mass often based on an Eddington-rate-capped Bondi-Hoyle-like accretion rate: $\dot M_{\rm BH}= (4\pi G^2 M_{\rm BH}^2 \rho) / (c_{\rm s}^2 + v_{\rm rel}^2)^{3/2}$, where $\rho$ and $c_{\rm s}$ are the gas density and gas sound speed, respectively, and $v_{\rm rel}$ denotes the relative velocity between the gas and the black hole. Depending on the numerical resolution this accretion rate is sometimes artificially increased, possibly in a density-dependent fashion, to compensate for the inability of simulations to resolve the multi-phase structure of gas\cite{Booth2009}. Many simulations also explored variations of the Bondi-Hoyle model to overcome its limitations. The Bondi model, for example, implicitly assumes that the
accreting gas has negligible angular momentum, which is most likely unrealistic. 
Some models therefore assume that black holes might be
primarily fed by gas driven to the centers by gravitational torques from non-axisymmetric
perturbations\cite{Shlosman1989}, which have more recently been explored in simulations\cite{Hopkins2011b,Bournaud2011,Gabor2013,AnglesAlcazar2013, AnglesAlcazar2017, Dave2019}.
Black holes also grow through mergers, which are modeled in cosmological simulations as well. Due to resolution limitations, general relativistic effects are not taken into account and it is assumed that the black holes of the two galaxies merge instantly once they come close enough, i.e. within their numerical accretion radius, which is typically calculated based on a local gas resolution element nearest neighbor search.\\[-0.3cm]

\noindent{\bf Feedback from active galactic nuclei:}
Active galactic nuclei are related to
observational phenomena associated with accreting supermassive black holes
including electromagnetic radiation, relativistic jets, and less-collimated non-relativistic outflows\cite{Krolik1999}. The resulting energy and momentum couple with the surrounding gas leading
to a regulation of black hole growth and star formation in more massive halos ($M \gsim 10^{12}\,{\rm M}_\odot$). This feedback is commonly divided in two modes that are implemented differently in simulations: quasar and radio mode. However, some galaxy formation models do not make this distinction arguing that
cosmological simulations lack the resolution to properly distinguish the two feedback modes, and to
limit the number of feedback channels to the minimum required to match the observational data\cite{Schaye2015}. Quasar mode feedback is associated with the radiatively efficient mode of black hole growth and is often implemented through energy or momentum injection assuming that the bolometric luminosity is proportional to the accretion rate, and a fixed fraction of this luminosity is deposited into the neighboring gas\cite{Springel2005, DiMatteo2005}. Recent
works have also implemented momentum-driven winds via radiation pressure on dust\cite{Debuhr2010,Costa2018,Barnes2018} and via broad-line-region winds\cite{Choi2012}. Radio mode feedback is caused by highly-collimated jets
of relativistic particles, which are often associated with X-ray bubbles with enough energy to offset cooling losses. Therefore, this feedback mode is assumed to be important for the regulation of star formation in massive galaxies. Radio mode feedback is often implemented as a second sub-resolution feedback channel once the accretion rate is below a critical value\cite{Sijacki2007,Weinberger2017}. Jets themselves cover an
enormous dynamic range, being launched
at several Schwarzschild radii, and propagating outwards to
tens of ${\rm kpc}$. Directly resolving them in detail in cosmological simulations is therefore currently not feasible.  The sub-resolution models for supermassive black holes are therefore still very uncertain since they have to bridge a very large scale gap between the actual accretion and feedback, and the scales that can be resolved with simulations.\\[-0.3cm]

\noindent{\bf Magnetic fields:} Magnetic fields permeate the Universe on all scales and impact the motion of ionized gas. Conversely, gas dynamics affects the topology and strength of magnetic fields. Cosmological simulations typically employ the ideal magnetohydrodynamics approach, which is a good approximation for cosmological magnetic fields. This approach assumes that the plasma is perfectly conducting and that relativistic effects, i.e. terms $\propto (v/c)^2$ such as the displacement current $c^{-1}\partial{\bf E} / \partial t$, are negligible. However, for other situations the ideal magnetohydrodynamics  approximation breaks down and non-ideal terms, such as ohmic resistivity, ambipolar diffusion and the Hall effect, must be taken into account. These effects are important especially at very small spatial scales, e.g. for individual star formation, causing a diffusion of the magnetic field. 
On large cosmological scales the impact of magnetic fields on the dynamics of gas is rather limited\cite{Marinacci2016}. However, magnetic fields are an essential constituent of the interstellar medium, providing both pressure support against gravity\cite{Ferriere2001} and influencing the propagation of cosmic rays \cite{Kotera2011}. Cosmological simulations including magnetic fields through the ideal magnetohydrodynamics are typically initialized with a certain magnetic seed field, since the approximations and assumptions of ideal magnetohydrodynamics do not permit the self-consistent generation of magnetic fields. Some simulations also consider source terms like the Biermann battery effect or field injection from stellar winds as the source for initial magnetic fields\cite{Donnert2009}. In most cases, the initial conditions of such cosmological simulations contain however a small seed field of the order of roughly $10^{-10}\,{\rm Gauss}$ at a redshift of around $z\sim100$. The simulation results are not sensitive to this seed field as long as its value is not significantly too large, close to violating observational constraints\cite{Marinacci2016,Marinacci2018}, or vanishingly small. The reason for this insensitivity lies in the strong amplification processes that occur during structure formation. 
This amplification typically occurs in two phases. At high redshifts,  a turbulent dynamo leads to an exponential amplification of the magnetic fields in halos. Once the initial turbulent amplification phase has saturated,
a second phase of magnetic field amplification starts leading to
a linear growth caused by a galactic dynamo\cite{Pakmor2017}. The numerical discretization of the ideal magnetohydrodynamics  equations is challenging because of the solenoidal constraint $\nabla \cdot {\bf B} = 0$. 
Two main families of discretization techniques exist: divergence cleaning schemes and constrained transport.
For the cleaning approach, source terms are added to the underlying magnetohydrodynamics  equations to correct for divergence errors\cite{Powell1999,Dedner2002}. 
Constrained transport discretizations\cite{Evans1988}, on the other hand, guarantee that the divergence is zero by construction. However, a more complex implementation is required in that case. For instance, either vector potentials\cite{Mocz2016}, Euler potentials\cite{Rosswog2007,Dolag2009} or staggered discretizations of the magnetic field components\cite{Stone1992,Londrillo2004,Fromang2006,Teyssier2006,Balsara2012} must be employed.\\

\footnotesize
\begin{tcolorbox}[enhanced,drop fuzzy shadow,colback=yellow!10!white,colframe=yellow!50!black,title={\bf Modeling cosmic magnetic fields}]
\vspace{-15pt}
\begin{flalign}
&\text{\bf Ideal magnetohydrodynamics equations:} & & \text{\bf MHD Maxwell equations:} \nonumber \\
&\frac{\partial \rho}{\partial t} + {\bf \nabla} \cdot (\rho {\bf v}) = 0 & & \nabla \times {\bf B} = \frac{4\pi}{c}{\bf J} & & \nabla\cdot{\bf B} = 0 \nonumber \\
&\frac{\partial \rho {\bf v}}{\partial t} + {\bf \nabla} \cdot (\rho {\bf v} \otimes {\bf v} + P\mathbb{1}) = \frac{{\bf J} \times {\bf B}}{c} & & \nonumber \frac{1}{c}\frac{\partial {\bf B}}{\partial t} + \nabla \times {\bf E} = 0 & & {\bf E} = -\frac{{\bf v} \times {\bf B}}{c} \nonumber\\
&\frac{\partial (\rho e + e_{\rm B})}{\partial t} + {\bf \nabla} \cdot \left[(\rho e + P) {\bf v} + c\frac{{\bf E} \times {\bf B}}{4\pi}\right] = 0  \nonumber 
\end{flalign}
\noindent\rule{\textwidth}{1pt}
\noindent The evolution of the magnetic field, ${\bf B}$, is given by the induction equation, $\partial{\bf B}/\partial t = \nabla\times({\bf v}\times {\bf B})$. Magnetic fields act on gas through the Lorentz force, ${\bf J} \times {\bf B} / c$ with the current density, ${\bf J} = c \nabla \times {\bf B} / (4\pi)$. The energy equation contains the magnetic energy density, $e_{\rm B} = ||{\bf B}||^2 / 8\pi$, and the Poynting vector, $c({\bf E}\times{\bf B} / 4\pi)$, in the flux part. 
\end{tcolorbox}
\normalsize

\noindent{\bf Cosmic rays:} Relativistic nuclei and electrons, known as cosmic rays, are another important component of the galactic ecosystem. They are accelerated through diffusive shock acceleration mostly in supernova remnants and jets of active galactic nuclei (first-order Fermi acceleration) and turbulence (second-order Fermi acceleration). Cosmic rays contribute to the pressure in the interstellar medium\cite{Ferriere2003,Cox2005}, provide an important heating channel\cite{Field1969,Wolfire1995}, and potentially play a role in driving galactic gas outflows\cite{Uhlig2012,Booth2013,Hanasz2013,Salem2014,Pakmor2016,Simpson2016,Ruszkowski2017,Farber2018,Girichidis2018,Jacob2018} due to their shallow equation of state ($P_{\rm cr} \propto \rho_{\rm cr}^{4/3}$), their long cooling time, and their ability to transfer energy to outflows outside of star-forming discs\cite{Pfrommer2017}. The propagation of cosmic rays is dictated by the strength and topology of the underlying magnetic fields.

Reliably modeling the propagation of cosmic rays therefore requires a detailed modeling of magnetic fields. To capture all these effects self-consistently, the injection, acceleration and the transport of cosmic rays, through anisotropic diffusion and streaming, must be included in simulations. This requires, in principle, a detailed knowledge of the cosmic ray energy spectrum to accurately estimate energy losses and heating rates. The discretization of the cosmic ray transport terms is difficult. For example, anisotropic diffusion requires discretization techniques that avoid the violation of the entropy condition by limiting the transverse fluxes\cite{Sharma2007,Pakmor2016,Kannan2016,Butsky2018}. 
Modeling cosmic ray streaming is particularly challenging because of the discontinuous dependence of the streaming velocity on the sign of the scalar product between the magnetic field and the cosmic ray pressure gradient in the one-moment formulation of cosmic ray hydrodynamics. This leads to unphysical oscillations of the solution and small time steps especially near cosmic ray pressure maxima if not addressed in form of regularization techniques -- such as replacing the sign function with the hyperbolic tangent function that ensures a smooth dependence of the streaming velocity on cosmic rays and gas properties\cite{Sharma2009,Ruszkowski2017,Butsky2018}, albeit at the expense of a dependence of the solution on a numerical parameter. An elegant solution of this problem is to replace the equation for the cosmic ray energy by two equations for cosmic ray energy and flux that are coupled to the MHD system of equations\cite{Jiang2018,Thomas2019}. This two-moment formulation can be derived from quasi-linear theory of cosmic ray transport and describes cosmic ray streaming and diffusion self-consistently with a hyperbolic set of equations, which also contains the evolution equations for Alfv\'{e}n waves that are self-generated by the streaming cosmic rays\cite{Thomas2019}.\\

\footnotesize
\begin{tcolorbox}[enhanced,drop fuzzy shadow,colback=yellow!10!white,colframe=yellow!50!black,title={\bf Modeling cosmic rays}]
\vspace{-15pt}
\begin{flalign}
&\text{\bf Ideal magnetohydrodynamics equations with cosmic rays:} & & \text{\bf MHD Maxwell equations:}\nonumber \\
& \frac{\partial \rho}{\partial t} + \nabla \cdot (\rho {\bf v}) = 0 & & \nabla \times {\bf B} = \frac{4\pi}{c}{\bf J} & \nabla\cdot{\bf B} = 0\nonumber \\
&\frac{\partial \rho {\bf v}}{\partial t} + {\bf \nabla} \cdot (\rho {\bf v} \otimes {\bf v} + P\mathbb{1}) = \frac{{\bf J} \times {\bf B}}{c} -{\bf \nabla} P_{\rm cr} & & \frac{1}{c}\frac{\partial {\bf B}}{\partial t} + \nabla \times {\bf E} = 0 & {\bf E} = -\frac{{\bf v} \times {\bf B}}{c}\nonumber \\
&\frac{\partial (\rho e + e_{\rm B})}{\partial t} + {\bf \nabla} \cdot \left[(\rho e + P) {\bf v} + c\frac{{\bf E} \times {\bf B}}{4\pi}\right] =- ({\bf v} + {\bf v_{\rm st}}) \cdot \nabla P_{\rm cr} + \Lambda_{\rm th} + \Gamma_{\rm th} \nonumber \\
& \text{\bf Cosmic rays energy density evolution:} \nonumber \\
& \frac{\partial \epsilon_{\rm cr}}{\partial t}
  +\nabla\cdot\left[ \epsilon_{\rm cr} ({\bf v}+ {\bf v}_{\rm st})
    - \kappa_\epsilon {\bf b}\left(\bf{b}\cdot\nabla \epsilon_{\rm cr}\right)\right] = -P_{\rm cr} \nabla\cdot ({\bf v} + {\bf v_{\rm st}}) + \Lambda_{\rm cr} + \Gamma_{\rm cr}. \nonumber
\end{flalign}
\noindent\rule{\textwidth}{1pt}
\noindent  Cosmic rays exhibit a force on the gas through  $\nabla P_{\rm cr}$. Their energy density is influenced by streaming with velocity ${\bf v}_{\rm st}$ ($\epsilon_{\rm cr} [{\bf v}+ {\bf v}_{\rm st}]$),
anisotropic diffusion with coefficient $\kappa_\epsilon$ ($\kappa_\epsilon {\bf b}\left[\bf{b}\cdot\nabla \epsilon_{\rm cr}\right])$, 
and adiabatic processes due to the compression of the Alfv\`en frame ($P_{\rm cr} \nabla\cdot [{\bf v} + {\bf v_{\rm st}}]$). The terms $\Lambda_{\rm th}$, $\Lambda_{\rm cr}$, $\Gamma_{\rm th}$ and $\Gamma_{\rm cr}$, represent non-adiabatic source and sink terms. 
\end{tcolorbox}
\normalsize

\noindent{\bf Radiation Hydrodynamics:} Radiation alters the thermal, kinetic, and chemical state of the gas. Radiation hydrodynamics simulations are required to capture this self-consistently. In the context of cosmological simulations, radiation hydrodynamics simulations have so far primarily been employed to study the epoch of reionization\cite{Gnedin2014,Ocvirk2018,Rosdahl2018}. These simulations are aimed at exploring the high redshift Universe and are typically not evolved towards the low redshift regime. Consequently, the employed galaxy formation models within these simulations can also not be tested against low redshift predictions. Only a limited number of simulations have studied the impact of radiation in the context of galaxy formation simulations\cite{Wise2012,Rosdahl2015}. The main reason for this lack of detailed radiation hydrodynamics studies is that numerical radiative transfer is challenging because of the high dimensionality caused by the frequency and directional dependencies of photon propagation. Even more challenging is the fact that in general the speed of light poses severe constraints on the timesteps of these simulations, which can however be circumvented to some degree through the application of a reduced speed of light approximation\cite{Gnedin2001,Gnedin2016, Deparis2019, Ocvirk2019}. The most common numerical methods for radiation hydrodynamics are ray-tracing, Monte Carlo, and moment-based methods. The ray-tracing method discretizes the radiative transfer equation along individual directions from each source. Long characteristic ray-tracing schemes\cite{Mihalas1984,Abel1999,Jaura2018}, cast rays from the source through the whole simulation domain, and the transport, absorption and emission of radiation is computed along each ray. 
Long characteristic schemes are accurate but computationally expensive, since they
scale as $\mathcal{O}(N_{\rm s} \times N_{\rm c}^p)$, where $N_{\rm s}$ is the number of sources, $N_{\rm c}$ is the number of underling discretization resolution elements, for example cells, and $p$ is a method- and geometry-dependent exponent\cite{Rijkhorst2006, Rosen2017}. Short characteristic methods\cite{Whalen2006,Trac2007,Pawlik2008,Petkova2011}, on the other hand, solve the radiative transport only along rays that connect nearby cells allowing an efficient parallelization and merging procedures to break the $\mathcal{O}(N_{\rm s} \times N_{\rm c}^p)$ scaling. Monte Carlo methods\cite{Ciardi2001, Oxley2003,Tasitsiomi2006,Semelin2007,Dullemond2012,Smith2019}, often only applied in post-processing, emit photon packets and propagate them probing the gas opacity, interaction lengths and scattering angles from underlying probability density functions thus stochastically solving the radiative transfer equation. One drawback of Monte Carlo schemes is that the signal-to-noise ratio improves only as the square root of the number of photon packets due to Poisson noise. Still, Monte Carlo is highly accurate and photon weighting, path-based estimators, and discrete diffusion schemes help overcome the efficiency barriers that inhibit convergence\cite{Lucy1999,Gentile2001,Densmore2007,SmithTsang2018}. Moment-based methods became popular over the last years due to superior scalability\cite{Levermore1984,Gonzalez2007, Rosdahl2013,Rosdahl2015,Kannan2018}. They are based on a fluid-like description of radiation fields by taking zeroth, first and second moments of the radiation specific intensity with respect to the angular variable. This defines a radiation energy density $E_\nu$, flux ${\bf F}_\nu$, pressure tensor ${\mathbb P}_\nu$ and hyperbolic conservation laws for the energy density and the radiation flux. Similar to the hydrodynamical case, where an equation of state is required to relate gas pressure and density, a non-unique closure relation is required to relate ${\mathbb P}_\nu$ to $E_\nu$ and ${\bf F}_\nu$. A widely used approach is to define ${\mathbb P}_\nu \equiv E_\nu {\mathbb D}$, where ${\mathbb D}$ is the Eddington tensor that can be estimated with different methods, for example through flux-limited diffusion\cite{Lucy1977,Krumholz2007}, the optically thin variable Eddington tensor approach\cite{Gnedin2001,Filantor2009,Petkova2009} or the M1 closure\cite{Levermore1984,Dubroca1999,Ripoll2001,Rosdahl2015,Kannan2018}. For the former methods ${\mathbb D}$ is estimated assuming that the gas between sources of radiation is always optically thick or thin. The M1 method, instead, computes the Eddington tensor by using local radiation quantities.\\

\footnotesize
\begin{tcolorbox}[enhanced,drop fuzzy shadow,colback=yellow!10!white,colframe=yellow!50!black,title={\bf Modeling cosmic radiation fields}]
\vspace{-15pt}
\begin{flalign}
&\text{\bf Radiation hydrodynamics equations:} && \text{\bf Radiative transfer equation:} && \nonumber \\
& \frac{\partial \rho}{\partial t} + {\bf \nabla} \cdot (\rho {\bf v}) = 0  && \frac{1}{c}\frac{\partial I_\nu}{\partial t} + {\bf n} \cdot \frac{\partial I_\nu}{\partial {\bf r}} = -\kappa_\nu I_\nu + j_\nu && \nonumber \\
& \frac{\partial (\rho {\bf v})}{\partial t} + {\bf \nabla} \cdot (\rho {\bf v} \otimes {\bf v} + P\mathbb{1}) = \Gamma_{\rm p} & & 
\nonumber \\
&\frac{\partial (\rho e)}{\partial t} + {\bf \nabla} \cdot (\rho e + P ){\bf v} = -\Lambda + \Gamma_{\rm E} & & 
\nonumber 
\end{flalign}
\noindent\rule{\textwidth}{1pt}
\noindent The radiative transfer equation relates the specific radiation intensity, $I_\nu$, with the absorption coefficient, $\kappa_\nu$, and the specific emissivity, $j_\nu$. The radiation direction of propagation is represented by the unit vector ${\bf n}$. $\Lambda$ is the cooling function, $\Gamma_{\rm p}$ and $\Gamma_{\rm E}$ are source terms that describe the transfer of momentum and energy from the radiation to the gas.
\end{tcolorbox}
\normalsize

\vspace{0.1cm}

\noindent{\bf Other physics:} Additional physical processes considered in some cosmological simulations of galaxy formation are, for example, dust physics\cite{McKinnon2016,Aoyama2017,Hou2017,McKinnon2017,Aoyama2018,McKinnon2018,Vogelsberger2018,Gjergo2018,Hou2019}, thermal conduction \cite{Ruszkowski2011b,Smith2013,Arth2014,Kannan2016,Yang2016,Kannan2017,Barnes2018}, and viscosity \cite{Parrish2012,Suzuki2013,ZuHone2015,Su2017}.
Dust has typically been neglected in galaxy formation simulations since it contributes only about $\sim1\%$ to the mass budget of the interstellar medium. However, dust plays an important role for the evolution of the interstellar medium affecting the thermochemistry and radiation processing. Therefore, recently galaxy formation began to incorporate first simple dust models to follow its production, growth and destruction in the interstellar medium. Most of these implementations treat dust as a passive scalar and model the processes affecting the dust population through effective rate equations.
Thermal conduction is another physical effect that is often neglected in cosmological simulations of galaxy formation. However, in hot plasmas of galaxy clusters, conduction can affect the thermodynamic properties of galaxy clusters as has recently been demonstrated\cite{Balbus2000, Quataert2008, Kannan2017, Barnes2019}. Simulating thermal conduction requires a precise numerical magnetohydrodynamics implementation to resolve the strength and topology of the magnetic field, and an efficient anisotropic diffusion solver to model the conduction\cite{Kannan2016}. 
\\[-0.4cm]

\noindent{\bf Caveats and limitations:} Simulations of the dark matter component typically boil down to implementing efficient $N$-body methods and parallelization schemes. Simulations of the baryonic matter component are however more challenging, since they require reliable hydrodynamics numerical schemes and well-posed sub-resolution models. These additional complications lead to some caveats and limitations of such simulations. \\[0.1cm]
\noindent {\bf \textit{Calibration:}}
The numerical implementation of baryonic physics is based on sub-resolution models due to the intrinsic resolution limitations of any simulation. These effective models depend on a certain number of adjustable parameters. Depending on the exact galaxy formation model implementation, these parameters can either be chosen based on physical arguments or they require a certain calibration procedure. The latter approach is often employed in large volume simulations, where the sub-resolution models are less detailed compared to those of zoom simulations. The calibration process consists of a parameter exploration for the effective models through a large number of simulations. These simulations typically cover a smaller volume compared to production simulations. The calibration is then based on a comparison to some key observables of the galaxy population like the star
formation rate density as a function of cosmic time, the galaxy stellar mass function at $z = 0$ and the present-day stellar-to-halo mass relation.  \\[0.1cm]
\noindent {\bf \textit{Numerical convergence:}} Cosmological simulations have to cover a wide range of spatial and time scales. This implies that simulations have to aim for the highest possible number of resolution elements. However, even state-of-the-art simulations cannot capture all relevant scales. Simulations are therefore often performed at different resolution levels to understand the exact dependence of the results on the number of resolution elements. A simulation prediction is then said to be converged once this prediction does not significantly change anymore if the numerical resolution is further increased.\\[0.2cm]
\noindent {\bf \textit{Diverging results:}} Various simulations now agree on a wide range of predictions. This is especially the case for predictions of the stellar content of galaxies and related observables. However, there is also a wide range of predictions that diverge among different simulations. For example, the characteristics of gas around galaxies are very sensitive to the feedback implementations used in the different galaxy formation models. This can lead to rather different outcomes for the thermodynamic structure of gas around galaxies. Such difference can then be used to differentiate and test galaxy formation models.

\subsection{Some Key Results of Hydrodynamical Simulations}

The results of hydrodynamical simulations can directly be confronted with observational data providing important tests for galaxy formation models. This often involves the construction of detailed mock observations based on the simulated data\cite{Torrey2015,Trayford2017}. Early simulations successfully reproduced properties of the intergalactic medium such as the column density distribution of the Lyman-$\alpha$ forest\cite{Hernquist1996}. Many simulations also focused on the formation of individual galaxies \cite{Katz1991,Navarro1991,Katz1992,Katz1992b,Navarro1997b}. However, such simulations suffered for a long time from, for example, inconsistent stellar masses, galaxy sizes, star formation histories and galaxy morphologies \cite{Navarro2000,Abadi2003,Scannapieco2008,Oppenheimer2010}. Only recently simulations began to produce realistic galaxies\cite{Vogelsberger2014,Schaye2015,Wang2015,Grand2017,Kaviraj2017,Springel2018,Hopkins2018}. However, different sub-resolution implementations of astrophysical processes remain a major source of uncertainties. Results of hydrodynamical simulations can be grouped into those for global properties for the whole galaxy population, and those for the properties of individual galaxies. \\[-0.3cm]

\noindent {\bf Global Properties:} 
Large volume simulations are ideally suited to explore global properties of the galaxy population due to their large statistical sample size. This enables direct comparisons to astronomical galaxy surveys. Table 2 presents some selected recent structure and galaxy formation simulations.\\[-0.4cm]

\noindent {\bf \textit{Stellar content of galaxies:}} 
One of the most fundamental properties of the galaxy population is the galaxy stellar mass function, which quantifies the comoving number density of galaxies as a function of galaxy stellar mass. 
Stellar mass functions are frequently described  by a Schechter function\cite{Schechter1976}
with parameter $M^*$, a characteristic mass scale above which the distribution is exponentially suppressed, a normalization $\phi^*$, and $\alpha^*$ setting the low-mass slope.
Observed low redshift parameters are roughly given by $\mathrm{log}(M_*/M_\odot)\approx11$,  $\mathrm{log}(\Phi^*/\mathrm{Mpc^{-3}})\approx-2.7$, $\alpha^*\approx-1.2$~\cite{Panter2007}. However, double Schechter functions provide an even better description of low redshift galaxy stellar mass functions\cite{Pozzetti2010,Baldry2012,Ilbert2013,Muzzin2013,Weigel2016}.
The halo mass function exhibits a steeper low-mass slope, $\approx -2$, than the galaxy stellar mass function and the exponential suppression occurs at a lower volume density.
Reproducing the observed stellar mass function therefore requires a strong suppression of star formation at both the low and high mass ends.
Galaxy formation models assume that supernova feedback flattens out the low-mass (${\rm M} \lsim 10^{12}\,{\rm M_\odot}$) slope by suppressing star formation\cite{Dekel1986, Larson1974, White1991} while the suppression of bright and high-mass (${\rm M} \gsim 10^{12}\,{\rm M_\odot}$) galaxies is regulated by feedback from active galactic nuclei.
Energetically plausible forms of supernova and active galactic nuclei feedback in simulations resulted in galaxy stellar mass functions that are consistent with observational data.
Simulation predictions are also often confronted with empirical constraints on the relationship between stellar mass and halo mass, which are derived based on various galaxy-halo mapping techniques\cite{Behroozi2010, Moster2013}. This ratio of stellar mass to halo mass peaks around halo masses of roughly $\sim 10^{12}\,{\rm M}_\odot$, where star formation is most efficient. For higher and lower halo masses, the star formation rates are reduced due to feedback processes. Modern large volume simulations reproduce the stellar to halo mass relationship at low and high redshifts reasonably well\cite{Schaye2015,Pillepich2018b}.\\[-0.4cm]

\noindent {\bf \textit{Gas around galaxies:}} 
One of the key advantages of hydrodynamical simulations compared to semi-analytic models (see Box 1) is their ability to make detailed predictions for the distribution and properties of gas around galaxies including the circumgalactic medium, the intracluster medium, and the intergalactic medium.
The circumgalactic and intergalactic media are quite diffuse ($n\sim10^{-3}-10^{-7}\mathrm{cm}^{-3}$) and cool ($T \sim 10^{4-6}\,{\rm K}$) and observations in emission, like Lyman-$\alpha$ and metal lines, are therefore rather challenging.
However, absorption line observations from background quasars can probe the distribution, enrichment, and ionization state of this gas.
One of the first successes of hydrodynamical simulations has been the reproduction of the declining trend of the number of absorbing clouds per unit redshift and linear interval of \ion{H}{i} column density with column density in the Lyman-$\alpha$ forest\cite{Hernquist1996}.
Reproducing properties of the circumgalactic medium, however, is significantly more challenging.
Observations of this gas indicate that it features a rich multi-phase structure where individual lines of sight simultaneously contain highly ionized, warm, and cool atomic species\cite{Werk2014,Werk2016}.  
The coolest and densest parts of this gas have spatial scales of $10-100\,{\rm pc}$ \cite{Stern2016}, although the coherence scale can reach up to $\sim 1\,{\rm kpc}$ \cite{Rubin2018}. These spatial scales are below the typical circumgalactic gas resolution limits of galaxy formation simulations. More recently, cosmological simulations with special circumgalactic gas refinement schemes have been employed to overcome some of the resolution limitations. Such simulations increase the numerical resolution in the circumgalactic gas reaching smaller spatial scales\cite{Peeples2018, vandeVoort2019, Suresh2019, Hummels2019}. At $z=2$ such simulations can reach a spatial resolution below $\sim 100\,{\rm pc}$\cite{Suresh2019}, and at $z=0$ below $\sim 1\,{\rm kpc}$ within the circumgalactic medium\cite{vandeVoort2019}.
In addition to resolution concerns, the circumgalactic medium is influenced by feedback-driven outflows from galaxies, whose characteristics are not yet properly understood and modeled.
The circumgalactic medium can therefore also be used to constrain feedback mechanisms. 
The intracluster medium can directly be observed via X-ray observations due to the much higher gas temperatures ($T\sim 10^{7-8}\, {\rm K}$).  
Many properties of the intracluster medium , like X-ray and Sunyaev-Zeldovich scaling relations or the iron distribution, can be accurately modeled in simulations\cite{Rasia2015,Planelles2014, Biffi2016, Vogelsberger2018b}. However, significant challenges remain for galaxy formation models to reproduce cluster entropy profiles and, in particular, distinct cool-core, and non cool-core clusters\cite{Barnes2017,Barnes2018b}.\\[-0.4cm]

\noindent {\bf \textit{Galaxy clustering:}} 
Galaxy clustering varies as a function of galaxy mass and galaxy properties, e.g. formation time, star formation rate, color.
Simulations now reproduce a number of features in the galaxy clustering signal including the mass dependent two-point correlation length\cite{Springel2018}, which increases with increasing masses\cite{Meneux2008,Foucaud2010,Wake2011}, the clustering signal for non- and star-forming galaxies\cite{Artale2017,Springel2018}, and the steepening of the power law slope $\gamma$ of the galaxy correlation function with declining redshift ($\gamma \sim 1.8$ at $z\simeq 0$ and $\gamma\sim 1.6$ at $z\simeq 1$ \cite{Marulli2013,Springel2018}).\\[-0.4cm]

\noindent {\bf \textit{Scaling relations:}} 
Galaxies exhibit a wide range of scaling relations linking various observables constituting another important test for galaxy formation models.
Modern large volume hydrodynamical simulations broadly reproduce many galaxy scaling relations including the mass-size\cite{Shen2003} , the supermassive black hole mass-stellar velocity dispersion relation\cite{Kormendy2013}, and the mass-metallicity \cite{Tremonti2004} relation\cite{Dave2017,DeRossi2017,Torrey2017,Torrey2018}. Also other galaxy characteristics like the color of galaxies as a function of galaxy stellar mass can now be reasonably well reproduced by cosmological simulations\cite{Trayford2015,Kaviraj2017,Nelson2018}. 
However, there are still points of tension including, for example, the magnitude of the scatter, the detailed shape, or the dependence on additional galaxy properties.  \\[-0.3cm]

\small
\begin{tcolorbox}[drop fuzzy shadow, left=0pt, title={\bf \hspace{0.1cm}Box 1: Semi-analytic modeling of galaxy formation}]
Studying baryonic physics through hydrodynamical simulations is
computationally expensive compared to dark matter-only $N$-body simulations. An alternative
approach is to model baryonic physics on top of $N$-body dark matter simulations through
analytic models. This combination of numerical dark matter-only simulations, and analytic models for the prescription of baryonic physics, is known as semi-analytic modeling\cite{Kauffmann1993,Somerville1999,Bower2006,Croton2006,Guo2011}.  These semi-analytic models track, for example, how much gas accretes onto halos, how much hot gas cools and
turns into stars, or how feedback processes remove cold gas from the galaxy or heat the halo
gas. The models are based on the merger history of dark matter halos extracted from  $N$-body simulations. The result of such a calculation is a predicted galaxy population that can be compared to observational data in a similar way as the output of full hydrodynamical simulations. The key advantage of semi-analytical models is their efficiency. It is therefore possible to perform a wide range of calculations, using different model variations. However, a disadvantage of semi-analytic models is that they are less self-consistent compared to hydrodynamical simulations. Furthermore, studying detailed gas properties, for example, the circumgalactic gas with these models is not directly possible since the gas component is not resolved.
\end{tcolorbox} 
\normalsize
\noindent {\bf Galaxy Properties:} 
The detailed properties of late-type disc-like and early-type spheroid-dominated galaxies have been studied extensively using simulations. \\[-0.4cm]

\noindent {\bf \textit{Properties of late-type galaxies:}} Simulating the formation of star-forming, late-type galaxies has been one of the most pressing challenges of computational galaxy formation. For a long time, simulations struggled to form galaxies with extended and rotationally-supported stellar and gaseous discs as observed in the Universe. These discs are expected to form through angular momentum conservation of the cooling gas in dark matter halos\cite{Fall1980, Mo1998}. However, realizing this mechanism in cosmological simulations turned out to be difficult, and early works produced galaxies dominated by a stellar spheroidal component, with a sub-dominant disc only\cite{Navarro2000, Scannapieco2009}. More efficient stellar feedback schemes were required to offset runaway radiative losses of the star-forming gas, the so-called overcooling catastrophe\cite{Balogh2001}, and to eject the low-angular momentum material responsible for the creation of the dominant stellar bulge , the so-called angular momentum catastrophe\cite{Brook2011}. The success of modern simulations in producing late-type disc galaxies is largely due to the ability of stellar feedback to regulate star formation efficiently\cite{Okamoto2005,Brooks2011,Guedes2011,Aumer2013b,Stinson2013a,Hopkins2014,Marinacci2014,Wang2015,Wetzel2016,Grand2017,Hopkins2018}. More recently, magnetic fields in late-type galaxies have also been studied to understand their topology and field strengths \cite{Pakmor2014,Beck2015,Rieder2016, Rieder2017,Rieder2017b,Pakmor2017}. Furthermore, the impact of cosmic rays in galaxies has been studied in more detail over the last years\cite{Pakmor2016,Ruszkowski2017}. These results indicate that cosmic rays are potentially important for driving galactic outflows. \\[-0.4cm]

\noindent {\bf \textit{Properties of early-type galaxies:}}
Simulations can also reproduce spheroid-dominated early-type systems, which broadly match the early formation history\cite{Naab2014}, scaling relations (e.g. the mass and size or velocity dispersion)\cite{Kobayashi2005, Feldmann2011}, and the metallicity distribution\cite{Kobayashi2004} of observed early-type galaxies.
The assembly of such large objects proceeds in two phases \cite{Oser2010,Huang2013,Rodriguez-Gomez2016,Clauwens2018}. At high redshift ($z\gsim 1.5$), galaxies grow predominantly in-situ by efficiently converting gas into stars. At later times, mass is predominantly gained through accretion of smaller substructures, i.e. mergers, which also considerably increases galaxy sizes.
Spatially-resolved spectral observations have shown that spheroid-dominated galaxies have diverse kinematics and shapes. The kinematics is usually described through the so-called spin parameter $\lambda_\mathrm{R}$. This quantity is used to split galaxies  into fast ($\lambda_\mathrm{R}>0.1$) and slow ($\lambda_\mathrm{R}<0.1$) rotator classes. The ellipticity $\epsilon$, instead, is used to define the spheroid's shape.
Simulations have played a major role in building a physical picture to explain the diversity in kinematics and shapes of spheroid-dominating galaxies that are produced based on their formation histories\cite{Naab2014,Lagos2018,Schulze2018}, in particular through gas dissipation, which builds rotationally-supported structures, and mergers, which set the late-time spin parameter.\\

\scriptsize
\begin{tcolorbox}[left=-4pt, title={\bf \hspace{0.1cm}Table 2: Recent structure and galaxy formation simulations}]
	\begin{tabular}{lllllll}
\hline\\[-2ex]
\multicolumn{1}{l}{\bf simulation} & \multicolumn{1}{l}{\bf volume} &  \multicolumn{1}{l}{\bf method\footnote{\scriptsize PM: particle-mesh; TreePM: tree + PM; FM: fast multipole; P$^3$M: particle-particle-particle-mesh; ML: multilevel; SPH: smoothed particle hydrodynamics; AMR: adaptive-mesh-refinement; MMFV: moving-mesh finite volume; MLFM: mesh-free finite mass}} & \multicolumn{1}{l}{\bf mass } & \multicolumn{1}{l}{\bf spatial} & \multicolumn{1}{l}{\bf primary} \\
\multicolumn{1}{l}{} & \multicolumn{1}{l}{} &  \multicolumn{1}{l}{} & \multicolumn{1}{l}{\bf resolution\footnote{\scriptsize highest resolution quoted (dark matter/gas)}} & \multicolumn{1}{l}{\bf resolution\footnote{\scriptsize for particle based codes, the minimum softening length is reported; for mesh codes, the minimum cell size is quoted (dark matter/gas)}} & \multicolumn{1}{l}{\bf reference} \\[0.5ex]
\hline\\[-2ex]
\multicolumn{1}{l}{} & \multicolumn{1}{l}{$[{\rm Mpc}^3]$} &  \multicolumn{1}{l}{} & \multicolumn{1}{l}{$[{\rm M}_\odot]$} & \multicolumn{1}{l}{$[{\rm kpc}]$} & \multicolumn{1}{l}{} \\[0.5ex]
\hline\\[-2ex]
{\bf dark matter-only}\\[0.5ex]
\hline\\[-2ex]
Millennium      & $685^3$   \hspace{1.1cm}     & TreePM    \hspace{1.1cm}            & $1.2\!\!\times\!\!10^9 / -$   \hspace{1.1cm}                       & $6.85/-$  \hspace{1cm}    & Springel et al.~(2005b) \cite{Millennium}\\
Millennium-2    & $137^3$        & TreePM                & $9.4\!\!\times\!\!10^6/-$                          & $1.37/-$      & Boylan-Kolchin et al.~(2009) \cite{MilleniumII} \\
Horizon 4$\pi$  & $2740^3$       & PM/ML                & $7.7\!\!\times\!\!10^9/-$                          & $10.41/-$      & Teyssier et al.~(2009) \cite{Teyssier2009} \\
Bolshoi         & $357^3$        & PM/ML                & $1.9\!\!\times\!\!10^8/-$              & $1.43/-$     & Klypin et al.~(2011) \cite{Klypin2011} \\
Full Universe Run & $29167^3$      & PM/ML                 & $1.4\!\!\times\!\!10^{12}/-$          & $55.6/-$     & Alimi et al.~(2012)\cite{Alimi2012}\\
Millennium-XXL  & $4110^3$       & TreePM                & $8.5\!\!\times\!\!10^9/-$                          & $13.7/-$      & Angulo et al.~(2012)\cite{Angulo2012}\\
MultiDark       & $1429^3$       & PM/ML                & $1.2\!\!\times\!\!10^{10}/-$              & $10/-$      & Prada et al.~(2012)\cite{Prada2012} \\
Dark Sky      & $11628^3$        & Tree/FM                & $5.7\!\!\times\!\!10^{10} / -$                          & $53.49/-$      & Skillman et al.~(2014) \cite{Skillman2014}\\
$\nu^2$GC     & $1647^3$              &     TreePM    & $3.2\!\!\times\!\!10^8 / -$  & $6.28/-$            & Ishiyama et al.~(2015) \cite{Ishiyama2015}\\
Q Continuum   & $1300^3$              &     TreePM/P$^3$M    & $1.5\!\!\times\!\!10^8 / -$  & $2.82/-$            & Heitmann et al.~(2015) \cite{Heitmann2015}\\
OuterRim      & $4225^3$        & TreePM/P$^3$M                & $2.6\!\!\times\!\!10^9 / -$                          & $6.0/-$      & Habib et al.~(2016) \cite{Habib2016}\\
EuclidFlagship      & $20000^{3}$        & Tree/FM                & $10^9 / -$                          & $5/-$      & Potter et al.~(2017) \cite{Potter2017}\\[0.5ex]
\hline\\[-2ex]
Aquarius        & zoom    & TreePM                & $1.7\!\!\times\!\!10^3/-$              & $0.02/-$      & Springel et al.~(2008) \cite{Springel2008}\\

Via Lactea II   & zoom    & Tree                & $4.1\!\!\times\!\!10^3/-$              & $0.04/-$      & Diemand et al.~(2008) \cite{Diemand2008}\\
GHALO           & zoom    & Tree                & $1.0\!\!\times\!\!10^3/-$              & $0.06/-$      & Stadel et al.~(2009) \cite{Stadel2009} \\
CLUES           & zoom    & TreePM                & $3.4\!\!\times\!\!10^5/-$              & $0.21/-$      & Libeskind et al.~(2010) \cite{Libeskind2010} \\
Phoenix         & zoom    & TreePM                & $8.7\!\!\times\!\!10^5/-$              & $0.21/-$      & Gao et al.~(2012) \cite{Gao2012} \\
ELVIS           & zoom    & TreePM                & $1.9\!\!\times\!\!10^5/-$              & $0.14/-$      & Garrison-Kimmel et al.~(2014) \cite{Garrison-Kimmel2014b} \\
COCO            & zoom    & TreePM                & $1.6\!\!\times\!\!10^5/-$              & $0.33/-$      & Hellwing et al. ~(2016) \cite{Hellwing2016} \\[0.5ex]
\hline\\[-2ex]
{\bf + baryons}\\[0.5ex]
\hline\\[-2ex]
Illustris       & $107^3$        & TreePM+MMFV     & $6.7\!\!\times\!\!10^6/1.3\!\!\times\!\!10^6$  & $1.42/0.71$      & Vogelsberger et al.~(2014) \cite{Vogelsberger2014} \\
Horizon-AGN     & $142^3$        & PM/ML+AMR    & $8.0\!\!\times\!\!10^7/1.0\!\!\times\!\!10^7$   & $1.0/1.0$      & Dubois et al.~(2014) \cite{Dubois2014} \\
EAGLE           & $100^3$        & TreePM+SPH    & $9.7\!\!\times\!\!10^6/1.8\!\!\times\!\!10^6$   & $0.7/0.7$      & Schaye et al.~(2015) \cite{Schaye2015} \\
MassiveBlack-2  & $143^3$        & TreePM+SPH    & $1.6\!\!\times\!\!10^7/3.2\!\!\times\!\!10^6$                           & $2.64/2.64$      & Khandai et al.~(2015) \cite{Khandai2015} \\
Bluetides\footnote{\scriptsize final redshift $z=8$; spatial resolution is in physical units at that redshift}       & $574^3$        & TreePM+SPH    & $1.7\!\!\times\!\!10^7/3.4\!\!\times\!\!10^6$                           & $0.24/0.24$      & Feng et al.~(2016) \cite{Feng2016} \\ 
Magneticum      & $68^3$        & TreePM+SPH    & $5.3\!\!\times\!\!10^7/1.1\!\!\times\!\!10^7$   & $1.4/0.7$-$1.4$      & Bocquet et al.~(2016) \cite{Bocquet2016} \\
MUFASA          & $74^3$        & TreePM+MLFM    & $9.6\!\!\times\!\!10^7/1.8\!\!\times\!\!10^7$                           & $0.74/0.74$      & Dave\'e et al.~(2016) \cite{Dave2016} \\
BAHAMAS          & $571^3$        & TreePM+SPH    & $5.5\!\!\times\!\!10^9/1.1\!\!\times\!\!10^9$                           & $0.25/0.25$      & McCarthy et al.~(2017) \cite{McCarthy2017} \\
Romulus25          & $25^3$        & Tree/FM+SPH    & $3.4\!\!\times\!\!10^5/2.1\!\!\times\!\!10^5$                           & $0.25/0.25$      & Tremmel et al.~(2017) \cite{Tremmel2017} \\
IllustrisTNG\footnote{\scriptsize IllustrisTNG consists of three main simulations: TNG50, TNG100, TNG300; numbers are quoted for TNG100}    & $111^3$        & TreePM+MMFV     & $7.5\!\!\times\!\!10^6/1.4\!\!\times\!\!10^6$                       & $0.74/0.19$      & Springel et al.~(2018) \cite{Springel2018}\\
Simba\footnote{\scriptsize numbers for largest volume simulation quoted}    & $147^3$        & TreePM+MLFM     & $1.4\!\!\times\!\!10^8/2.7\!\!\times\!\!10^7$                       & $0.74/0.74$      & Dav\'e et al.~(2019) \cite{Dave2019}\\[0.5ex]
\hline\\[-2ex]
Eris           & zoom    & Tree+SPH    & $9.8\!\!\times\!\!10^4/2\!\!\times\!\!10^4$  & $0.12/0.12$      & Guedes et al.~(2011) \cite{Guedes2011} \\
VELA           & zoom    & PM/ML + AMR  & $8.3\!\!\times\!\!10^4/1.9\!\!\times\!\!10^5$  & $0.03/0.03$\footnote{\scriptsize in physical units at $z = 3$}      & Ceverino et al.~(2014) \cite{Ceverino2014} \\
NIHAO           & zoom    & Tree+SPH    & $3.4\!\!\times\!\!10^3/6.2\!\!\times\!\!10^2$  & $0.12/0.05$      & Wang et al.~(2015) \cite{Wang2015} \\
APOSTLE         & zoom    & TreePM+SPH    & $5.0\!\!\times\!\!10^4/1.0\!\!\times\!\!10^4$  & $0.13/0.13$      & Sawala et al.~(2016) \cite{Sawala2016} \\
Latte/FIRE          & zoom    & TreePM+MLFM    & $3.5\!\!\times\!\!10^4/7.1\!\!\times\!\!10^3$  & $0.02/0.001$      & Wetzel et al.~(2016) \cite{Wetzel2016} \\
Auriga          & zoom    & TreePM+MMFV     & $4.0\!\!\times\!\!10^4/6.0\!\!\times\!\!10^3$  & $0.18/0.18$\footnote{\scriptsize for baryons the minimum physical softening is reported}     & Grand et al.~(2017) \cite{Grand2017} \\
MACSIS & zoom & TreePM+SPH & $6.4\!\!\times\!\!10^9/1.2\!\!\times\!\!10^9$ & $5.77/5.77$  & Barnes et al.~(2017) \cite{Barnes2017b}\\
Cluster-EAGLE   & zoom    & TreePM+SPH    & $9.7\!\!\times\!\!10^6/1.8\!\!\times\!\!10^6$  & $0.7/0.7$  & Barnes et al.~(2017) \cite{Barnes2017} \\
		The Three Hundred Project & zoom    & TreePM+SPH     & $1.9\!\!\times\!\!10^9/3.5\!\!\times\!\!10^8$  & $9.59/9.59$     & Cui et al.~(2018) \cite{Cui2018} \\
FABLE          & zoom    & TreePM+MMFV     & $8.1\!\!\times\!\!10^7/1.5\!\!\times\!\!10^7$  & $4.15/4.15$     & Henden et al.~(2018) \cite{Henden2018} \\
RomulusC        & zoom    & Tree/FM+SPH    & $3.4\!\!\times\!\!10^5/2.1\!\!\times\!\!10^5$  & $0.25/0.25$      & Tremmel et al.~(2019) \cite{Tremmel2019} \\
\hline
\end{tabular}
\end{tcolorbox}
\normalsize

\section{Simulations of Alternative Cosmological Models}

Cosmological simulations of galaxy formation have also been used to explore alternative cosmological models. At the most basic level the cosmological model can be altered in three different ways: alternative forms of dark matter, alternative forms of dark energy, or alternative forms of gravity. We note that many simulations of alternative cosmological models typically only consider the dark matter component and do not model baryons. However, these simulations then neglect the important backreaction between baryons and dark matter. Similarly, simulations including baryons are also now important to infer cosmological parameters. For example, DESI, LSST and Euclid will rely on models based on galaxy formation simulations to achieve their forecasted precision. Future explorations of alternative cosmologies have to consider and include these effects by also modeling the baryon component.\\

\small
\begin{tcolorbox}[drop fuzzy shadow, left=0pt, title={\bf \hspace{0.1cm}Box 2: Small-scale problems of cold dark matter}]
The cold dark matter paradigm correctly describes the large-scale distribution of galaxies. On sub-galactic scales
however, some problems have been identified over the last decades\cite{Bullock2017}. Among the
most relevant challenges are: the under-abundance of dwarf galaxies in the Milky Way and in the field (the missing satellites problem\cite{Klypin1999,Moore1999,Zavala2009, Papastergis2011, Klypin2015}), the inconsistency of inner dark
matter density profiles in low surface brightness and dwarf
galaxies (the cusp-core problem\cite{deBlok1997, Walker2011}), the deficit of dark matter in the inner regions of massive dwarf galaxies (the too-big-to-fail problem\cite{BoylanKolchin2011, Papastergis2015}), and the
large variety of shapes of dwarf rotation curves (the diversity problem\cite{Oman2015}). Most of these problems have been found by contrasting
dark matter-only simulations with observations, which do not take into account the
complex baryonic dark matter interactions. It is therefore possible that these challenges can be solved through the
proper modeling of baryonic physics. For instance, the existence of dark
matter cores can potentially be explained by the gravitational transfer of energy from supernovae into the orbits of dark matter particles\cite{Navarro1996b, Governato2012,Onorbe2015,Chan2015,Read2016}.
Alternatively, these discrepancies between observations and cold dark matter simulations can also be explored through alternative dark matter models. These small-scale problems have therefore generated significant interest in the exploration of alternative dark matter scenarios. 
\end{tcolorbox}
\normalsize

\subsection{Alternative Forms of Dark Matter}
A wide range of alternative dark matter models have been proposed over the last decades. However, not all of these models have been studied in detail through simulations. Mostly three main classes of alternative dark matter models have been simulated: warm dark matter, self-interacting dark matter, and fuzzy dark matter. Many of these models have been invoked to address small-scale problems of the cold dark matter paradigm (see Box 2).\\[-0.3cm]

\noindent {\bf Warm Dark Matter:} Cold dark matter models exhibit a high-$k$ cut-off in the initial power spectrum due to free-streaming or collisional damping.
For a canonical weakly interactive massive particle this cut-off is of the order of $1$ comoving parsec corresponding to a mass scale of $10^{-6}\,{\rm M}_\odot$\cite{Green2004}. Warm dark matter models, on the other hand, have an effective free-streaming length
$\lambda_{\rm fs}$ that scales inversely with particle mass\cite{Bode2001}.
For recent cosmologies\cite{Planck2016}, this relation is approximately $\lambda_{\rm fs}=33 (m_{\rm WDM}/1\,{\rm keV})^{-1.11}\,{\rm kpc}$ and the corresponding free-streaming mass is $M_{\rm fs}=2 \times 10^{7} (m_{\rm WDM}/1\,{\rm keV})^{-3.33}\,{\rm M}_\odot$. The reduction of small-scale power within warm dark matter models has two consequences: first, a reduction of low mass halos, and second a reduction of the central density of halos. Simulations of warm dark matter models are typically carried out with the same numerical methods as cold dark matter simulations, but with modified initial conditions. However, the power spectrum cut-off leads to artificial and numerical discreteness effects in $N$-body simulations\cite{Wang2007}. Special care is then required to avoid a contamination of results in that case. More recently alternative methods based on phase-space tessellation techniques have been employed to study warm dark matter models avoiding these numerical artifacts\cite{Angulo2013}.\\[-0.3cm]

\noindent {\bf Self-Interacting Dark Matter:} Dark matter models that involve dark matter self-interactions\cite{Carlson1992,Spergel2000} have also been explored extensively. Self-interactions are commonly quantified in terms of the
cross section per unit particle mass, $\sigma/m$. Models with constant and velocity-dependent cross sections have both been studied with simulations\cite{vogelsberger2012}. The
high central dark matter densities observed in clusters exclude self-interacting dark matter models with $\sigma/m \gtrsim
0.5\,{\rm cm}^2/{\rm g}$ for these cluster mass scales.  
Recently, more general self-interacting dark matter models have been suggested. Those have both truncated power spectra and  self-interactions\cite{CyrRacine2016,Vogelsberger2016}. Such models affect the internal structure of dark matter halos through the scattering of particles that cause the formation of density cores. On the other hand, the truncated power spectra also lead, similar to warm dark matter models, to a suppression of halo substructure. 
Various recent simulations have demonstrated that models with $\sigma/m \approx 0.5 - 10\,{\rm cm}^2 /{\rm g}$
produce dark matter cores in dwarf galaxies with sizes $\sim  0.3 - 1.5\,{\rm kpc}$ and central densities
$2 - 0.2 \times 10^8\,{\rm M}_\odot/{\rm kpc}^3 = 7.4 - 0.74\,{\rm GeV}/{\rm cm}^3$
that can alleviate some cold dark matter small-scale problems\cite{Vogelsberger2012b,Peter2013,Fry2015,Elbert2015}. Simulations of self-interacting dark matter are based on the $N$-body approach coupled to a local Monte Carlo-based probabilistic scattering scheme to model particle self-interactions. \\[-0.3cm]

\noindent{\bf Fuzzy Dark Matter:} An ultralight bosonic scalar field is a completely different alternative to the cold dark matter paradigm\cite{Hui2017}, where a
bosonic fluid with a particle mass of $m \sim 10^{-22}\,{\rm eV}$ suppresses small-scale structure  owing to macroscopic quantum properties\cite{Lee1996,Hu2000,Peebles2000} with a typical de Broglie wavelength of  $\lambda_{\rm DB} \sim 1\,{\rm kpc}$\cite{Chavanis2011,Suarez2014}. The dark matter fluid forms in this case
a cosmological Bose-Einstein condensate\cite{Matos2009,Lundgren2010,Robles2013}.
Such an ultralight scalar field of spin-0 at zero temperature is
described in the non-relativistic limit by the Schr\"odinger-Poisson equations \cite{Seidel1990,Sin1994,Lee1996,Hu2000}: $i\hbar \partial \psi/\partial t = -\hbar^2/2 m {\boldsymbol \nabla}^2 \psi + m V \psi$ and ${\boldsymbol \nabla }^2 V = 4 \pi G (\rho - \bar\rho)$, where $\rho = |\psi|^2$ is the fluid density, 
$\bar \rho$
is the mean density, and $V$ is the potential. One consequence of the macroscopic quantum behavior of the fluid is that the
fluid admits stable, minimum-energy soliton configurations
forming at the centers of self-gravitating halos. These ${\rm kpc}$-scale soliton cores offer one possible solution to the cusp-core problem of cold dark matter.
Numerically, the Schr\"odinger-Poisson equations can, for example, be solved through adaptive spectral methods or through a reformulation into a hydrodynamics problem, that can be solved with hydrodynamical discretization techniques, based on the  Madelung formulation\cite{Mocz2015, Nori2018}.

\subsection{Alternative Forms of Dark Energy}

Cosmological simulations must include at least a cosmological constant to account for the accelerated
expansion of the Universe. A wide range of alternative dark energy models have, however, been considered
in the literature\cite{Copeland2006} and a number of these have also been studied with simulations\cite{Baldi2012}.
\\[-0.3cm]

\noindent{\bf Dynamical dark energy:} The most simple extension in the dark energy sector is to assume a dark energy density that is time dependent but still spatially 
homogeneous -- at least on sub-horizon scales. This behavior can, for example, be obtained in scalar field models
of dark energy\cite{Baldi2012}. Cosmic structure growth is then only affected via an altered background expansion.
The only change required to perform cosmological simulations of such models is
then to modify the calculation of the Hubble expansion rate in the numerical integration\cite{Linder2003,Grossi2009}.
As the growth function is different than in $\Lambda$CDM, extra care is also required when choosing the
amplitude of matter density fluctuations in the initial conditions, i.e. taking into account at what redshift
observational constraints on the amount of fluctuations are aimed to be matched. For example, models
with a higher dark energy density at early times suppress structure growth and have hence a lower amplitude
of fluctuations at redshift zero for the same scalar amplitude in the cosmic microwave background\cite{Linder2003,Grossi2009}.
Dynamical dark energy can have a surprisingly large impact on galaxy properties in simulations\cite{Penzo2014}. In practice, this results in degeneracies
between cosmology and the feedback physics that is required to match observations.
\\[-0.3cm]

\noindent{\bf Inhomogeneous dark energy:} Models of dark energy that exhibit sizable spatial
fluctuations within the horizon represent the next level of complexity. For such models, and even more so for the coupled dark energy models, a clear distinction between dark energy and modified gravity is often not possible as
accelerations arising from spatial fluctuations in the dark energy field can also be interpreted
as modifications to the laws of gravity. Relatively little simulation work has been done on models
in which inhomogeneous dark energy interacts with matter only gravitationally, such as, for example, in the
clustering dark energy scenario\cite{Sefusatti2011}.
\\[-0.3cm]

\noindent{\bf Coupled dark energy:} In the hope to alleviate the puzzle of the similar energy density of matter
and dark energy at the present cosmic epoch, additional non-gravitational couplings between these sectors have been proposed\cite{Amendola2001}.
Such a coupling of dark energy to matter could either be universal, i.e. involving all
matter species, or non-universal, with dark energy, for example, coupling only to dark matter but not to baryons. Models 
with a universal coupling typically require a screening mechanism that hides its effects in dense environments
like the solar system, where experimental tests of gravity tightly constrain a direct coupling to baryons.
In contrast, models with a coupling only to dark matter are observationally much less constrained. In both cases, 
growing perturbations in the matter density field can naturally give rise to corresponding fluctuations
in the coupled dark energy field.
Coupled dark energy scenarios have been widely studied with simulations, either avoiding\cite{Maccio2004,Baldi2010} or including\cite{Li2011b, Li2011} a treatment of the
spatial fluctuations of dark energy. In the former case, the main effects of coupling terms are a time
dependence of the gravitating particle mass of the coupled matter species, as well as a velocity dependent friction
term. Accounting for the spatial fluctuation additionally results in a fifth force proportional to the gradient of the dark energy field. These effects have, for example, been found to lower the concentrations and
baryon fraction of halos\cite{Baldi2010}, thereby reducing potential tensions compared to a $\Lambda$CDM cosmology.

\subsection{Alternative Forms of Gravity}
\label{sec:mod_grav}

While general relativity has been tested to high precision within the solar system, constraints on galactic and intergalactic scales are much weaker. Indeed, additional components that have so far not been directly observed, dark matter and dark energy, need to be added to allow a viable description of cosmology by general relativity. As an alternative, modifications of the laws of gravity have been proposed, which could make at least one of these components obsolete.\\[-0.3cm]

\noindent{\bf Modified gravity as an alternative to dark matter:} Dark matter models successfully explain observations on many different scales, including the cosmic microwave background, the Lyman-$\alpha$ forest, the clustering of galaxies, and the internal dynamics of galaxies and galaxy clusters. Most work aimed at replacing the role of dark matter by a modification of the laws of gravity has focused only on a subset of these areas. For example, modified Newtonian dynamics\cite{Milgrom1983,Famaey2012}, a change in Newton's second law at small acceleration values ($\vec{F}=m \mu(|\vec{a}|/a_0) \vec{a}$, with $a_0 \sim 10^{-10} \mathrm{m} \, \mathrm{s}^{-2}$ and $\mu(x) \rightarrow 1$ for $x \gg 1$ and $\mu(x) \rightarrow x$ for $x \ll 1$), or alternatively a change in Poisson's equation of Newtonian gravity ($\vec{\nabla}\cdot(\mu(|\vec{a}|/a_0) \vec{a}) = 4 \pi \rho$), has been proposed to account for the flat rotation curves of galaxies at large radii. Since this modified Poisson's equation is non-linear,  gravity algorithms that are based on the principle of linear force superposition such as direct summation, tree and Fourier transform-based schemes are not suitable to simulate these kinds of models. Simulations have therefore been performed with the multigrid method with the full approximation scheme\cite{Brandt1977}. The non-linear partial differential equation is then discretized on a grid with a finite difference representation of the differential operator and iteratively solved using Gauss-Seidel relaxation. Since modified Newtonian dynamics is not a relativistic theory, relativistic extensions of it have also been proposed, for example, tensor-vector-scalar gravity (TeVeS)\cite{Bekenstein2004,Skordis2008}. Here gravity is mediated by a tensor (metric), vector and scalar field. However, these models have not been widely studied in full cosmological simulations yet. Models without a dark matter component such as modified Newtonian dynamics also naturally account for the tight observed relation between the gravitational acceleration inferred from galaxy rotation curves and that expected from the observed baryonic mass\cite{McGaugh2016,Lelli2017}. However, galaxy formation simulations within the $\Lambda$CDM framework can also produce a sufficiently tight relation\cite{Keller2017,Ludlow2017,Dutton2019}.
\\[-0.3cm]

\noindent{\bf Modified gravity as an alternative to dark energy:} Dark energy has only been observed through its impact on the background expansion of the Universe. Replacing dark energy with a modification to the laws of gravity is, compared to replacing dark matter, easier. In fact, a cosmological constant in the Einstein field equations can also be interpreted as modified gravity rather than an unexpectedly small zero-point energy of a quantum field. In the literature, a wide range of much more sophisticated modified gravity theories have been considered. 
While many of these can account for the observed accelerated expansion of the Universe, Occam's razor would typically disfavor them compared to a cosmological constant in the absence of observational evidence beyond the observed background expansion. Cosmological simulations have been widely used to investigate the observational signatures of such extended gravity models to guide observational searches for potential modifications of gravity over a wide range of scales. Many modified gravity models that exhibit interesting behavior on, for example, galactic and intergalactic scales have been designed such that they approach general relativity in dense environments such as the solar system to avoid violating experimental constraints. Such screening mechanisms typically involve non-linear partial differential equations, which renders them numerically challenging and requires tailored techniques\cite{Llinares2018}. Most schemes resort to the multigrid method with the full approximation scheme\cite{Brandt1977}, e.g. employed on an adaptively refining mesh\cite{Li2012,Puchwein2013,Llinares2014}. With such methods cosmological simulations have been carried out for a number of screened modified gravity models, including Chameleon-$f(R)$, DGP, symmetron, dilaton, and Galileon gravity\cite{Brax2012,Barreira2013,Winther2015}. 
Most such studies focused on collisionless simulations. Semi-analytical galaxy formation models combined with Chameleon-$f(R)$ gravity demonstrated that the gravity modification effects on basic properties such as galaxy stellar mass functions and cosmic star formation rate densities are rather small and comparable to the uncertainties of the semi-analytical models\cite{Fontanot2013}. Clustering signals and relative velocities of halo pairs can, however, change notably\cite{Jennings2012, Fontanot2013}. Post-processing $\Lambda$CDM galaxy formation simulations with a modified gravity solver suggests that there should be characteristic changes in the internal kinematics of galaxies such as features in their rotation curves near the screening threshold\cite{Naik2018}, which can also result in degeneracies with the core/cusp problem\cite{Lombriser2015}. Fully self-consistent simulation studies of galaxy formation
in such screened modified gravity models have only started very recently \cite{Arnold2019}. Such simulations should in principle also take into account effects that modified gravity has on stellar physics\cite{Davis2012}.

\section{Conclusions and Outlook}

Cosmological simulations of galaxy formation play a crucial role for our understanding of galaxy formation. 
Especially, the last years have seen enormous progress on two fronts: large volume simulations modeling large samples of galaxies, and zoom simulations with refined galaxy formation models that resolve the physical processes in more detail. 
Modern galaxy formation simulations reproduce now a plethora of observational results and create virtual universes that are to first order nearly identical to the real Universe. At the same time, these simulations are also employed to explore alternative cosmological models with modifications to the nature of dark matter, dark energy and gravity. This progress in the field of galaxy formation simulations has mostly been driven by a better understanding of galaxy formation physics, refined numerical methods, and the steady growth of computing power. \\

\noindent Cosmological simulations of galaxy formation use a variety of different numerical methods, and different implementations of galaxy formation physics. Despite these differences, such simulations have now converged on a wide range of predictions for the evolution of galaxies. It therefore seems that the basic physical mechanisms that shape the galaxy population have been identified, and that their current modeling is sufficient to produce realistic galaxy populations. However, these physical processes are implemented through still rather crude sub-resolution models. Sub-resolution models aim to capture the relevant physics through an effective description. In fact, cosmological simulations will always have to rely on these sub-resolution models since truly ab initio cosmological simulations of galaxy formation are and will remain impossible. 
One danger associated with the application of sub-resolution models is the belief that the reproduction of large amounts of observational data automatically implies a correct and physically
plausible effective model and therefore detailed understanding of galaxy formation. This is problematic since sub-resolution models contain per construction a certain number of adjustable and degenerate parameters, and at the same time do not really capture the detailed physics at play but only provide an effective coarse description. Caution is therefore required to not over-interpret some of the recent successes generated by these models. \\

\noindent One of the next goals of computational galaxy formation is to understand which  detailed physical processes drive the outcomes of effective physical models. For example, many simulations employ rather crude and incomplete models for the generation of galactic outflows without a detailed modeling of the driving process. Future simulations should aim at understanding these processes in more detail to illuminate the true physical processes at work going beyond the crude effective models to gain more fundamental insights. This will also lead to a better understanding of what physics actually drives the overall behavior of currently existing coarse-grained effective sub-resolution models. While constructing new models and simulations, it is important to keep in mind that the major goal of simulations is not primarily to fit observed data, but rather to gain insights into galaxy formation physics. Advances in this direction benefit often more from failures of certain ideas or models, rather than a perfect reproduction of observational data that is to some degree subject to the calibration of free model parameters and the coarse-grained nature of the employed models. Another frontier of cosmological galaxy formation simulations is the desire to provide large volume simulations that match the statistical sample sizes of upcoming large observational surveys. This requires very large volume simulations with well-understood sub-resolution models. The development and better understanding of refined sub-resolution models, the desire to achieve higher numerical resolution, and simulations with larger volumes represent the main challenges of cosmological simulations of the next decade.

\section*{Acknowledgements}
We thank David Barnes, Mike Boylan-Kolchin, Lars Hernquist, Rahul Kannan, Hui Li, Stephanie O'Neil, R\"udiger Pakmor, Christoph Pfrommer, Laura Sales, Aaron Smith, Volker Springel, and Rainer Weinberger for useful comments. We also thank the reviewers for helpful feedback. MV acknowledges support through an MIT RSC award, a Kavli Research Investment Fund, NASA ATP grant NNX17AG29G, and NSF grants AST-1814053 and AST-1814259. FM acknowledges support through the Program ``Rita Levi
Montalcini'' of the Italian MIUR.

\clearpage

\tiny
\begin{multicols}{2}

\begin{thebibliography}{100}
\expandafter\ifx\csname url\endcsname\relax
  \def\url#1{\texttt{#1}}\fi
\expandafter\ifx\csname urlprefix\endcsname\relax\def\urlprefix{URL }\fi
\expandafter\ifx\csname doiprefix\endcsname\relax\def\doiprefix{DOI }\fi
\providecommand{\bibinfo}[2]{#2}
\providecommand{\eprint}[2][]{\url{#2}}

\bibitem{Somerville2015}
\bibinfo{author}{{Somerville}, R.~S.} \& \bibinfo{author}{{Dav{\'e}}, R.}
\newblock \bibinfo{journal}{\bibinfo{title}{{Physical Models of Galaxy
  Formation in a Cosmological Framework}}}.
\newblock {\emph{\JournalTitle{\araa}}} \textbf{\bibinfo{volume}{53}},
  \bibinfo{pages}{51--113} (\bibinfo{year}{2015}).
\newblock \doiprefix 10.1146/annurev-astro-082812-140951.
\newblock \eprint{1412.2712}.

\bibitem{Naab2017}
\bibinfo{author}{{Naab}, T.} \& \bibinfo{author}{{Ostriker}, J.~P.}
\newblock \bibinfo{journal}{\bibinfo{title}{{Theoretical Challenges in Galaxy
  Formation}}}.
\newblock {\emph{\JournalTitle{\araa}}} \textbf{\bibinfo{volume}{55}},
  \bibinfo{pages}{59--109} (\bibinfo{year}{2017}).
\newblock \doiprefix 10.1146/annurev-astro-081913-040019.
\newblock \eprint{1612.06891}.

\bibitem{Planck2016}
\bibinfo{author}{{Planck Collaboration}} \emph{et~al.}
\newblock \bibinfo{journal}{\bibinfo{title}{{Planck 2015 results. XIII.
  Cosmological parameters}}}.
\newblock {\emph{\JournalTitle{\aap}}} \textbf{\bibinfo{volume}{594}},
  \bibinfo{pages}{A13} (\bibinfo{year}{2016}).
\newblock \doiprefix 10.1051/0004-6361/201525830.
\newblock \eprint{1502.01589}.

\bibitem{Seljak1996}
\bibinfo{author}{{Seljak}, U.} \& \bibinfo{author}{{Zaldarriaga}, M.}
\newblock \bibinfo{journal}{\bibinfo{title}{{A Line-of-Sight Integration
  Approach to Cosmic Microwave Background Anisotropies}}}.
\newblock {\emph{\JournalTitle{\apj}}} \textbf{\bibinfo{volume}{469}},
  \bibinfo{pages}{437} (\bibinfo{year}{1996}).
\newblock \doiprefix 10.1086/177793.
\newblock \eprint{astro-ph/9603033}.

\bibitem{Peacock1997}
\bibinfo{author}{{Peacock}, J.~A.}
\newblock \bibinfo{journal}{\bibinfo{title}{{The evolution of galaxy
  clustering}}}.
\newblock {\emph{\JournalTitle{\mnras}}} \textbf{\bibinfo{volume}{284}},
  \bibinfo{pages}{885--898} (\bibinfo{year}{1997}).
\newblock \doiprefix 10.1093/mnras/284.4.885.
\newblock \eprint{astro-ph/9608151}.

\bibitem{Eisenstein1998}
\bibinfo{author}{{Eisenstein}, D.~J.} \& \bibinfo{author}{{Hu}, W.}
\newblock \bibinfo{journal}{\bibinfo{title}{{Baryonic Features in the Matter
  Transfer Function}}}.
\newblock {\emph{\JournalTitle{\apj}}} \textbf{\bibinfo{volume}{496}},
  \bibinfo{pages}{605--614} (\bibinfo{year}{1998}).
\newblock \doiprefix 10.1086/305424.
\newblock \eprint{astro-ph/9709112}.

\bibitem{Eisenstein1999}
\bibinfo{author}{{Eisenstein}, D.~J.} \& \bibinfo{author}{{Hu}, W.}
\newblock \bibinfo{journal}{\bibinfo{title}{{Power Spectra for Cold Dark Matter
  and Its Variants}}}.
\newblock {\emph{\JournalTitle{\apj}}} \textbf{\bibinfo{volume}{511}},
  \bibinfo{pages}{5--15} (\bibinfo{year}{1999}).
\newblock \doiprefix 10.1086/306640.
\newblock \eprint{astro-ph/9710252}.

\bibitem{Baugh1995}
\bibinfo{author}{{Baugh}, C.~M.}, \bibinfo{author}{{Gaztanaga}, E.} \&
  \bibinfo{author}{{Efstathiou}, G.}
\newblock \bibinfo{journal}{\bibinfo{title}{{A comparison of the evolution of
  density fields in perturbation theory and numerical simulations - II.
  Counts-in-cells analysis}}}.
\newblock {\emph{\JournalTitle{\mnras}}} \textbf{\bibinfo{volume}{274}},
  \bibinfo{pages}{1049--1070} (\bibinfo{year}{1995}).
\newblock \doiprefix 10.1093/mnras/274.4.1049.
\newblock \eprint{astro-ph/9408057}.

\bibitem{White1996}
\bibinfo{author}{{White}, S.~D.~M.}
\newblock \bibinfo{title}{{Cosmology and large scale structure}}.
\newblock In \bibinfo{editor}{{Schaeffer}, R.}, \bibinfo{editor}{{Silk}, J.},
  \bibinfo{editor}{{Spiro}, M.} \& \bibinfo{editor}{{Zinn-Justin}, J.} (eds.)
  \emph{\bibinfo{booktitle}{Cosmology and Large Scale Structure}},
  \bibinfo{pages}{349--430} (\bibinfo{year}{1996}).

\bibitem{Zeldovich1970}
\bibinfo{author}{{Zel'dovich}, Y.~B.}
\newblock \bibinfo{journal}{\bibinfo{title}{{Gravitational instability: An
  approximate theory for large density perturbations.}}}
\newblock {\emph{\JournalTitle{\aap}}} \textbf{\bibinfo{volume}{5}},
  \bibinfo{pages}{84--89} (\bibinfo{year}{1970}).

\bibitem{Bertschinger2001}
\bibinfo{author}{{Bertschinger}, E.}
\newblock \bibinfo{journal}{\bibinfo{title}{{Multiscale Gaussian Random Fields
  and Their Application to Cosmological Simulations}}}.
\newblock {\emph{\JournalTitle{\apjs}}} \textbf{\bibinfo{volume}{137}},
  \bibinfo{pages}{1--20} (\bibinfo{year}{2001}).
\newblock \doiprefix 10.1086/322526.
\newblock \eprint{astro-ph/0103301}.

\bibitem{Jenkins2010}
\bibinfo{author}{{Jenkins}, A.}
\newblock \bibinfo{journal}{\bibinfo{title}{{Second-order Lagrangian
  perturbation theory initial conditions for resimulations}}}.
\newblock {\emph{\JournalTitle{\mnras}}} \textbf{\bibinfo{volume}{403}},
  \bibinfo{pages}{1859--1872} (\bibinfo{year}{2010}).
\newblock \doiprefix 10.1111/j.1365-2966.2010.16259.x.
\newblock \eprint{0910.0258}.

\bibitem{Hahn2011}
\bibinfo{author}{{Hahn}, O.} \& \bibinfo{author}{{Abel}, T.}
\newblock \bibinfo{journal}{\bibinfo{title}{{Multi-scale initial conditions for
  cosmological simulations}}}.
\newblock {\emph{\JournalTitle{\mnras}}} \textbf{\bibinfo{volume}{415}},
  \bibinfo{pages}{2101--2121} (\bibinfo{year}{2011}).
\newblock \doiprefix 10.1111/j.1365-2966.2011.18820.x.
\newblock \eprint{1103.6031}.

\bibitem{Garrison2016}
\bibinfo{author}{{Garrison}, L.~H.}, \bibinfo{author}{{Eisenstein}, D.~J.},
  \bibinfo{author}{{Ferrer}, D.}, \bibinfo{author}{{Metchnik}, M.~V.} \&
  \bibinfo{author}{{Pinto}, P.~A.}
\newblock \bibinfo{journal}{\bibinfo{title}{{Improving initial conditions for
  cosmological N-body simulations}}}.
\newblock {\emph{\JournalTitle{\mnras}}} \textbf{\bibinfo{volume}{461}},
  \bibinfo{pages}{4125--4145} (\bibinfo{year}{2016}).
\newblock \doiprefix 10.1093/mnras/stw1594.
\newblock \eprint{1605.02333}.

\bibitem{Hoffman1991}
\bibinfo{author}{{Hoffman}, Y.} \& \bibinfo{author}{{Ribak}, E.}
\newblock \bibinfo{journal}{\bibinfo{title}{{Constrained realizations of
  Gaussian fields - A simple algorithm}}}.
\newblock {\emph{\JournalTitle{\apjl}}} \textbf{\bibinfo{volume}{380}},
  \bibinfo{pages}{L5--L8} (\bibinfo{year}{1991}).
\newblock \doiprefix 10.1086/186160.

\bibitem{Salmon1996}
\bibinfo{author}{{Salmon}, J.}
\newblock \bibinfo{journal}{\bibinfo{title}{{Generation of Correlated and
  Constrained Gaussian Stochastic Processes for N-Body Simulations}}}.
\newblock {\emph{\JournalTitle{\apj}}} \textbf{\bibinfo{volume}{460}},
  \bibinfo{pages}{59} (\bibinfo{year}{1996}).
\newblock \doiprefix 10.1086/176952.

\bibitem{Price2007}
\bibinfo{author}{{Price}, D.~J.} \& \bibinfo{author}{{Monaghan}, J.~J.}
\newblock \bibinfo{journal}{\bibinfo{title}{{An energy-conserving formalism for
  adaptive gravitational force softening in smoothed particle hydrodynamics and
  N-body codes}}}.
\newblock {\emph{\JournalTitle{\mnras}}} \textbf{\bibinfo{volume}{374}},
  \bibinfo{pages}{1347--1358} (\bibinfo{year}{2007}).
\newblock \doiprefix 10.1111/j.1365-2966.2006.11241.x.
\newblock \eprint{astro-ph/0610872}.

\bibitem{Barnes1986}
\bibinfo{author}{{Barnes}, J.} \& \bibinfo{author}{{Hut}, P.}
\newblock \bibinfo{journal}{\bibinfo{title}{{A hierarchical O(N log N)
  force-calculation algorithm}}}.
\newblock {\emph{\JournalTitle{\nat}}} \textbf{\bibinfo{volume}{324}},
  \bibinfo{pages}{446--449} (\bibinfo{year}{1986}).
\newblock \doiprefix 10.1038/324446a0.

\bibitem{Dehnen2000}
\bibinfo{author}{{Dehnen}, W.}
\newblock \bibinfo{journal}{\bibinfo{title}{{A Very Fast and
  Momentum-conserving Tree Code}}}.
\newblock {\emph{\JournalTitle{\apjl}}} \textbf{\bibinfo{volume}{536}},
  \bibinfo{pages}{L39--L42} (\bibinfo{year}{2000}).
\newblock \doiprefix 10.1086/312724.
\newblock \eprint{astro-ph/0003209}.

\bibitem{Greengard1987}
\bibinfo{author}{{Greengard}, L.} \& \bibinfo{author}{{Rokhlin}, V.}
\newblock \bibinfo{journal}{\bibinfo{title}{{A fast algorithm for particle
  simulations}}}.
\newblock {\emph{\JournalTitle{Journal of Computational Physics}}}
  \textbf{\bibinfo{volume}{73}}, \bibinfo{pages}{325--348}
  (\bibinfo{year}{1987}).
\newblock \doiprefix 10.1016/0021-9991(87)90140-9.

\bibitem{Hernquist1991}
\bibinfo{author}{{Hernquist}, L.}, \bibinfo{author}{{Bouchet}, F.~R.} \&
  \bibinfo{author}{{Suto}, Y.}
\newblock \bibinfo{journal}{\bibinfo{title}{{Application of the Ewald method to
  cosmological N-body simulations}}}.
\newblock {\emph{\JournalTitle{\apjs}}} \textbf{\bibinfo{volume}{75}},
  \bibinfo{pages}{231--240} (\bibinfo{year}{1991}).
\newblock \doiprefix 10.1086/191530.

\bibitem{Ewald1921}
\bibinfo{author}{{Ewald}, P.~P.}
\newblock \bibinfo{journal}{\bibinfo{title}{{Die Berechnung optischer und
  elektrostatischer Gitterpotentiale}}}.
\newblock {\emph{\JournalTitle{Annalen der Physik}}}
  \textbf{\bibinfo{volume}{369}}, \bibinfo{pages}{253--287}
  (\bibinfo{year}{1921}).
\newblock \doiprefix 10.1002/andp.19213690304.

\bibitem{Hockney1981}
\bibinfo{author}{{Hockney}, R.~W.} \& \bibinfo{author}{{Eastwood}, J.~W.}
\newblock \emph{\bibinfo{title}{{Computer Simulation Using Particles}}}
  (\bibinfo{publisher}{New York: McGraw-Hill}, \bibinfo{year}{1981}).

\bibitem{Brandt1977}
\bibinfo{author}{{Brandt}, A.}
\newblock \bibinfo{journal}{\bibinfo{title}{{Multi-level adaptive solutions to
  boundary-value problems}}}.
\newblock {\emph{\JournalTitle{Math. Comp.}}} \textbf{\bibinfo{volume}{31}},
  \bibinfo{pages}{333--390} (\bibinfo{year}{1977}).
\newblock \doiprefix 10.1090/S0025-5718-1977-0431719-X.

\bibitem{Efstathiou1985}
\bibinfo{author}{{Efstathiou}, G.}, \bibinfo{author}{{Davis}, M.},
  \bibinfo{author}{{White}, S.~D.~M.} \& \bibinfo{author}{{Frenk}, C.~S.}
\newblock \bibinfo{journal}{\bibinfo{title}{{Numerical techniques for large
  cosmological N-body simulations}}}.
\newblock {\emph{\JournalTitle{\apjs}}} \textbf{\bibinfo{volume}{57}},
  \bibinfo{pages}{241--260} (\bibinfo{year}{1985}).
\newblock \doiprefix 10.1086/191003.

\bibitem{Bode2003}
\bibinfo{author}{{Bode}, P.} \& \bibinfo{author}{{Ostriker}, J.~P.}
\newblock \bibinfo{journal}{\bibinfo{title}{{Tree Particle-Mesh: An Adaptive,
  Efficient, and Parallel Code for Collisionless Cosmological Simulation}}}.
\newblock {\emph{\JournalTitle{\apjs}}} \textbf{\bibinfo{volume}{145}},
  \bibinfo{pages}{1--13} (\bibinfo{year}{2003}).
\newblock \doiprefix 10.1086/345538.
\newblock \eprint{astro-ph/0302065}.

\bibitem{Kravtsov1997}
\bibinfo{author}{{Kravtsov}, A.~V.}, \bibinfo{author}{{Klypin}, A.~A.} \&
  \bibinfo{author}{{Khokhlov}, A.~M.}
\newblock \bibinfo{journal}{\bibinfo{title}{{Adaptive Refinement Tree: A New
  High-Resolution N-Body Code for Cosmological Simulations}}}.
\newblock {\emph{\JournalTitle{\apjs}}} \textbf{\bibinfo{volume}{111}},
  \bibinfo{pages}{73--94} (\bibinfo{year}{1997}).
\newblock \doiprefix 10.1086/313015.
\newblock \eprint{astro-ph/9701195}.

\bibitem{Wang2007}
\bibinfo{author}{{Wang}, J.} \& \bibinfo{author}{{White}, S.~D.~M.}
\newblock \bibinfo{journal}{\bibinfo{title}{{Discreteness effects in
  simulations of hot/warm dark matter}}}.
\newblock {\emph{\JournalTitle{\mnras}}} \textbf{\bibinfo{volume}{380}},
  \bibinfo{pages}{93--103} (\bibinfo{year}{2007}).
\newblock \doiprefix 10.1111/j.1365-2966.2007.12053.x.
\newblock \eprint{astro-ph/0702575}.

\bibitem{Widrow1993}
\bibinfo{author}{{Widrow}, L.~M.} \& \bibinfo{author}{{Kaiser}, N.}
\newblock \bibinfo{journal}{\bibinfo{title}{{Using the Schroedinger Equation to
  Simulate Collisionless Matter}}}.
\newblock {\emph{\JournalTitle{\apjl}}} \textbf{\bibinfo{volume}{416}},
  \bibinfo{pages}{L71} (\bibinfo{year}{1993}).
\newblock \doiprefix 10.1086/187073.

\bibitem{Schaller2014}
\bibinfo{author}{{Schaller}, M.}, \bibinfo{author}{{Becker}, C.},
  \bibinfo{author}{{Ruchayskiy}, O.}, \bibinfo{author}{{Boyarsky}, A.} \&
  \bibinfo{author}{{Shaposhnikov}, M.}
\newblock \bibinfo{journal}{\bibinfo{title}{{A new framework for numerical
  simulations of structure formation}}}.
\newblock {\emph{\JournalTitle{\mnras}}} \textbf{\bibinfo{volume}{442}},
  \bibinfo{pages}{3073--3095} (\bibinfo{year}{2014}).
\newblock \doiprefix 10.1093/mnras/stu1069.
\newblock \eprint{1310.5102}.

\bibitem{Uhlemann2014}
\bibinfo{author}{{Uhlemann}, C.}, \bibinfo{author}{{Kopp}, M.} \&
  \bibinfo{author}{{Haugg}, T.}
\newblock \bibinfo{journal}{\bibinfo{title}{{Schr{\"o}dinger method as N-body
  double and UV completion of dust}}}.
\newblock {\emph{\JournalTitle{Phys. Rev. D}}} \textbf{\bibinfo{volume}{90}},
  \bibinfo{pages}{023517} (\bibinfo{year}{2014}).
\newblock \doiprefix 10.1103/PhysRevD.90.023517.
\newblock \eprint{1403.5567}.

\bibitem{Colombi2014}
\bibinfo{author}{{Colombi}, S.} \& \bibinfo{author}{{Touma}, J.}
\newblock \bibinfo{journal}{\bibinfo{title}{{Vlasov-Poisson in 1D:
  waterbags}}}.
\newblock {\emph{\JournalTitle{\mnras}}} \textbf{\bibinfo{volume}{441}},
  \bibinfo{pages}{2414--2432} (\bibinfo{year}{2014}).
\newblock \doiprefix 10.1093/mnras/stu739.

\bibitem{Colombi2015}
\bibinfo{author}{{Colombi}, S.}
\newblock \bibinfo{journal}{\bibinfo{title}{{Vlasov-Poisson in 1D for initially
  cold systems: post-collapse Lagrangian perturbation theory}}}.
\newblock {\emph{\JournalTitle{\mnras}}} \textbf{\bibinfo{volume}{446}},
  \bibinfo{pages}{2902--2920} (\bibinfo{year}{2015}).
\newblock \doiprefix 10.1093/mnras/stu2308.
\newblock \eprint{1411.4165}.

\bibitem{Vogelsberger2008}
\bibinfo{author}{{Vogelsberger}, M.}, \bibinfo{author}{{White}, S.~D.~M.},
  \bibinfo{author}{{Helmi}, A.} \& \bibinfo{author}{{Springel}, V.}
\newblock \bibinfo{journal}{\bibinfo{title}{{The fine-grained phase-space
  structure of cold dark matter haloes}}}.
\newblock {\emph{\JournalTitle{\mnras}}} \textbf{\bibinfo{volume}{385}},
  \bibinfo{pages}{236--254} (\bibinfo{year}{2008}).
\newblock \doiprefix 10.1111/j.1365-2966.2007.12746.x.
\newblock \eprint{0711.1105}.

\bibitem{Vogelsberger2011}
\bibinfo{author}{{Vogelsberger}, M.} \& \bibinfo{author}{{White}, S.~D.~M.}
\newblock \bibinfo{journal}{\bibinfo{title}{{Streams and caustics: the
  fine-grained structure of {$\Lambda$} cold dark matter haloes}}}.
\newblock {\emph{\JournalTitle{\mnras}}} \textbf{\bibinfo{volume}{413}},
  \bibinfo{pages}{1419--1438} (\bibinfo{year}{2011}).
\newblock \doiprefix 10.1111/j.1365-2966.2011.18224.x.
\newblock \eprint{1002.3162}.

\bibitem{Hahn2013}
\bibinfo{author}{{Hahn}, O.}, \bibinfo{author}{{Abel}, T.} \&
  \bibinfo{author}{{Kaehler}, R.}
\newblock \bibinfo{journal}{\bibinfo{title}{{A new approach to simulating
  collisionless dark matter fluids}}}.
\newblock {\emph{\JournalTitle{\mnras}}} \textbf{\bibinfo{volume}{434}},
  \bibinfo{pages}{1171--1191} (\bibinfo{year}{2013}).
\newblock \doiprefix 10.1093/mnras/stt1061.
\newblock \eprint{1210.6652}.

\bibitem{Yoshikawa2013}
\bibinfo{author}{{Yoshikawa}, K.}, \bibinfo{author}{{Yoshida}, N.} \&
  \bibinfo{author}{{Umemura}, M.}
\newblock \bibinfo{journal}{\bibinfo{title}{{Direct Integration of the
  Collisionless Boltzmann Equation in Six-dimensional Phase Space:
  Self-gravitating Systems}}}.
\newblock {\emph{\JournalTitle{\apj}}} \textbf{\bibinfo{volume}{762}},
  \bibinfo{pages}{116} (\bibinfo{year}{2013}).
\newblock \doiprefix 10.1088/0004-637X/762/2/116.
\newblock \eprint{1206.6152}.

\bibitem{Teyssier2002}
\bibinfo{author}{{Teyssier}, R.}
\newblock \bibinfo{journal}{\bibinfo{title}{{Cosmological hydrodynamics with
  adaptive mesh refinement. A new high resolution code called RAMSES}}}.
\newblock {\emph{\JournalTitle{\aap}}} \textbf{\bibinfo{volume}{385}},
  \bibinfo{pages}{337--364} (\bibinfo{year}{2002}).
\newblock \doiprefix 10.1051/0004-6361:20011817.
\newblock \eprint{arXiv:astro-ph/0111367}.

\bibitem{Springel2005b}
\bibinfo{author}{{Springel}, V.}
\newblock \bibinfo{journal}{\bibinfo{title}{{The cosmological simulation code
  GADGET-2}}}.
\newblock {\emph{\JournalTitle{MNRAS}}} \textbf{\bibinfo{volume}{364}},
  \bibinfo{pages}{1105--1134} (\bibinfo{year}{2005}).
\newblock \doiprefix 10.1111/j.1365-2966.2005.09655.x.
\newblock \eprint{arXiv:astro-ph/0505010}.

\bibitem{Springel2010}
\bibinfo{author}{{Springel}, V.}
\newblock \bibinfo{journal}{\bibinfo{title}{{E pur si muove: Galilean-invariant
  cosmological hydrodynamical simulations on a moving mesh}}}.
\newblock {\emph{\JournalTitle{\mnras}}} \textbf{\bibinfo{volume}{401}},
  \bibinfo{pages}{791--851} (\bibinfo{year}{2010}).
\newblock \doiprefix 10.1111/j.1365-2966.2009.15715.x.
\newblock \eprint{0901.4107}.

\bibitem{Bryan2014}
\bibinfo{author}{{Bryan}, G.~L.} \emph{et~al.}
\newblock \bibinfo{journal}{\bibinfo{title}{{ENZO: An Adaptive Mesh Refinement
  Code for Astrophysics}}}.
\newblock {\emph{\JournalTitle{\apjs}}} \textbf{\bibinfo{volume}{211}},
  \bibinfo{pages}{19} (\bibinfo{year}{2014}).
\newblock \doiprefix 10.1088/0067-0049/211/2/19.
\newblock \eprint{1307.2265}.

\bibitem{Jetley2008}
\bibinfo{author}{Jetley, P.}, \bibinfo{author}{Kale, L.~V.},
  \bibinfo{author}{Gioachin, F.}, \bibinfo{author}{Quinn, T.} \&
  \bibinfo{author}{Mendes, C.}
\newblock \bibinfo{title}{Massively parallel cosmological simulations with
  changa}.
\newblock In \emph{\bibinfo{booktitle}{2008 IEEE International Parallel \&
  Distributed Processing Symposium}}, \bibinfo{pages}{1--12}
  (\bibinfo{publisher}{IEEE Computer Society}, \bibinfo{address}{Los Alamitos,
  CA, USA}, \bibinfo{year}{2008}).
\newblock \doiprefix 10.1109/IPDPS.2008.4536319.

\bibitem{Gioachin2010}
\bibinfo{author}{Gioachin, F.}, \bibinfo{author}{Kal\'e, L.~V.},
  \bibinfo{author}{Quinn, T.~R.}, \bibinfo{author}{Jetley, P.} \&
  \bibinfo{author}{Wesolowski, L.}
\newblock \bibinfo{title}{Scaling hierarchical n-body simulations on gpu
  clusters}.
\newblock In \emph{\bibinfo{booktitle}{SC Conference}}, \bibinfo{pages}{1--11}
  (\bibinfo{publisher}{IEEE Computer Society}, \bibinfo{address}{Los Alamitos,
  CA, USA}, \bibinfo{year}{2010}).
\newblock \doiprefix 10.1109/SC.2010.49.

\bibitem{Menon2015}
\bibinfo{author}{{Menon}, H.} \emph{et~al.}
\newblock \bibinfo{journal}{\bibinfo{title}{{Adaptive techniques for clustered
  N-body cosmological simulations}}}.
\newblock {\emph{\JournalTitle{Computational Astrophysics and Cosmology}}}
  \textbf{\bibinfo{volume}{2}}, \bibinfo{pages}{1} (\bibinfo{year}{2015}).
\newblock \doiprefix 10.1186/s40668-015-0007-9.
\newblock \eprint{1409.1929}.

\bibitem{Hopkins2015}
\bibinfo{author}{{Hopkins}, P.~F.}
\newblock \bibinfo{journal}{\bibinfo{title}{{A new class of accurate, mesh-free
  hydrodynamic simulation methods}}}.
\newblock {\emph{\JournalTitle{\mnras}}} \textbf{\bibinfo{volume}{450}},
  \bibinfo{pages}{53--110} (\bibinfo{year}{2015}).
\newblock \doiprefix 10.1093/mnras/stv195.
\newblock \eprint{1409.7395}.

\bibitem{Habib2016}
\bibinfo{author}{{Habib}, S.} \emph{et~al.}
\newblock \bibinfo{journal}{\bibinfo{title}{{HACC: Simulating sky surveys on
  state-of-the-art supercomputing architectures}}}.
\newblock {\emph{\JournalTitle{\na}}} \textbf{\bibinfo{volume}{42}},
  \bibinfo{pages}{49--65} (\bibinfo{year}{2016}).
\newblock \doiprefix 10.1016/j.newast.2015.06.003.
\newblock \eprint{1410.2805}.

\bibitem{Potter2017}
\bibinfo{author}{{Potter}, D.}, \bibinfo{author}{{Stadel}, J.} \&
  \bibinfo{author}{{Teyssier}, R.}
\newblock \bibinfo{journal}{\bibinfo{title}{{PKDGRAV3: beyond trillion particle
  cosmological simulations for the next era of galaxy surveys}}}.
\newblock {\emph{\JournalTitle{Computational Astrophysics and Cosmology}}}
  \textbf{\bibinfo{volume}{4}}, \bibinfo{pages}{2} (\bibinfo{year}{2017}).
\newblock \doiprefix 10.1186/s40668-017-0021-1.
\newblock \eprint{1609.08621}.

\bibitem{Wadsley2017}
\bibinfo{author}{{Wadsley}, J.~W.}, \bibinfo{author}{{Keller}, B.~W.} \&
  \bibinfo{author}{{Quinn}, T.~R.}
\newblock \bibinfo{journal}{\bibinfo{title}{{Gasoline2: a modern smoothed
  particle hydrodynamics code}}}.
\newblock {\emph{\JournalTitle{\mnras}}} \textbf{\bibinfo{volume}{471}},
  \bibinfo{pages}{2357--2369} (\bibinfo{year}{2017}).
\newblock \doiprefix 10.1093/mnras/stx1643.
\newblock \eprint{1707.03824}.

\bibitem{Schaller2018}
\bibinfo{author}{{Schaller}, M.}, \bibinfo{author}{{Gonnet}, P.},
  \bibinfo{author}{{Chalk}, A. B.~G.} \& \bibinfo{author}{{Draper}, P.~W.}
\newblock \bibinfo{title}{{SWIFT: SPH With Inter-dependent Fine-grained
  Tasking}} (\bibinfo{year}{2018}).
\newblock \eprint{1805.020}.

\bibitem{Press1974}
\bibinfo{author}{{Press}, W.~H.} \& \bibinfo{author}{{Schechter}, P.}
\newblock \bibinfo{journal}{\bibinfo{title}{{Formation of Galaxies and Clusters
  of Galaxies by Self-Similar Gravitational Condensation}}}.
\newblock {\emph{\JournalTitle{\apj}}} \textbf{\bibinfo{volume}{187}},
  \bibinfo{pages}{425--438} (\bibinfo{year}{1974}).
\newblock \doiprefix 10.1086/152650.

\bibitem{White1976}
\bibinfo{author}{{White}, S.~D.~M.}
\newblock \bibinfo{journal}{\bibinfo{title}{{The dynamics of rich clusters of
  galaxies}}}.
\newblock {\emph{\JournalTitle{\mnras}}} \textbf{\bibinfo{volume}{177}},
  \bibinfo{pages}{717--733} (\bibinfo{year}{1976}).
\newblock \doiprefix 10.1093/mnras/177.3.717.

\bibitem{Aarseth1979}
\bibinfo{author}{{Aarseth}, S.~J.}, \bibinfo{author}{{Gott}, J.~R., III} \&
  \bibinfo{author}{{Turner}, E.~L.}
\newblock \bibinfo{journal}{\bibinfo{title}{{N-body simulations of galaxy
  clustering. I - Initial conditions and galaxy collapse times}}}.
\newblock {\emph{\JournalTitle{\apj}}} \textbf{\bibinfo{volume}{228}},
  \bibinfo{pages}{664--683} (\bibinfo{year}{1979}).
\newblock \doiprefix 10.1086/156892.

\bibitem{Efstathiou1979}
\bibinfo{author}{{Efstathiou}, G.}
\newblock \bibinfo{journal}{\bibinfo{title}{{The clustering of galaxies and its
  dependence upon Omega}}}.
\newblock {\emph{\JournalTitle{\mnras}}} \textbf{\bibinfo{volume}{187}},
  \bibinfo{pages}{117--127} (\bibinfo{year}{1979}).
\newblock \doiprefix 10.1093/mnras/187.2.117.

\bibitem{Skillman2014}
\bibinfo{author}{{Skillman}, S.~W.} \emph{et~al.}
\newblock \bibinfo{journal}{\bibinfo{title}{{Dark Sky Simulations: Early Data
  Release}}}.
\newblock {\emph{\JournalTitle{arXiv e-prints}}}  (\bibinfo{year}{2014}).
\newblock \eprint{1407.2600}.

\bibitem{Gnedin2004}
\bibinfo{author}{{Gnedin}, O.~Y.}, \bibinfo{author}{{Kravtsov}, A.~V.},
  \bibinfo{author}{{Klypin}, A.~A.} \& \bibinfo{author}{{Nagai}, D.}
\newblock \bibinfo{journal}{\bibinfo{title}{{Response of Dark Matter Halos to
  Condensation of Baryons: Cosmological Simulations and Improved Adiabatic
  Contraction Model}}}.
\newblock {\emph{\JournalTitle{\apj}}} \textbf{\bibinfo{volume}{616}},
  \bibinfo{pages}{16--26} (\bibinfo{year}{2004}).
\newblock \doiprefix 10.1086/424914.
\newblock \eprint{astro-ph/0406247}.

\bibitem{DiCintio2014b}
\bibinfo{author}{{Di Cintio}, A.} \emph{et~al.}
\newblock \bibinfo{journal}{\bibinfo{title}{{The dependence of dark matter
  profiles on the stellar-to-halo mass ratio: a prediction for cusps versus
  cores}}}.
\newblock {\emph{\JournalTitle{\mnras}}} \textbf{\bibinfo{volume}{437}},
  \bibinfo{pages}{415--423} (\bibinfo{year}{2014}).
\newblock \doiprefix 10.1093/mnras/stt1891.
\newblock \eprint{1306.0898}.

\bibitem{Zhu2016}
\bibinfo{author}{{Zhu}, Q.} \emph{et~al.}
\newblock \bibinfo{journal}{\bibinfo{title}{{Baryonic impact on the dark matter
  distribution in Milky Way-sized galaxies and their satellites}}}.
\newblock {\emph{\JournalTitle{\mnras}}} \textbf{\bibinfo{volume}{458}},
  \bibinfo{pages}{1559--1580} (\bibinfo{year}{2016}).
\newblock \doiprefix 10.1093/mnras/stw374.

\bibitem{Benitez-Llambay2018}
\bibinfo{author}{{Ben{\'\i}tez-Llambay}, A.}, \bibinfo{author}{{Frenk}, C.~S.},
  \bibinfo{author}{{Ludlow}, A.~D.} \& \bibinfo{author}{{Navarro}, J.~F.}
\newblock \bibinfo{journal}{\bibinfo{title}{{Baryon-induced dark matter cores
  in the EAGLE simulations}}}.
\newblock {\emph{\JournalTitle{\mnras}}} \textbf{\bibinfo{volume}{488}},
  \bibinfo{pages}{2387--2404} (\bibinfo{year}{2019}).
\newblock \doiprefix 10.1093/mnras/stz1890.
\newblock \eprint{1810.04186}.

\bibitem{Bose2018}
\bibinfo{author}{{Bose}, S.} \emph{et~al.}
\newblock \bibinfo{journal}{\bibinfo{title}{{No cores in dark matter-dominated
  dwarf galaxies with bursty star formation histories}}}.
\newblock {\emph{\JournalTitle{\mnras}}} \textbf{\bibinfo{volume}{486}},
  \bibinfo{pages}{4790--4804} (\bibinfo{year}{2019}).
\newblock \doiprefix 10.1093/mnras/stz1168.
\newblock \eprint{1810.03635}.

\bibitem{Katz2018}
\bibinfo{author}{{Katz}, H.} \emph{et~al.}
\newblock \bibinfo{journal}{\bibinfo{title}{{Stellar feedback and the energy
  budget of late-type Galaxies: missing baryons and core creation}}}.
\newblock {\emph{\JournalTitle{\mnras}}} \textbf{\bibinfo{volume}{480}},
  \bibinfo{pages}{4287--4301} (\bibinfo{year}{2018}).
\newblock \doiprefix 10.1093/mnras/sty2129.

\bibitem{Read2019}
\bibinfo{author}{{Read}, J.~I.}, \bibinfo{author}{{Walker}, M.~G.} \&
  \bibinfo{author}{{Steger}, P.}
\newblock \bibinfo{journal}{\bibinfo{title}{{Dark matter heats up in dwarf
  galaxies}}}.
\newblock {\emph{\JournalTitle{\mnras}}}  (\bibinfo{year}{2019}).
\newblock \doiprefix 10.1093/mnras/sty3404.
\newblock \eprint{1808.06634}.

\bibitem{Diemer2014}
\bibinfo{author}{{Diemer}, B.} \& \bibinfo{author}{{Kravtsov}, A.~V.}
\newblock \bibinfo{journal}{\bibinfo{title}{{Dependence of the Outer Density
  Profiles of Halos on Their Mass Accretion Rate}}}.
\newblock {\emph{\JournalTitle{\apj}}} \textbf{\bibinfo{volume}{789}},
  \bibinfo{pages}{1} (\bibinfo{year}{2014}).
\newblock \doiprefix 10.1088/0004-637X/789/1/1.
\newblock \eprint{1401.1216}.

\bibitem{More2015}
\bibinfo{author}{{More}, S.}, \bibinfo{author}{{Diemer}, B.} \&
  \bibinfo{author}{{Kravtsov}, A.~V.}
\newblock \bibinfo{journal}{\bibinfo{title}{{The Splashback Radius as a
  Physical Halo Boundary and the Growth of Halo Mass}}}.
\newblock {\emph{\JournalTitle{\apj}}} \textbf{\bibinfo{volume}{810}},
  \bibinfo{pages}{36} (\bibinfo{year}{2015}).
\newblock \doiprefix 10.1088/0004-637X/810/1/36.
\newblock \eprint{1504.05591}.

\bibitem{Fillmore1984}
\bibinfo{author}{{Fillmore}, J.~A.} \& \bibinfo{author}{{Goldreich}, P.}
\newblock \bibinfo{journal}{\bibinfo{title}{{Self-similar gravitational
  collapse in an expanding universe}}}.
\newblock {\emph{\JournalTitle{\apj}}} \textbf{\bibinfo{volume}{281}},
  \bibinfo{pages}{1--8} (\bibinfo{year}{1984}).
\newblock \doiprefix 10.1086/162070.

\bibitem{Bertschinger1985}
\bibinfo{author}{{Bertschinger}, E.}
\newblock \bibinfo{journal}{\bibinfo{title}{{Self-similar secondary infall and
  accretion in an Einstein-de Sitter universe}}}.
\newblock {\emph{\JournalTitle{\apjs}}} \textbf{\bibinfo{volume}{58}},
  \bibinfo{pages}{39--65} (\bibinfo{year}{1985}).
\newblock \doiprefix 10.1086/191028.

\bibitem{Diemer2013}
\bibinfo{author}{{Diemer}, B.}, \bibinfo{author}{{More}, S.} \&
  \bibinfo{author}{{Kravtsov}, A.~V.}
\newblock \bibinfo{journal}{\bibinfo{title}{{The Pseudo-evolution of Halo
  Mass}}}.
\newblock {\emph{\JournalTitle{\apj}}} \textbf{\bibinfo{volume}{766}},
  \bibinfo{pages}{25} (\bibinfo{year}{2013}).
\newblock \doiprefix 10.1088/0004-637X/766/1/25.
\newblock \eprint{1207.0816}.

\bibitem{Davis1985}
\bibinfo{author}{{Davis}, M.}, \bibinfo{author}{{Efstathiou}, G.},
  \bibinfo{author}{{Frenk}, C.~S.} \& \bibinfo{author}{{White}, S.~D.~M.}
\newblock \bibinfo{journal}{\bibinfo{title}{{The evolution of large-scale
  structure in a universe dominated by cold dark matter}}}.
\newblock {\emph{\JournalTitle{\apj}}} \textbf{\bibinfo{volume}{292}},
  \bibinfo{pages}{371--394} (\bibinfo{year}{1985}).
\newblock \doiprefix 10.1086/163168.

\bibitem{Springel2001}
\bibinfo{author}{{Springel}, V.}, \bibinfo{author}{{White}, S.~D.~M.},
  \bibinfo{author}{{Tormen}, G.} \& \bibinfo{author}{{Kauffmann}, G.}
\newblock \bibinfo{journal}{\bibinfo{title}{{Populating a cluster of galaxies -
  I. Results at z=0}}}.
\newblock {\emph{\JournalTitle{\mnras}}} \textbf{\bibinfo{volume}{328}},
  \bibinfo{pages}{726--750} (\bibinfo{year}{2001}).
\newblock \doiprefix 10.1046/j.1365-8711.2001.04912.x.
\newblock \eprint{astro-ph/0012055}.

\bibitem{Behroozi2013}
\bibinfo{author}{{Behroozi}, P.~S.}, \bibinfo{author}{{Wechsler}, R.~H.} \&
  \bibinfo{author}{{Wu}, H.-Y.}
\newblock \bibinfo{journal}{\bibinfo{title}{{The ROCKSTAR Phase-space Temporal
  Halo Finder and the Velocity Offsets of Cluster Cores}}}.
\newblock {\emph{\JournalTitle{\apj}}} \textbf{\bibinfo{volume}{762}},
  \bibinfo{pages}{109} (\bibinfo{year}{2013}).
\newblock \doiprefix 10.1088/0004-637X/762/2/109.
\newblock \eprint{1110.4372}.

\bibitem{Bond1991}
\bibinfo{author}{{Bond}, J.~R.}, \bibinfo{author}{{Cole}, S.},
  \bibinfo{author}{{Efstathiou}, G.} \& \bibinfo{author}{{Kaiser}, N.}
\newblock \bibinfo{journal}{\bibinfo{title}{{Excursion set mass functions for
  hierarchical Gaussian fluctuations}}}.
\newblock {\emph{\JournalTitle{\apj}}} \textbf{\bibinfo{volume}{379}},
  \bibinfo{pages}{440--460} (\bibinfo{year}{1991}).
\newblock \doiprefix 10.1086/170520.

\bibitem{Jenkins2001}
\bibinfo{author}{{Jenkins}, A.} \emph{et~al.}
\newblock \bibinfo{journal}{\bibinfo{title}{{The mass function of dark matter
  haloes}}}.
\newblock {\emph{\JournalTitle{\mnras}}} \textbf{\bibinfo{volume}{321}},
  \bibinfo{pages}{372--384} (\bibinfo{year}{2001}).
\newblock \doiprefix 10.1046/j.1365-8711.2001.04029.x.
\newblock \eprint{astro-ph/0005260}.

\bibitem{Sheth2002}
\bibinfo{author}{{Sheth}, R.~K.} \& \bibinfo{author}{{Tormen}, G.}
\newblock \bibinfo{journal}{\bibinfo{title}{{An excursion set model of
  hierarchical clustering: ellipsoidal collapse and the moving barrier}}}.
\newblock {\emph{\JournalTitle{\mnras}}} \textbf{\bibinfo{volume}{329}},
  \bibinfo{pages}{61--75} (\bibinfo{year}{2002}).
\newblock \doiprefix 10.1046/j.1365-8711.2002.04950.x.
\newblock \eprint{astro-ph/0105113}.

\bibitem{White2002}
\bibinfo{author}{{White}, M.}
\newblock \bibinfo{journal}{\bibinfo{title}{{The Mass Function}}}.
\newblock {\emph{\JournalTitle{\apjs}}} \textbf{\bibinfo{volume}{143}},
  \bibinfo{pages}{241--255} (\bibinfo{year}{2002}).
\newblock \doiprefix 10.1086/342752.
\newblock \eprint{astro-ph/0207185}.

\bibitem{Reed2003}
\bibinfo{author}{{Reed}, D.} \emph{et~al.}
\newblock \bibinfo{journal}{\bibinfo{title}{{Evolution of the mass function of
  dark matter haloes}}}.
\newblock {\emph{\JournalTitle{\mnras}}} \textbf{\bibinfo{volume}{346}},
  \bibinfo{pages}{565--572} (\bibinfo{year}{2003}).
\newblock \doiprefix 10.1046/j.1365-2966.2003.07113.x.
\newblock \eprint{astro-ph/0301270}.

\bibitem{Warren2006}
\bibinfo{author}{{Warren}, M.~S.}, \bibinfo{author}{{Abazajian}, K.},
  \bibinfo{author}{{Holz}, D.~E.} \& \bibinfo{author}{{Teodoro}, L.}
\newblock \bibinfo{journal}{\bibinfo{title}{{Precision Determination of the
  Mass Function of Dark Matter Halos}}}.
\newblock {\emph{\JournalTitle{\apj}}} \textbf{\bibinfo{volume}{646}},
  \bibinfo{pages}{881--885} (\bibinfo{year}{2006}).
\newblock \doiprefix 10.1086/504962.
\newblock \eprint{astro-ph/0506395}.

\bibitem{Reed2007}
\bibinfo{author}{{Reed}, D.~S.}, \bibinfo{author}{{Governato}, F.},
  \bibinfo{author}{{Quinn}, T.}, \bibinfo{author}{{Stadel}, J.} \&
  \bibinfo{author}{{Lake}, G.}
\newblock \bibinfo{journal}{\bibinfo{title}{{The age dependence of galaxy
  clustering}}}.
\newblock {\emph{\JournalTitle{\mnras}}} \textbf{\bibinfo{volume}{378}},
  \bibinfo{pages}{777--784} (\bibinfo{year}{2007}).
\newblock \doiprefix 10.1111/j.1365-2966.2007.11826.x.
\newblock \eprint{astro-ph/0602003}.

\bibitem{Tinker2008}
\bibinfo{author}{{Tinker}, J.} \emph{et~al.}
\newblock \bibinfo{journal}{\bibinfo{title}{{Toward a Halo Mass Function for
  Precision Cosmology: The Limits of Universality}}}.
\newblock {\emph{\JournalTitle{\apj}}} \textbf{\bibinfo{volume}{688}},
  \bibinfo{pages}{709--728} (\bibinfo{year}{2008}).
\newblock \doiprefix 10.1086/591439.
\newblock \eprint{0803.2706}.

\bibitem{Angulo2013}
\bibinfo{author}{{Angulo}, R.~E.}, \bibinfo{author}{{Hahn}, O.} \&
  \bibinfo{author}{{Abel}, T.}
\newblock \bibinfo{journal}{\bibinfo{title}{{The warm dark matter halo mass
  function below the cut-off scale}}}.
\newblock {\emph{\JournalTitle{\mnras}}} \textbf{\bibinfo{volume}{434}},
  \bibinfo{pages}{3337--3347} (\bibinfo{year}{2013}).
\newblock \doiprefix 10.1093/mnras/stt1246.
\newblock \eprint{1304.2406}.

\bibitem{Schneider2015}
\bibinfo{author}{{Schneider}, A.}
\newblock \bibinfo{journal}{\bibinfo{title}{{Structure formation with
  suppressed small-scale perturbations}}}.
\newblock {\emph{\JournalTitle{\mnras}}} \textbf{\bibinfo{volume}{451}},
  \bibinfo{pages}{3117--3130} (\bibinfo{year}{2015}).
\newblock \doiprefix 10.1093/mnras/stv1169.
\newblock \eprint{1412.2133}.

\bibitem{Eke1996}
\bibinfo{author}{{Eke}, V.~R.}, \bibinfo{author}{{Cole}, S.} \&
  \bibinfo{author}{{Frenk}, C.~S.}
\newblock \bibinfo{journal}{\bibinfo{title}{{Cluster evolution as a diagnostic
  for Omega}}}.
\newblock {\emph{\JournalTitle{\mnras}}} \textbf{\bibinfo{volume}{282}}
  (\bibinfo{year}{1996}).
\newblock \doiprefix 10.1093/mnras/282.1.263.
\newblock \eprint{astro-ph/9601088}.

\bibitem{Sheth2001}
\bibinfo{author}{{Sheth}, R.~K.}, \bibinfo{author}{{Mo}, H.~J.} \&
  \bibinfo{author}{{Tormen}, G.}
\newblock \bibinfo{journal}{\bibinfo{title}{{Ellipsoidal collapse and an
  improved model for the number and spatial distribution of dark matter
  haloes}}}.
\newblock {\emph{\JournalTitle{\mnras}}} \textbf{\bibinfo{volume}{323}},
  \bibinfo{pages}{1--12} (\bibinfo{year}{2001}).
\newblock \doiprefix 10.1046/j.1365-8711.2001.04006.x.
\newblock \eprint{astro-ph/9907024}.

\bibitem{Angulo2012}
\bibinfo{author}{{Angulo}, R.~E.} \emph{et~al.}
\newblock \bibinfo{journal}{\bibinfo{title}{{Scaling relations for galaxy
  clusters in the Millennium-XXL simulation}}}.
\newblock {\emph{\JournalTitle{\mnras}}} \textbf{\bibinfo{volume}{426}},
  \bibinfo{pages}{2046--2062} (\bibinfo{year}{2012}).
\newblock \doiprefix 10.1111/j.1365-2966.2012.21830.x.
\newblock \eprint{1203.3216}.

\bibitem{Watson2013}
\bibinfo{author}{{Watson}, W.~A.} \emph{et~al.}
\newblock \bibinfo{journal}{\bibinfo{title}{{The halo mass function through the
  cosmic ages}}}.
\newblock {\emph{\JournalTitle{\mnras}}} \textbf{\bibinfo{volume}{433}},
  \bibinfo{pages}{1230--1245} (\bibinfo{year}{2013}).
\newblock \doiprefix 10.1093/mnras/stt791.
\newblock \eprint{1212.0095}.

\bibitem{Heitmann2015}
\bibinfo{author}{{Heitmann}, K.} \emph{et~al.}
\newblock \bibinfo{journal}{\bibinfo{title}{{The Q Continuum Simulation:
  Harnessing the Power of GPU Accelerated Supercomputers}}}.
\newblock {\emph{\JournalTitle{\apjs}}} \textbf{\bibinfo{volume}{219}},
  \bibinfo{pages}{34} (\bibinfo{year}{2015}).
\newblock \doiprefix 10.1088/0067-0049/219/2/34.
\newblock \eprint{1411.3396}.

\bibitem{Bocquet2016}
\bibinfo{author}{{Bocquet}, S.}, \bibinfo{author}{{Saro}, A.},
  \bibinfo{author}{{Dolag}, K.} \& \bibinfo{author}{{Mohr}, J.~J.}
\newblock \bibinfo{journal}{\bibinfo{title}{{Halo mass function: baryon impact,
  fitting formulae, and implications for cluster cosmology}}}.
\newblock {\emph{\JournalTitle{\mnras}}} \textbf{\bibinfo{volume}{456}},
  \bibinfo{pages}{2361--2373} (\bibinfo{year}{2016}).
\newblock \doiprefix 10.1093/mnras/stv2657.
\newblock \eprint{1502.07357}.

\bibitem{Springel2006}
\bibinfo{author}{{Springel}, V.}, \bibinfo{author}{{Frenk}, C.~S.} \&
  \bibinfo{author}{{White}, S.~D.~M.}
\newblock \bibinfo{journal}{\bibinfo{title}{{The large-scale structure of the
  Universe}}}.
\newblock {\emph{\JournalTitle{\nat}}} \textbf{\bibinfo{volume}{440}},
  \bibinfo{pages}{1137--1144} (\bibinfo{year}{2006}).
\newblock \doiprefix 10.1038/nature04805.
\newblock \eprint{arXiv:astro-ph/0604561}.

\bibitem{Springel2018}
\bibinfo{author}{{Springel}, V.} \emph{et~al.}
\newblock \bibinfo{journal}{\bibinfo{title}{{First results from the
  IllustrisTNG simulations: matter and galaxy clustering}}}.
\newblock {\emph{\JournalTitle{\mnras}}} \textbf{\bibinfo{volume}{475}},
  \bibinfo{pages}{676--698} (\bibinfo{year}{2018}).
\newblock \doiprefix 10.1093/mnras/stx3304.
\newblock \eprint{1707.03397}.

\bibitem{Cooray2002}
\bibinfo{author}{{Cooray}, A.} \& \bibinfo{author}{{Sheth}, R.}
\newblock \bibinfo{journal}{\bibinfo{title}{{Halo models of large scale
  structure}}}.
\newblock {\emph{\JournalTitle{\physrep}}} \textbf{\bibinfo{volume}{372}},
  \bibinfo{pages}{1--129} (\bibinfo{year}{2002}).
\newblock \doiprefix 10.1016/S0370-1573(02)00276-4.
\newblock \eprint{astro-ph/0206508}.

\bibitem{Benson2000}
\bibinfo{author}{{Benson}, A.~J.}, \bibinfo{author}{{Cole}, S.},
  \bibinfo{author}{{Frenk}, C.~S.}, \bibinfo{author}{{Baugh}, C.~M.} \&
  \bibinfo{author}{{Lacey}, C.~G.}
\newblock \bibinfo{journal}{\bibinfo{title}{{The nature of galaxy bias and
  clustering}}}.
\newblock {\emph{\JournalTitle{\mnras}}} \textbf{\bibinfo{volume}{311}},
  \bibinfo{pages}{793--808} (\bibinfo{year}{2000}).
\newblock \doiprefix 10.1046/j.1365-8711.2000.03101.x.
\newblock \eprint{astro-ph/9903343}.

\bibitem{Navarro1996}
\bibinfo{author}{{Navarro}, J.~F.}, \bibinfo{author}{{Frenk}, C.~S.} \&
  \bibinfo{author}{{White}, S.~D.~M.}
\newblock \bibinfo{journal}{\bibinfo{title}{{The Structure of Cold Dark Matter
  Halos}}}.
\newblock {\emph{\JournalTitle{\apj}}} \textbf{\bibinfo{volume}{462}},
  \bibinfo{pages}{563} (\bibinfo{year}{1996}).
\newblock \doiprefix 10.1086/177173.
\newblock \eprint{astro-ph/9508025}.

\bibitem{Navarro1997}
\bibinfo{author}{{Navarro}, J.~F.}, \bibinfo{author}{{Frenk}, C.~S.} \&
  \bibinfo{author}{{White}, S.~D.~M.}
\newblock \bibinfo{journal}{\bibinfo{title}{{A Universal Density Profile from
  Hierarchical Clustering}}}.
\newblock {\emph{\JournalTitle{\apj}}} \textbf{\bibinfo{volume}{490}},
  \bibinfo{pages}{493--508} (\bibinfo{year}{1997}).
\newblock \doiprefix 10.1086/304888.
\newblock \eprint{astro-ph/9611107}.

\bibitem{Moore1999}
\bibinfo{author}{{Moore}, B.} \emph{et~al.}
\newblock \bibinfo{journal}{\bibinfo{title}{{Dark Matter Substructure within
  Galactic Halos}}}.
\newblock {\emph{\JournalTitle{\apjl}}} \textbf{\bibinfo{volume}{524}},
  \bibinfo{pages}{L19--L22} (\bibinfo{year}{1999}).
\newblock \doiprefix 10.1086/312287.
\newblock \eprint{astro-ph/9907411}.

\bibitem{Navarro2004}
\bibinfo{author}{{Navarro}, J.~F.} \emph{et~al.}
\newblock \bibinfo{journal}{\bibinfo{title}{{The inner structure of
  {$\Lambda$}CDM haloes - III. Universality and asymptotic slopes}}}.
\newblock {\emph{\JournalTitle{\mnras}}} \textbf{\bibinfo{volume}{349}},
  \bibinfo{pages}{1039--1051} (\bibinfo{year}{2004}).
\newblock \doiprefix 10.1111/j.1365-2966.2004.07586.x.
\newblock \eprint{astro-ph/0311231}.

\bibitem{Einasto1965}
\bibinfo{author}{{Einasto}, J.}
\newblock \bibinfo{journal}{\bibinfo{title}{{On the Construction of a Composite
  Model for the Galaxy and on the Determination of the System of Galactic
  Parameters}}}.
\newblock {\emph{\JournalTitle{Trudy Astrofizicheskogo Instituta Alma-Ata}}}
  \textbf{\bibinfo{volume}{5}}, \bibinfo{pages}{87--100}
  (\bibinfo{year}{1965}).

\bibitem{Ludlow2014}
\bibinfo{author}{{Ludlow}, A.~D.} \emph{et~al.}
\newblock \bibinfo{journal}{\bibinfo{title}{{The mass-concentration-redshift
  relation of cold dark matter haloes}}}.
\newblock {\emph{\JournalTitle{\mnras}}} \textbf{\bibinfo{volume}{441}},
  \bibinfo{pages}{378--388} (\bibinfo{year}{2014}).
\newblock \doiprefix 10.1093/mnras/stu483.
\newblock \eprint{1312.0945}.

\bibitem{Ludlow2016}
\bibinfo{author}{{Ludlow}, A.~D.} \emph{et~al.}
\newblock \bibinfo{journal}{\bibinfo{title}{{The mass-concentration-redshift
  relation of cold and warm dark matter haloes}}}.
\newblock {\emph{\JournalTitle{\mnras}}} \textbf{\bibinfo{volume}{460}},
  \bibinfo{pages}{1214--1232} (\bibinfo{year}{2016}).
\newblock \doiprefix 10.1093/mnras/stw1046.
\newblock \eprint{1601.02624}.

\bibitem{Klypin2016}
\bibinfo{author}{{Klypin}, A.}, \bibinfo{author}{{Yepes}, G.},
  \bibinfo{author}{{Gottl{\"o}ber}, S.}, \bibinfo{author}{{Prada}, F.} \&
  \bibinfo{author}{{He{\ss}}, S.}
\newblock \bibinfo{journal}{\bibinfo{title}{{MultiDark simulations: the story
  of dark matter halo concentrations and density profiles}}}.
\newblock {\emph{\JournalTitle{\mnras}}} \textbf{\bibinfo{volume}{457}},
  \bibinfo{pages}{4340--4359} (\bibinfo{year}{2016}).
\newblock \doiprefix 10.1093/mnras/stw248.
\newblock \eprint{1411.4001}.

\bibitem{Jing2002}
\bibinfo{author}{{Jing}, Y.~P.}
\newblock \bibinfo{journal}{\bibinfo{title}{{Intrinsic correlation of halo
  ellipticity and its implications for large-scale weak lensing surveys}}}.
\newblock {\emph{\JournalTitle{\mnras}}} \textbf{\bibinfo{volume}{335}},
  \bibinfo{pages}{L89--L93} (\bibinfo{year}{2002}).
\newblock \doiprefix 10.1046/j.1365-8711.2002.05899.x.
\newblock \eprint{astro-ph/0206098}.

\bibitem{Allgood2006}
\bibinfo{author}{{Allgood}, B.} \emph{et~al.}
\newblock \bibinfo{journal}{\bibinfo{title}{{The shape of dark matter haloes:
  dependence on mass, redshift, radius and formation}}}.
\newblock {\emph{\JournalTitle{\mnras}}} \textbf{\bibinfo{volume}{367}},
  \bibinfo{pages}{1781--1796} (\bibinfo{year}{2006}).
\newblock \doiprefix 10.1111/j.1365-2966.2006.10094.x.
\newblock \eprint{astro-ph/0508497}.

\bibitem{Bett2007}
\bibinfo{author}{{Bett}, P.} \emph{et~al.}
\newblock \bibinfo{journal}{\bibinfo{title}{{The spin and shape of dark matter
  haloes in the Millennium simulation of a {$\Lambda$} cold dark matter
  universe}}}.
\newblock {\emph{\JournalTitle{\mnras}}} \textbf{\bibinfo{volume}{376}},
  \bibinfo{pages}{215--232} (\bibinfo{year}{2007}).
\newblock \doiprefix 10.1111/j.1365-2966.2007.11432.x.
\newblock \eprint{astro-ph/0608607}.

\bibitem{Diemand2007}
\bibinfo{author}{{Diemand}, J.}, \bibinfo{author}{{Kuhlen}, M.} \&
  \bibinfo{author}{{Madau}, P.}
\newblock \bibinfo{journal}{\bibinfo{title}{{Formation and Evolution of Galaxy
  Dark Matter Halos and Their Substructure}}}.
\newblock {\emph{\JournalTitle{\apj}}} \textbf{\bibinfo{volume}{667}},
  \bibinfo{pages}{859--877} (\bibinfo{year}{2007}).
\newblock \doiprefix 10.1086/520573.
\newblock \eprint{astro-ph/0703337}.

\bibitem{Navarro2010}
\bibinfo{author}{{Navarro}, J.~F.} \emph{et~al.}
\newblock \bibinfo{journal}{\bibinfo{title}{{The diversity and similarity of
  simulated cold dark matter haloes}}}.
\newblock {\emph{\JournalTitle{\mnras}}} \textbf{\bibinfo{volume}{402}},
  \bibinfo{pages}{21--34} (\bibinfo{year}{2010}).
\newblock \doiprefix 10.1111/j.1365-2966.2009.15878.x.
\newblock \eprint{0810.1522}.

\bibitem{Ghigna1998}
\bibinfo{author}{{Ghigna}, S.} \emph{et~al.}
\newblock \bibinfo{journal}{\bibinfo{title}{{Dark matter haloes within
  clusters}}}.
\newblock {\emph{\JournalTitle{\mnras}}} \textbf{\bibinfo{volume}{300}},
  \bibinfo{pages}{146--162} (\bibinfo{year}{1998}).
\newblock \doiprefix 10.1046/j.1365-8711.1998.01918.x.
\newblock \eprint{astro-ph/9801192}.

\bibitem{Springel2008}
\bibinfo{author}{{Springel}, V.} \emph{et~al.}
\newblock \bibinfo{journal}{\bibinfo{title}{{The Aquarius Project: the
  subhaloes of galactic haloes}}}.
\newblock {\emph{\JournalTitle{\mnras}}} \textbf{\bibinfo{volume}{391}},
  \bibinfo{pages}{1685--1711} (\bibinfo{year}{2008}).
\newblock \doiprefix 10.1111/j.1365-2966.2008.14066.x.
\newblock \eprint{0809.0898}.

\bibitem{Gao2012}
\bibinfo{author}{{Gao}, L.} \emph{et~al.}
\newblock \bibinfo{journal}{\bibinfo{title}{{The Phoenix Project: the dark side
  of rich Galaxy clusters}}}.
\newblock {\emph{\JournalTitle{\mnras}}} \textbf{\bibinfo{volume}{425}},
  \bibinfo{pages}{2169--2186} (\bibinfo{year}{2012}).
\newblock \doiprefix 10.1111/j.1365-2966.2012.21564.x.
\newblock \eprint{1201.1940}.

\bibitem{Springel2008b}
\bibinfo{author}{{Springel}, V.} \emph{et~al.}
\newblock \bibinfo{journal}{\bibinfo{title}{{Prospects for detecting
  supersymmetric dark matter in the Galactic halo}}}.
\newblock {\emph{\JournalTitle{\nat}}} \textbf{\bibinfo{volume}{456}},
  \bibinfo{pages}{73--76} (\bibinfo{year}{2008}).
\newblock \doiprefix 10.1038/nature07411.
\newblock \eprint{0809.0894}.

\bibitem{Gao2004}
\bibinfo{author}{{Gao}, L.}, \bibinfo{author}{{White}, S.~D.~M.},
  \bibinfo{author}{{Jenkins}, A.}, \bibinfo{author}{{Stoehr}, F.} \&
  \bibinfo{author}{{Springel}, V.}
\newblock \bibinfo{journal}{\bibinfo{title}{{The subhalo populations of
  {$\Lambda$}CDM dark haloes}}}.
\newblock {\emph{\JournalTitle{\mnras}}} \textbf{\bibinfo{volume}{355}},
  \bibinfo{pages}{819--834} (\bibinfo{year}{2004}).
\newblock \doiprefix 10.1111/j.1365-2966.2004.08360.x.
\newblock \eprint{astro-ph/0404589}.

\bibitem{Colella1984}
\bibinfo{author}{{Colella}, P.} \& \bibinfo{author}{{Woodward}, P.~R.}
\newblock \bibinfo{journal}{\bibinfo{title}{{The Piecewise Parabolic Method
  (PPM) for Gas-Dynamical Simulations}}}.
\newblock {\emph{\JournalTitle{Journal of Computational Physics}}}
  \textbf{\bibinfo{volume}{54}}, \bibinfo{pages}{174--201}
  (\bibinfo{year}{1984}).
\newblock \doiprefix 10.1016/0021-9991(84)90143-8.

\bibitem{Woodward1986}
\bibinfo{author}{{Woodward}, P.~R.}
\newblock \bibinfo{title}{{PPM: Piecewise-Parabolic Methods for Astrophysical
  Fluid Dynamics}}.
\newblock In \bibinfo{editor}{{Winkler}, K.-H.~A.} \&
  \bibinfo{editor}{{Norman}, M.~L.} (eds.) \emph{\bibinfo{booktitle}{NATO
  Advanced Science Institutes (ASI) Series C}}, vol. \bibinfo{volume}{188} of
  \emph{\bibinfo{series}{NATO Advanced Science Institutes (ASI) Series C}},
  \bibinfo{pages}{245} (\bibinfo{year}{1986}).

\bibitem{Berger1984}
\bibinfo{author}{{Berger}, M.~J.} \& \bibinfo{author}{{Oliger}, J.}
\newblock \bibinfo{journal}{\bibinfo{title}{{Adaptive Mesh Refinement for
  Hyperbolic Partial Differential Equations}}}.
\newblock {\emph{\JournalTitle{Journal of Computational Physics}}}
  \textbf{\bibinfo{volume}{53}}, \bibinfo{pages}{484--512}
  (\bibinfo{year}{1984}).
\newblock \doiprefix 10.1016/0021-9991(84)90073-1.

\bibitem{Berger1989}
\bibinfo{author}{{Berger}, M.~J.} \& \bibinfo{author}{{Colella}, P.}
\newblock \bibinfo{journal}{\bibinfo{title}{{Local adaptive mesh refinement for
  shock hydrodynamics}}}.
\newblock {\emph{\JournalTitle{Journal of Computational Physics}}}
  \textbf{\bibinfo{volume}{82}}, \bibinfo{pages}{64--84}
  (\bibinfo{year}{1989}).
\newblock \doiprefix 10.1016/0021-9991(89)90035-1.

\bibitem{Klein1994}
\bibinfo{author}{{Klein}, R.~I.}, \bibinfo{author}{{McKee}, C.~F.} \&
  \bibinfo{author}{{Colella}, P.}
\newblock \bibinfo{journal}{\bibinfo{title}{{On the hydrodynamic interaction of
  shock waves with interstellar clouds. 1: Nonradiative shocks in small
  clouds}}}.
\newblock {\emph{\JournalTitle{\apj}}} \textbf{\bibinfo{volume}{420}},
  \bibinfo{pages}{213--236} (\bibinfo{year}{1994}).
\newblock \doiprefix 10.1086/173554.

\bibitem{Bryan1995}
\bibinfo{author}{{Bryan}, G.~L.} \& \bibinfo{author}{{Norman}, M.~L.}
\newblock \bibinfo{title}{{Simulating X-ray Clusters with Adaptive Mesh
  Refinement}}.
\newblock In \emph{\bibinfo{booktitle}{American Astronomical Society Meeting
  Abstracts}}, vol.~\bibinfo{volume}{27} of \emph{\bibinfo{series}{Bulletin of
  the American Astronomical Society}}, \bibinfo{pages}{1421}
  (\bibinfo{year}{1995}).

\bibitem{Cockburn1989}
\bibinfo{author}{{Cockburn}, B.}, \bibinfo{author}{{Lin}, S.-Y.} \&
  \bibinfo{author}{{Shu}, C.-W.}
\newblock \bibinfo{journal}{\bibinfo{title}{{TVB Runge Kutta Local Projection
  Discontinuous Galerkin Finite Element Method for Conservation Laws III:
  One-Dimensional Systems}}}.
\newblock {\emph{\JournalTitle{Journal of Computational Physics}}}
  \textbf{\bibinfo{volume}{84}}, \bibinfo{pages}{90--113}
  (\bibinfo{year}{1989}).
\newblock \doiprefix 10.1016/0021-9991(89)90183-6.

\bibitem{Mocz2014b}
\bibinfo{author}{{Mocz}, P.}, \bibinfo{author}{{Vogelsberger}, M.},
  \bibinfo{author}{{Sijacki}, D.}, \bibinfo{author}{{Pakmor}, R.} \&
  \bibinfo{author}{{Hernquist}, L.}
\newblock \bibinfo{journal}{\bibinfo{title}{{A discontinuous Galerkin method
  for solving the fluid and magnetohydrodynamic equations in astrophysical
  simulations}}}.
\newblock {\emph{\JournalTitle{\mnras}}} \textbf{\bibinfo{volume}{437}},
  \bibinfo{pages}{397--414} (\bibinfo{year}{2014}).
\newblock \doiprefix 10.1093/mnras/stt1890.
\newblock \eprint{1305.5536}.

\bibitem{Guillet2019}
\bibinfo{author}{{Guillet}, T.}, \bibinfo{author}{{Pakmor}, R.},
  \bibinfo{author}{{Springel}, V.}, \bibinfo{author}{{Chandrashekar}, P.} \&
  \bibinfo{author}{{Klingenberg}, C.}
\newblock \bibinfo{journal}{\bibinfo{title}{{High-order Magnetohydrodynamics
  for Astrophysics with an Adaptive Mesh Refinement Discontinuous Galerkin
  Scheme}}}.
\newblock {\emph{\JournalTitle{\mnras}}}  (\bibinfo{year}{2019}).
\newblock \doiprefix 10.1093/mnras/stz314.
\newblock \eprint{1806.02343}.

\bibitem{Lucy1977}
\bibinfo{author}{{Lucy}, L.~B.}
\newblock \bibinfo{journal}{\bibinfo{title}{{A numerical approach to the
  testing of the fission hypothesis}}}.
\newblock {\emph{\JournalTitle{\aj}}} \textbf{\bibinfo{volume}{82}},
  \bibinfo{pages}{1013--1024} (\bibinfo{year}{1977}).
\newblock \doiprefix 10.1086/112164.

\bibitem{Gingold1977}
\bibinfo{author}{{Gingold}, R.~A.} \& \bibinfo{author}{{Monaghan}, J.~J.}
\newblock \bibinfo{journal}{\bibinfo{title}{{Smoothed particle hydrodynamics -
  Theory and application to non-spherical stars}}}.
\newblock {\emph{\JournalTitle{\mnras}}} \textbf{\bibinfo{volume}{181}},
  \bibinfo{pages}{375--389} (\bibinfo{year}{1977}).
\newblock \doiprefix 10.1093/mnras/181.3.375.

\bibitem{SpringelSPH}
\bibinfo{author}{{Springel}, V.}
\newblock \bibinfo{journal}{\bibinfo{title}{{Smoothed Particle Hydrodynamics in
  Astrophysics}}}.
\newblock {\emph{\JournalTitle{\araa}}} \textbf{\bibinfo{volume}{48}},
  \bibinfo{pages}{391--430} (\bibinfo{year}{2010}).
\newblock \doiprefix 10.1146/annurev-astro-081309-130914.
\newblock \eprint{1109.2219}.

\bibitem{Price2012}
\bibinfo{author}{{Price}, D.~J.}
\newblock \bibinfo{journal}{\bibinfo{title}{{Smoothed particle hydrodynamics
  and magnetohydrodynamics}}}.
\newblock {\emph{\JournalTitle{Journal of Computational Physics}}}
  \textbf{\bibinfo{volume}{231}}, \bibinfo{pages}{759--794}
  (\bibinfo{year}{2012}).
\newblock \doiprefix 10.1016/j.jcp.2010.12.011.
\newblock \eprint{1012.1885}.

\bibitem{Read2012}
\bibinfo{author}{{Read}, J.~I.} \& \bibinfo{author}{{Hayfield}, T.}
\newblock \bibinfo{journal}{\bibinfo{title}{{SPHS: smoothed particle
  hydrodynamics with a higher order dissipation switch}}}.
\newblock {\emph{\JournalTitle{\mnras}}} \textbf{\bibinfo{volume}{422}},
  \bibinfo{pages}{3037--3055} (\bibinfo{year}{2012}).
\newblock \doiprefix 10.1111/j.1365-2966.2012.20819.x.
\newblock \eprint{1111.6985}.

\bibitem{Hopkins2013}
\bibinfo{author}{{Hopkins}, P.~F.}
\newblock \bibinfo{journal}{\bibinfo{title}{{A general class of Lagrangian
  smoothed particle hydrodynamics methods and implications for fluid mixing
  problems}}}.
\newblock {\emph{\JournalTitle{\mnras}}} \textbf{\bibinfo{volume}{428}},
  \bibinfo{pages}{2840--2856} (\bibinfo{year}{2013}).
\newblock \doiprefix 10.1093/mnras/sts210.
\newblock \eprint{1206.5006}.

\bibitem{Keller2014}
\bibinfo{author}{{Keller}, B.~W.}, \bibinfo{author}{{Wadsley}, J.},
  \bibinfo{author}{{Benincasa}, S.~M.} \& \bibinfo{author}{{Couchman},
  H.~M.~P.}
\newblock \bibinfo{journal}{\bibinfo{title}{{A superbubble feedback model for
  galaxy simulations}}}.
\newblock {\emph{\JournalTitle{\mnras}}} \textbf{\bibinfo{volume}{442}},
  \bibinfo{pages}{3013--3025} (\bibinfo{year}{2014}).
\newblock \doiprefix 10.1093/mnras/stu1058.
\newblock \eprint{1405.2625}.

\bibitem{Schaye2015}
\bibinfo{author}{{Schaye}, J.} \emph{et~al.}
\newblock \bibinfo{journal}{\bibinfo{title}{{The EAGLE project: simulating the
  evolution and assembly of galaxies and their environments}}}.
\newblock {\emph{\JournalTitle{\mnras}}} \textbf{\bibinfo{volume}{446}},
  \bibinfo{pages}{521--554} (\bibinfo{year}{2015}).
\newblock \doiprefix 10.1093/mnras/stu2058.
\newblock \eprint{1407.7040}.

\bibitem{Wang2015}
\bibinfo{author}{{Wang}, L.} \emph{et~al.}
\newblock \bibinfo{journal}{\bibinfo{title}{{NIHAO project - I. Reproducing the
  inefficiency of galaxy formation across cosmic time with a large sample of
  cosmological hydrodynamical simulations}}}.
\newblock {\emph{\JournalTitle{\mnras}}} \textbf{\bibinfo{volume}{454}},
  \bibinfo{pages}{83--94} (\bibinfo{year}{2015}).
\newblock \doiprefix 10.1093/mnras/stv1937.
\newblock \eprint{1503.04818}.

\bibitem{Barnes2017}
\bibinfo{author}{{Barnes}, D.~J.} \emph{et~al.}
\newblock \bibinfo{journal}{\bibinfo{title}{{The Cluster-EAGLE project: global
  properties of simulated clusters with resolved galaxies}}}.
\newblock {\emph{\JournalTitle{\mnras}}} \textbf{\bibinfo{volume}{471}},
  \bibinfo{pages}{1088--1106} (\bibinfo{year}{2017}).
\newblock \doiprefix 10.1093/mnras/stx1647.
\newblock \eprint{1703.10907}.

\bibitem{Gnedin1995}
\bibinfo{author}{{Gnedin}, N.~Y.}
\newblock \bibinfo{journal}{\bibinfo{title}{{Softened Lagrangian hydrodynamics
  for cosmology}}}.
\newblock {\emph{\JournalTitle{\apjs}}} \textbf{\bibinfo{volume}{97}},
  \bibinfo{pages}{231--257} (\bibinfo{year}{1995}).
\newblock \doiprefix 10.1086/192141.

\bibitem{Pen1998}
\bibinfo{author}{{Pen}, U.-L.}
\newblock \bibinfo{journal}{\bibinfo{title}{{A High-Resolution Adaptive Moving
  Mesh Hydrodynamic Algorithm}}}.
\newblock {\emph{\JournalTitle{\apjs}}} \textbf{\bibinfo{volume}{115}},
  \bibinfo{pages}{19--34} (\bibinfo{year}{1998}).
\newblock \doiprefix 10.1086/313074.
\newblock \eprint{astro-ph/9704258}.

\bibitem{Duffell2011}
\bibinfo{author}{{Duffell}, P.~C.} \& \bibinfo{author}{{MacFadyen}, A.~I.}
\newblock \bibinfo{journal}{\bibinfo{title}{{TESS: A Relativistic Hydrodynamics
  Code on a Moving Voronoi Mesh}}}.
\newblock {\emph{\JournalTitle{\apjs}}} \textbf{\bibinfo{volume}{197}},
  \bibinfo{pages}{15} (\bibinfo{year}{2011}).
\newblock \doiprefix 10.1088/0067-0049/197/2/15.
\newblock \eprint{1104.3562}.

\bibitem{Vandenbroucke2016}
\bibinfo{author}{{Vandenbroucke}, B.} \& \bibinfo{author}{{De Rijcke}, S.}
\newblock \bibinfo{journal}{\bibinfo{title}{{The moving mesh code SHADOWFAX}}}.
\newblock {\emph{\JournalTitle{Astronomy and Computing}}}
  \textbf{\bibinfo{volume}{16}}, \bibinfo{pages}{109--130}
  (\bibinfo{year}{2016}).
\newblock \doiprefix 10.1016/j.ascom.2016.05.001.
\newblock \eprint{1605.03576}.

\bibitem{Sutherland1993}
\bibinfo{author}{{Sutherland}, R.~S.} \& \bibinfo{author}{{Dopita}, M.~A.}
\newblock \bibinfo{journal}{\bibinfo{title}{{Cooling functions for low-density
  astrophysical plasmas}}}.
\newblock {\emph{\JournalTitle{\apjs}}} \textbf{\bibinfo{volume}{88}},
  \bibinfo{pages}{253--327} (\bibinfo{year}{1993}).
\newblock \doiprefix 10.1086/191823.

\bibitem{Wiersma2009}
\bibinfo{author}{{Wiersma}, R.~P.~C.}, \bibinfo{author}{{Schaye}, J.} \&
  \bibinfo{author}{{Smith}, B.~D.}
\newblock \bibinfo{journal}{\bibinfo{title}{{The effect of photoionization on
  the cooling rates of enriched, astrophysical plasmas}}}.
\newblock {\emph{\JournalTitle{\mnras}}} \textbf{\bibinfo{volume}{393}},
  \bibinfo{pages}{99--107} (\bibinfo{year}{2009}).
\newblock \doiprefix 10.1111/j.1365-2966.2008.14191.x.
\newblock \eprint{0807.3748}.

\bibitem{Haardt2012}
\bibinfo{author}{{Haardt}, F.} \& \bibinfo{author}{{Madau}, P.}
\newblock \bibinfo{journal}{\bibinfo{title}{{Radiative Transfer in a Clumpy
  Universe. IV. New Synthesis Models of the Cosmic UV/X-Ray Background}}}.
\newblock {\emph{\JournalTitle{\apj}}} \textbf{\bibinfo{volume}{746}},
  \bibinfo{pages}{125} (\bibinfo{year}{2012}).
\newblock \doiprefix 10.1088/0004-637X/746/2/125.
\newblock \eprint{1105.2039}.

\bibitem{Tielens2010}
\bibinfo{author}{{Tielens}, A.~G.~G.~M.}
\newblock \emph{\bibinfo{title}{{The Physics and Chemistry of the Interstellar
  Medium}}} (\bibinfo{publisher}{Cambridge, UK: Cambridge University Press},
  \bibinfo{year}{2010}).

\bibitem{Springel2003}
\bibinfo{author}{{Springel}, V.} \& \bibinfo{author}{{Hernquist}, L.}
\newblock \bibinfo{journal}{\bibinfo{title}{{Cosmological smoothed particle
  hydrodynamics simulations: a hybrid multiphase model for star formation}}}.
\newblock {\emph{\JournalTitle{\mnras}}} \textbf{\bibinfo{volume}{339}},
  \bibinfo{pages}{289--311} (\bibinfo{year}{2003}).
\newblock \doiprefix 10.1046/j.1365-8711.2003.06206.x.
\newblock \eprint{arXiv:astro-ph/0206393}.

\bibitem{Agertz2011}
\bibinfo{author}{{Agertz}, O.}, \bibinfo{author}{{Teyssier}, R.} \&
  \bibinfo{author}{{Moore}, B.}
\newblock \bibinfo{journal}{\bibinfo{title}{{The formation of disc galaxies in
  a {$\Lambda$}CDM universe}}}.
\newblock {\emph{\JournalTitle{\mnras}}} \textbf{\bibinfo{volume}{410}},
  \bibinfo{pages}{1391--1408} (\bibinfo{year}{2011}).
\newblock \doiprefix 10.1111/j.1365-2966.2010.17530.x.
\newblock \eprint{1004.0005}.

\bibitem{DallaVecchia2012}
\bibinfo{author}{{Dalla Vecchia}, C.} \& \bibinfo{author}{{Schaye}, J.}
\newblock \bibinfo{journal}{\bibinfo{title}{{Simulating galactic outflows with
  thermal supernova feedback}}}.
\newblock {\emph{\JournalTitle{\mnras}}} \textbf{\bibinfo{volume}{426}},
  \bibinfo{pages}{140--158} (\bibinfo{year}{2012}).
\newblock \doiprefix 10.1111/j.1365-2966.2012.21704.x.
\newblock \eprint{1203.5667}.

\bibitem{Hopkins2012}
\bibinfo{author}{{Hopkins}, P.~F.}, \bibinfo{author}{{Quataert}, E.} \&
  \bibinfo{author}{{Murray}, N.}
\newblock \bibinfo{journal}{\bibinfo{title}{{The structure of the interstellar
  medium of star-forming galaxies}}}.
\newblock {\emph{\JournalTitle{\mnras}}} \textbf{\bibinfo{volume}{421}},
  \bibinfo{pages}{3488--3521} (\bibinfo{year}{2012}).
\newblock \doiprefix 10.1111/j.1365-2966.2012.20578.x.
\newblock \eprint{1110.4636}.

\bibitem{Agertz2013}
\bibinfo{author}{{Agertz}, O.}, \bibinfo{author}{{Kravtsov}, A.~V.},
  \bibinfo{author}{{Leitner}, S.~N.} \& \bibinfo{author}{{Gnedin}, N.~Y.}
\newblock \bibinfo{journal}{\bibinfo{title}{{Toward a Complete Accounting of
  Energy and Momentum from Stellar Feedback in Galaxy Formation Simulations}}}.
\newblock {\emph{\JournalTitle{\apj}}} \textbf{\bibinfo{volume}{770}},
  \bibinfo{pages}{25} (\bibinfo{year}{2013}).
\newblock \doiprefix 10.1088/0004-637X/770/1/25.
\newblock \eprint{1210.4957}.

\bibitem{Rosdahl2015}
\bibinfo{author}{{Rosdahl}, J.}, \bibinfo{author}{{Schaye}, J.},
  \bibinfo{author}{{Teyssier}, R.} \& \bibinfo{author}{{Agertz}, O.}
\newblock \bibinfo{journal}{\bibinfo{title}{{Galaxies that shine:
  radiation-hydrodynamical simulations of disc galaxies}}}.
\newblock {\emph{\JournalTitle{\mnras}}} \textbf{\bibinfo{volume}{451}},
  \bibinfo{pages}{34--58} (\bibinfo{year}{2015}).
\newblock \doiprefix 10.1093/mnras/stv937.
\newblock \eprint{1501.04632}.

\bibitem{Emerick2019}
\bibinfo{author}{{Emerick}, A.}, \bibinfo{author}{{Bryan}, G.~L.} \&
  \bibinfo{author}{{Mac Low}, M.-M.}
\newblock \bibinfo{journal}{\bibinfo{title}{{Simulating an isolated dwarf
  galaxy with multichannel feedback and chemical yields from individual
  stars}}}.
\newblock {\emph{\JournalTitle{\mnras}}} \textbf{\bibinfo{volume}{482}},
  \bibinfo{pages}{1304--1329} (\bibinfo{year}{2019}).
\newblock \doiprefix 10.1093/mnras/sty2689.
\newblock \eprint{1807.07182}.

\bibitem{Krumholz2011}
\bibinfo{author}{{Krumholz}, M.~R.} \& \bibinfo{author}{{Gnedin}, N.~Y.}
\newblock \bibinfo{journal}{\bibinfo{title}{{A Comparison of Methods for
  Determining the Molecular Content of Model Galaxies}}}.
\newblock {\emph{\JournalTitle{\apj}}} \textbf{\bibinfo{volume}{729}},
  \bibinfo{pages}{36} (\bibinfo{year}{2011}).
\newblock \doiprefix 10.1088/0004-637X/729/1/36.
\newblock \eprint{1011.4065}.

\bibitem{Bigiel2011}
\bibinfo{author}{{Bigiel}, F.} \emph{et~al.}
\newblock \bibinfo{journal}{\bibinfo{title}{{A Constant Molecular Gas Depletion
  Time in Nearby Disk Galaxies}}}.
\newblock {\emph{\JournalTitle{\apjl}}} \textbf{\bibinfo{volume}{730}},
  \bibinfo{pages}{L13} (\bibinfo{year}{2011}).
\newblock \doiprefix 10.1088/2041-8205/730/2/L13.
\newblock \eprint{1102.1720}.

\bibitem{Krumholz2012}
\bibinfo{author}{{Krumholz}, M.~R.}, \bibinfo{author}{{Dekel}, A.} \&
  \bibinfo{author}{{McKee}, C.~F.}
\newblock \bibinfo{journal}{\bibinfo{title}{{A Universal, Local Star Formation
  Law in Galactic Clouds, nearby Galaxies, High-redshift Disks, and
  Starbursts}}}.
\newblock {\emph{\JournalTitle{\apj}}} \textbf{\bibinfo{volume}{745}},
  \bibinfo{pages}{69} (\bibinfo{year}{2012}).
\newblock \doiprefix 10.1088/0004-637X/745/1/69.
\newblock \eprint{1109.4150}.

\bibitem{Hopkins2014}
\bibinfo{author}{{Hopkins}, P.~F.} \emph{et~al.}
\newblock \bibinfo{journal}{\bibinfo{title}{{Galaxies on FIRE (Feedback In
  Realistic Environments): stellar feedback explains cosmologically inefficient
  star formation}}}.
\newblock {\emph{\JournalTitle{\mnras}}} \textbf{\bibinfo{volume}{445}},
  \bibinfo{pages}{581--603} (\bibinfo{year}{2014}).
\newblock \doiprefix 10.1093/mnras/stu1738.
\newblock \eprint{1311.2073}.

\bibitem{Hopkins2018}
\bibinfo{author}{{Hopkins}, P.~F.} \emph{et~al.}
\newblock \bibinfo{journal}{\bibinfo{title}{{FIRE-2 simulations: physics versus
  numerics in galaxy formation}}}.
\newblock {\emph{\JournalTitle{\mnras}}} \textbf{\bibinfo{volume}{480}},
  \bibinfo{pages}{800--863} (\bibinfo{year}{2018}).
\newblock \doiprefix 10.1093/mnras/sty1690.
\newblock \eprint{1702.06148}.

\bibitem{Stinson2006}
\bibinfo{author}{{Stinson}, G.} \emph{et~al.}
\newblock \bibinfo{journal}{\bibinfo{title}{{Star formation and feedback in
  smoothed particle hydrodynamic simulations - I. Isolated galaxies}}}.
\newblock {\emph{\JournalTitle{\mnras}}} \textbf{\bibinfo{volume}{373}},
  \bibinfo{pages}{1074--1090} (\bibinfo{year}{2006}).
\newblock \doiprefix 10.1111/j.1365-2966.2006.11097.x.
\newblock \eprint{arXiv:astro-ph/0602350}.

\bibitem{Teyssier2013}
\bibinfo{author}{{Teyssier}, R.}, \bibinfo{author}{{Pontzen}, A.},
  \bibinfo{author}{{Dubois}, Y.} \& \bibinfo{author}{{Read}, J.~I.}
\newblock \bibinfo{journal}{\bibinfo{title}{{Cusp-core transformations in dwarf
  galaxies: observational predictions}}}.
\newblock {\emph{\JournalTitle{\mnras}}} \textbf{\bibinfo{volume}{429}},
  \bibinfo{pages}{3068--3078} (\bibinfo{year}{2013}).
\newblock \doiprefix 10.1093/mnras/sts563.
\newblock \eprint{1206.4895}.

\bibitem{Vogelsberger2014}
\bibinfo{author}{{Vogelsberger}, M.} \emph{et~al.}
\newblock \bibinfo{journal}{\bibinfo{title}{{Properties of galaxies reproduced
  by a hydrodynamic simulation}}}.
\newblock {\emph{\JournalTitle{\nat}}} \textbf{\bibinfo{volume}{509}},
  \bibinfo{pages}{177--182} (\bibinfo{year}{2014}).
\newblock \doiprefix 10.1038/nature13316.
\newblock \eprint{1405.1418}.

\bibitem{Pillepich2018}
\bibinfo{author}{{Pillepich}, A.} \emph{et~al.}
\newblock \bibinfo{journal}{\bibinfo{title}{{Simulating galaxy formation with
  the IllustrisTNG model}}}.
\newblock {\emph{\JournalTitle{\mnras}}} \textbf{\bibinfo{volume}{473}},
  \bibinfo{pages}{4077--4106} (\bibinfo{year}{2018}).
\newblock \doiprefix 10.1093/mnras/stx2656.
\newblock \eprint{1703.02970}.

\bibitem{Semenov2017}
\bibinfo{author}{{Semenov}, V.~A.}, \bibinfo{author}{{Kravtsov}, A.~V.} \&
  \bibinfo{author}{{Gnedin}, N.~Y.}
\newblock \bibinfo{journal}{\bibinfo{title}{{The Physical Origin of Long Gas
  Depletion Times in Galaxies}}}.
\newblock {\emph{\JournalTitle{\apj}}} \textbf{\bibinfo{volume}{845}},
  \bibinfo{pages}{133} (\bibinfo{year}{2017}).
\newblock \doiprefix 10.3847/1538-4357/aa8096.
\newblock \eprint{1704.04239}.

\bibitem{Trebitsch2017}
\bibinfo{author}{{Trebitsch}, M.}, \bibinfo{author}{{Blaizot}, J.},
  \bibinfo{author}{{Rosdahl}, J.}, \bibinfo{author}{{Devriendt}, J.} \&
  \bibinfo{author}{{Slyz}, A.}
\newblock \bibinfo{journal}{\bibinfo{title}{{Fluctuating feedback-regulated
  escape fraction of ionizing radiation in low-mass, high-redshift galaxies}}}.
\newblock {\emph{\JournalTitle{\mnras}}} \textbf{\bibinfo{volume}{470}},
  \bibinfo{pages}{224--239} (\bibinfo{year}{2017}).
\newblock \doiprefix 10.1093/mnras/stx1060.
\newblock \eprint{1705.00941}.

\bibitem{Rosdahl2018}
\bibinfo{author}{{Rosdahl}, J.} \emph{et~al.}
\newblock \bibinfo{journal}{\bibinfo{title}{{The SPHINX cosmological
  simulations of the first billion years: the impact of binary stars on
  reionization}}}.
\newblock {\emph{\JournalTitle{\mnras}}} \textbf{\bibinfo{volume}{479}},
  \bibinfo{pages}{994--1016} (\bibinfo{year}{2018}).
\newblock \doiprefix 10.1093/mnras/sty1655.
\newblock \eprint{1801.07259}.

\bibitem{Gnedin2011}
\bibinfo{author}{{Gnedin}, N.~Y.} \& \bibinfo{author}{{Kravtsov}, A.~V.}
\newblock \bibinfo{journal}{\bibinfo{title}{{Environmental Dependence of the
  Kennicutt-Schmidt Relation in Galaxies}}}.
\newblock {\emph{\JournalTitle{\apj}}} \textbf{\bibinfo{volume}{728}},
  \bibinfo{pages}{88} (\bibinfo{year}{2011}).
\newblock \doiprefix 10.1088/0004-637X/728/2/88.
\newblock \eprint{1004.0003}.

\bibitem{Christensen2012}
\bibinfo{author}{{Christensen}, C.} \emph{et~al.}
\newblock \bibinfo{journal}{\bibinfo{title}{{Implementing molecular hydrogen in
  hydrodynamic simulations of galaxy formation}}}.
\newblock {\emph{\JournalTitle{\mnras}}} \textbf{\bibinfo{volume}{425}},
  \bibinfo{pages}{3058--3076} (\bibinfo{year}{2012}).
\newblock \doiprefix 10.1111/j.1365-2966.2012.21628.x.
\newblock \eprint{1205.5567}.

\bibitem{Feldmann2012}
\bibinfo{author}{{Feldmann}, R.}, \bibinfo{author}{{Gnedin}, N.~Y.} \&
  \bibinfo{author}{{Kravtsov}, A.~V.}
\newblock \bibinfo{journal}{\bibinfo{title}{{The X-factor in Galaxies. II. The
  Molecular-hydrogen-Star-formation Relation}}}.
\newblock {\emph{\JournalTitle{\apj}}} \textbf{\bibinfo{volume}{758}},
  \bibinfo{pages}{127} (\bibinfo{year}{2012}).
\newblock \doiprefix 10.1088/0004-637X/758/2/127.
\newblock \eprint{1204.3910}.

\bibitem{Kuhlen2012}
\bibinfo{author}{{Kuhlen}, M.}, \bibinfo{author}{{Krumholz}, M.~R.},
  \bibinfo{author}{{Madau}, P.}, \bibinfo{author}{{Smith}, B.~D.} \&
  \bibinfo{author}{{Wise}, J.}
\newblock \bibinfo{journal}{\bibinfo{title}{{Dwarf Galaxy Formation with
  H$_{2}$-regulated Star Formation}}}.
\newblock {\emph{\JournalTitle{\apj}}} \textbf{\bibinfo{volume}{749}},
  \bibinfo{pages}{36} (\bibinfo{year}{2012}).
\newblock \doiprefix 10.1088/0004-637X/749/1/36.
\newblock \eprint{1105.2376}.

\bibitem{Monaco2012}
\bibinfo{author}{{Monaco}, P.}, \bibinfo{author}{{Murante}, G.},
  \bibinfo{author}{{Borgani}, S.} \& \bibinfo{author}{{Dolag}, K.}
\newblock \bibinfo{journal}{\bibinfo{title}{{Schmidt-Kennicutt relations in SPH
  simulations of disc galaxies with effective thermal feedback from
  supernovae}}}.
\newblock {\emph{\JournalTitle{\mnras}}} \textbf{\bibinfo{volume}{421}},
  \bibinfo{pages}{2485--2497} (\bibinfo{year}{2012}).
\newblock \doiprefix 10.1111/j.1365-2966.2012.20482.x.
\newblock \eprint{1109.0484}.

\bibitem{Li2017}
\bibinfo{author}{{Li}, H.} \emph{et~al.}
\newblock \bibinfo{journal}{\bibinfo{title}{{Star Cluster Formation in
  Cosmological Simulations. I. Properties of Young Clusters}}}.
\newblock {\emph{\JournalTitle{\apj}}} \textbf{\bibinfo{volume}{834}},
  \bibinfo{pages}{69} (\bibinfo{year}{2017}).
\newblock \doiprefix 10.3847/1538-4357/834/1/69.
\newblock \eprint{1608.03244}.

\bibitem{Wiersma2009a}
\bibinfo{author}{{Wiersma}, R.~P.~C.}, \bibinfo{author}{{Schaye}, J.},
  \bibinfo{author}{{Theuns}, T.}, \bibinfo{author}{{Dalla Vecchia}, C.} \&
  \bibinfo{author}{{Tornatore}, L.}
\newblock \bibinfo{journal}{\bibinfo{title}{{Chemical enrichment in
  cosmological, smoothed particle hydrodynamics simulations}}}.
\newblock {\emph{\JournalTitle{\mnras}}} \textbf{\bibinfo{volume}{399}},
  \bibinfo{pages}{574--600} (\bibinfo{year}{2009}).
\newblock \doiprefix 10.1111/j.1365-2966.2009.15331.x.
\newblock \eprint{0902.1535}.

\bibitem{Vogelsberger2013}
\bibinfo{author}{{Vogelsberger}, M.} \emph{et~al.}
\newblock \bibinfo{journal}{\bibinfo{title}{{A model for cosmological
  simulations of galaxy formation physics}}}.
\newblock {\emph{\JournalTitle{\mnras}}} \textbf{\bibinfo{volume}{436}},
  \bibinfo{pages}{3031--3067} (\bibinfo{year}{2013}).
\newblock \doiprefix 10.1093/mnras/stt1789.
\newblock \eprint{1305.2913}.

\bibitem{Naiman2018}
\bibinfo{author}{{Naiman}, J.~P.} \emph{et~al.}
\newblock \bibinfo{journal}{\bibinfo{title}{{First results from the
  IllustrisTNG simulations: a tale of two elements - chemical evolution of
  magnesium and europium}}}.
\newblock {\emph{\JournalTitle{\mnras}}} \textbf{\bibinfo{volume}{477}},
  \bibinfo{pages}{1206--1224} (\bibinfo{year}{2018}).
\newblock \doiprefix 10.1093/mnras/sty618.
\newblock \eprint{1707.03401}.

\bibitem{Oppenheimer2010}
\bibinfo{author}{{Oppenheimer}, B.~D.} \emph{et~al.}
\newblock \bibinfo{journal}{\bibinfo{title}{{Feedback and recycled wind
  accretion: assembling the z = 0 galaxy mass function}}}.
\newblock {\emph{\JournalTitle{\mnras}}} \textbf{\bibinfo{volume}{406}},
  \bibinfo{pages}{2325--2338} (\bibinfo{year}{2010}).
\newblock \doiprefix 10.1111/j.1365-2966.2010.16872.x.
\newblock \eprint{0912.0519}.

\bibitem{Stinson2013a}
\bibinfo{author}{{Stinson}, G.~S.} \emph{et~al.}
\newblock \bibinfo{journal}{\bibinfo{title}{{Making Galaxies In a Cosmological
  Context: the need for early stellar feedback}}}.
\newblock {\emph{\JournalTitle{\mnras}}} \textbf{\bibinfo{volume}{428}},
  \bibinfo{pages}{129--140} (\bibinfo{year}{2013a}).
\newblock \doiprefix 10.1093/mnras/sts028.
\newblock \eprint{1208.0002}.

\bibitem{Smith2018}
\bibinfo{author}{{Smith}, M.~C.}, \bibinfo{author}{{Sijacki}, D.} \&
  \bibinfo{author}{{Shen}, S.}
\newblock \bibinfo{journal}{\bibinfo{title}{{Cosmological simulations of
  dwarfs: the need for ISM physics beyond SN feedback alone}}}.
\newblock {\emph{\JournalTitle{\mnras}}} \textbf{\bibinfo{volume}{485}},
  \bibinfo{pages}{3317--3333} (\bibinfo{year}{2019}).
\newblock \doiprefix 10.1093/mnras/stz599.
\newblock \eprint{1807.04288}.

\bibitem{Behroozi2010}
\bibinfo{author}{{Behroozi}, P.~S.}, \bibinfo{author}{{Conroy}, C.} \&
  \bibinfo{author}{{Wechsler}, R.~H.}
\newblock \bibinfo{journal}{\bibinfo{title}{{A Comprehensive Analysis of
  Uncertainties Affecting the Stellar Mass-Halo Mass Relation for 0 < z < 4}}}.
\newblock {\emph{\JournalTitle{\apj}}} \textbf{\bibinfo{volume}{717}},
  \bibinfo{pages}{379--403} (\bibinfo{year}{2010}).
\newblock \doiprefix 10.1088/0004-637X/717/1/379.
\newblock \eprint{1001.0015}.

\bibitem{Moster2013}
\bibinfo{author}{{Moster}, B.~P.}, \bibinfo{author}{{Naab}, T.} \&
  \bibinfo{author}{{White}, S.~D.~M.}
\newblock \bibinfo{journal}{\bibinfo{title}{{Galactic star formation and
  accretion histories from matching galaxies to dark matter haloes}}}.
\newblock {\emph{\JournalTitle{\mnras}}} \textbf{\bibinfo{volume}{428}},
  \bibinfo{pages}{3121--3138} (\bibinfo{year}{2013}).
\newblock \doiprefix 10.1093/mnras/sts261.
\newblock \eprint{1205.5807}.

\bibitem{Muratov2015}
\bibinfo{author}{{Muratov}, A.~L.} \emph{et~al.}
\newblock \bibinfo{journal}{\bibinfo{title}{{Gusty, gaseous flows of FIRE:
  galactic winds in cosmological simulations with explicit stellar feedback}}}.
\newblock {\emph{\JournalTitle{\mnras}}} \textbf{\bibinfo{volume}{454}},
  \bibinfo{pages}{2691--2713} (\bibinfo{year}{2015}).
\newblock \doiprefix 10.1093/mnras/stv2126.
\newblock \eprint{1501.03155}.

\bibitem{Gehren1984}
\bibinfo{author}{{Gehren}, T.}, \bibinfo{author}{{Fried}, J.},
  \bibinfo{author}{{Wehinger}, P.~A.} \& \bibinfo{author}{{Wyckoff}, S.}
\newblock \bibinfo{journal}{\bibinfo{title}{{Host galaxies of quasars and their
  association with galaxy clusters}}}.
\newblock {\emph{\JournalTitle{\apj}}} \textbf{\bibinfo{volume}{278}},
  \bibinfo{pages}{11--27} (\bibinfo{year}{1984}).
\newblock \doiprefix 10.1086/161763.

\bibitem{Kormendy1995}
\bibinfo{author}{{Kormendy}, J.} \& \bibinfo{author}{{Richstone}, D.}
\newblock \bibinfo{journal}{\bibinfo{title}{{Inward Bound---The Search For
  Supermassive Black Holes In Galactic Nuclei}}}.
\newblock {\emph{\JournalTitle{\araa}}} \textbf{\bibinfo{volume}{33}},
  \bibinfo{pages}{581} (\bibinfo{year}{1995}).
\newblock \doiprefix 10.1146/annurev.aa.33.090195.003053.

\bibitem{Filippenko2003}
\bibinfo{author}{{Filippenko}, A.~V.} \& \bibinfo{author}{{Ho}, L.~C.}
\newblock \bibinfo{journal}{\bibinfo{title}{{A Low-Mass Central Black Hole in
  the Bulgeless Seyfert 1 Galaxy NGC 4395}}}.
\newblock {\emph{\JournalTitle{\apjl}}} \textbf{\bibinfo{volume}{588}},
  \bibinfo{pages}{L13--L16} (\bibinfo{year}{2003}).
\newblock \doiprefix 10.1086/375361.
\newblock \eprint{astro-ph/0303429}.

\bibitem{Shields2008}
\bibinfo{author}{{Shields}, J.~C.} \emph{et~al.}
\newblock \bibinfo{journal}{\bibinfo{title}{{An Accreting Black Hole in the
  Nuclear Star Cluster of the Bulgeless Galaxy NGC 1042}}}.
\newblock {\emph{\JournalTitle{\apj}}} \textbf{\bibinfo{volume}{682}},
  \bibinfo{pages}{104--109} (\bibinfo{year}{2008}).
\newblock \doiprefix 10.1086/589680.
\newblock \eprint{0804.4024}.

\bibitem{Reines2011}
\bibinfo{author}{{Reines}, A.~E.}, \bibinfo{author}{{Sivakoff}, G.~R.},
  \bibinfo{author}{{Johnson}, K.~E.} \& \bibinfo{author}{{Brogan}, C.~L.}
\newblock \bibinfo{journal}{\bibinfo{title}{{An actively accreting massive
  black hole in the dwarf starburst galaxy Henize2-10}}}.
\newblock {\emph{\JournalTitle{\nat}}} \textbf{\bibinfo{volume}{470}},
  \bibinfo{pages}{66--68} (\bibinfo{year}{2011}).
\newblock \doiprefix 10.1038/nature09724.
\newblock \eprint{1101.1309}.

\bibitem{Moran2014}
\bibinfo{author}{{Moran}, E.~C.}, \bibinfo{author}{{Shahinyan}, K.},
  \bibinfo{author}{{Sugarman}, H.~R.}, \bibinfo{author}{{V{\'e}lez}, D.~O.} \&
  \bibinfo{author}{{Eracleous}, M.}
\newblock \bibinfo{journal}{\bibinfo{title}{{Black Holes At the Centers of
  Nearby Dwarf Galaxies}}}.
\newblock {\emph{\JournalTitle{\aj}}} \textbf{\bibinfo{volume}{148}},
  \bibinfo{pages}{136} (\bibinfo{year}{2014}).
\newblock \doiprefix 10.1088/0004-6256/148/6/136.
\newblock \eprint{1408.4451}.

\bibitem{Booth2009}
\bibinfo{author}{{Booth}, C.~M.} \& \bibinfo{author}{{Schaye}, J.}
\newblock \bibinfo{journal}{\bibinfo{title}{{Cosmological simulations of the
  growth of supermassive black holes and feedback from active galactic nuclei:
  method and tests}}}.
\newblock {\emph{\JournalTitle{\mnras}}} \textbf{\bibinfo{volume}{398}},
  \bibinfo{pages}{53--74} (\bibinfo{year}{2009}).
\newblock \doiprefix 10.1111/j.1365-2966.2009.15043.x.
\newblock \eprint{0904.2572}.

\bibitem{Shlosman1989}
\bibinfo{author}{{Shlosman}, I.}, \bibinfo{author}{{Frank}, J.} \&
  \bibinfo{author}{{Begelman}, M.~C.}
\newblock \bibinfo{journal}{\bibinfo{title}{{Bars within bars - A mechanism for
  fuelling active galactic nuclei}}}.
\newblock {\emph{\JournalTitle{\nat}}} \textbf{\bibinfo{volume}{338}},
  \bibinfo{pages}{45--47} (\bibinfo{year}{1989}).
\newblock \doiprefix 10.1038/338045a0.

\bibitem{Hopkins2011b}
\bibinfo{author}{{Hopkins}, P.~F.} \& \bibinfo{author}{{Quataert}, E.}
\newblock \bibinfo{journal}{\bibinfo{title}{{An analytic model of angular
  momentum transport by gravitational torques: from galaxies to massive black
  holes}}}.
\newblock {\emph{\JournalTitle{\mnras}}} \textbf{\bibinfo{volume}{415}},
  \bibinfo{pages}{1027--1050} (\bibinfo{year}{2011}).
\newblock \doiprefix 10.1111/j.1365-2966.2011.18542.x.
\newblock \eprint{1007.2647}.

\bibitem{Bournaud2011}
\bibinfo{author}{{Bournaud}, F.} \emph{et~al.}
\newblock \bibinfo{journal}{\bibinfo{title}{{Black Hole Growth and Active
  Galactic Nuclei Obscuration by Instability-driven Inflows in High-redshift
  Disk Galaxies Fed by Cold Streams}}}.
\newblock {\emph{\JournalTitle{\apjl}}} \textbf{\bibinfo{volume}{741}},
  \bibinfo{pages}{L33} (\bibinfo{year}{2011}).
\newblock \doiprefix 10.1088/2041-8205/741/2/L33.
\newblock \eprint{1107.1483}.

\bibitem{Gabor2013}
\bibinfo{author}{{Gabor}, J.~M.} \& \bibinfo{author}{{Bournaud}, F.}
\newblock \bibinfo{journal}{\bibinfo{title}{{Simulations of supermassive black
  hole growth in high-redshift disc galaxies}}}.
\newblock {\emph{\JournalTitle{\mnras}}} \textbf{\bibinfo{volume}{434}},
  \bibinfo{pages}{606--620} (\bibinfo{year}{2013}).
\newblock \doiprefix 10.1093/mnras/stt1046.
\newblock \eprint{1306.2954}.

\bibitem{AnglesAlcazar2013}
\bibinfo{author}{{Angl{\'e}s-Alc{\'a}zar}, D.}, \bibinfo{author}{{{\"O}zel},
  F.} \& \bibinfo{author}{{Dav{\'e}}, R.}
\newblock \bibinfo{journal}{\bibinfo{title}{{Black Hole-Galaxy Correlations
  without Self-regulation}}}.
\newblock {\emph{\JournalTitle{\apj}}} \textbf{\bibinfo{volume}{770}},
  \bibinfo{pages}{5} (\bibinfo{year}{2013}).
\newblock \doiprefix 10.1088/0004-637X/770/1/5.
\newblock \eprint{1303.5058}.

\bibitem{AnglesAlcazar2017}
\bibinfo{author}{{Angl{\'e}s-Alc{\'a}zar}, D.} \emph{et~al.}
\newblock \bibinfo{journal}{\bibinfo{title}{{Black holes on FIRE: stellar
  feedback limits early feeding of galactic nuclei}}}.
\newblock {\emph{\JournalTitle{\mnras}}} \textbf{\bibinfo{volume}{472}},
  \bibinfo{pages}{L109--L114} (\bibinfo{year}{2017}).
\newblock \doiprefix 10.1093/mnrasl/slx161.
\newblock \eprint{1707.03832}.

\bibitem{Dave2019}
\bibinfo{author}{{Dav{\'e}}, R.} \emph{et~al.}
\newblock \bibinfo{journal}{\bibinfo{title}{{SIMBA: Cosmological simulations
  with black hole growth and feedback}}}.
\newblock {\emph{\JournalTitle{\mnras}}} \textbf{\bibinfo{volume}{486}},
  \bibinfo{pages}{2827--2849} (\bibinfo{year}{2019}).
\newblock \doiprefix 10.1093/mnras/stz937.
\newblock \eprint{1901.10203}.

\bibitem{Krolik1999}
\bibinfo{author}{{Krolik}, J.~H.}
\newblock \emph{\bibinfo{title}{{Active galactic nuclei : from the central
  black hole to the galactic environment}}} (\bibinfo{publisher}{Princeton,
  N.~J.~: Princeton University Press}, \bibinfo{year}{1999}).

\bibitem{Springel2005}
\bibinfo{author}{{Springel}, V.}, \bibinfo{author}{{Di Matteo}, T.} \&
  \bibinfo{author}{{Hernquist}, L.}
\newblock \bibinfo{journal}{\bibinfo{title}{{Modelling feedback from stars and
  black holes in galaxy mergers}}}.
\newblock {\emph{\JournalTitle{\mnras}}} \textbf{\bibinfo{volume}{361}},
  \bibinfo{pages}{776--794} (\bibinfo{year}{2005}).
\newblock \doiprefix 10.1111/j.1365-2966.2005.09238.x.
\newblock \eprint{astro-ph/0411108}.

\bibitem{DiMatteo2005}
\bibinfo{author}{{Di Matteo}, T.}, \bibinfo{author}{{Springel}, V.} \&
  \bibinfo{author}{{Hernquist}, L.}
\newblock \bibinfo{journal}{\bibinfo{title}{{Energy input from quasars
  regulates the growth and activity of black holes and their host galaxies}}}.
\newblock {\emph{\JournalTitle{\nat}}} \textbf{\bibinfo{volume}{433}},
  \bibinfo{pages}{604--607} (\bibinfo{year}{2005}).
\newblock \doiprefix 10.1038/nature03335.
\newblock \eprint{astro-ph/0502199}.

\bibitem{Debuhr2010}
\bibinfo{author}{{Debuhr}, J.}, \bibinfo{author}{{Quataert}, E.},
  \bibinfo{author}{{Ma}, C.-P.} \& \bibinfo{author}{{Hopkins}, P.}
\newblock \bibinfo{journal}{\bibinfo{title}{{Self-regulated black hole growth
  via momentum deposition in galaxy merger simulations}}}.
\newblock {\emph{\JournalTitle{\mnras}}} \textbf{\bibinfo{volume}{406}},
  \bibinfo{pages}{L55--L59} (\bibinfo{year}{2010}).
\newblock \doiprefix 10.1111/j.1745-3933.2010.00881.x.
\newblock \eprint{0909.2872}.

\bibitem{Costa2018}
\bibinfo{author}{{Costa}, T.}, \bibinfo{author}{{Rosdahl}, J.},
  \bibinfo{author}{{Sijacki}, D.} \& \bibinfo{author}{{Haehnelt}, M.~G.}
\newblock \bibinfo{journal}{\bibinfo{title}{{Driving gas shells with radiation
  pressure on dust in radiation-hydrodynamic simulations}}}.
\newblock {\emph{\JournalTitle{\mnras}}} \textbf{\bibinfo{volume}{473}},
  \bibinfo{pages}{4197--4219} (\bibinfo{year}{2018}).
\newblock \doiprefix 10.1093/mnras/stx2598.
\newblock \eprint{1703.05766}.

\bibitem{Barnes2018}
\bibinfo{author}{{Barnes}, D.~J.}, \bibinfo{author}{{Kannan}, R.},
  \bibinfo{author}{{Vogelsberger}, M.} \& \bibinfo{author}{{Marinacci}, F.}
\newblock \bibinfo{journal}{\bibinfo{title}{{Radiative AGN feedback on a moving
  mesh: the impact of the galactic disc and dust physics on outflow
  properties}}}.
\newblock {\emph{\JournalTitle{arXiv e-prints}}}  (\bibinfo{year}{2018}).
\newblock \eprint{1812.01611}.

\bibitem{Choi2012}
\bibinfo{author}{{Choi}, E.}, \bibinfo{author}{{Ostriker}, J.~P.},
  \bibinfo{author}{{Naab}, T.} \& \bibinfo{author}{{Johansson}, P.~H.}
\newblock \bibinfo{journal}{\bibinfo{title}{{Radiative and Momentum-based
  Mechanical Active Galactic Nucleus Feedback in a Three-dimensional Galaxy
  Evolution Code}}}.
\newblock {\emph{\JournalTitle{\apj}}} \textbf{\bibinfo{volume}{754}},
  \bibinfo{pages}{125} (\bibinfo{year}{2012}).
\newblock \doiprefix 10.1088/0004-637X/754/2/125.
\newblock \eprint{1205.2082}.

\bibitem{Sijacki2007}
\bibinfo{author}{{Sijacki}, D.}, \bibinfo{author}{{Springel}, V.},
  \bibinfo{author}{{Di Matteo}, T.} \& \bibinfo{author}{{Hernquist}, L.}
\newblock \bibinfo{journal}{\bibinfo{title}{{A unified model for AGN feedback
  in cosmological simulations of structure formation}}}.
\newblock {\emph{\JournalTitle{\mnras}}} \textbf{\bibinfo{volume}{380}},
  \bibinfo{pages}{877--900} (\bibinfo{year}{2007}).
\newblock \doiprefix 10.1111/j.1365-2966.2007.12153.x.
\newblock \eprint{0705.2238}.

\bibitem{Weinberger2017}
\bibinfo{author}{{Weinberger}, R.} \emph{et~al.}
\newblock \bibinfo{journal}{\bibinfo{title}{{Simulating galaxy formation with
  black hole driven thermal and kinetic feedback}}}.
\newblock {\emph{\JournalTitle{\mnras}}} \textbf{\bibinfo{volume}{465}},
  \bibinfo{pages}{3291--3308} (\bibinfo{year}{2017}).
\newblock \doiprefix 10.1093/mnras/stw2944.
\newblock \eprint{1607.03486}.

\bibitem{Marinacci2016}
\bibinfo{author}{{Marinacci}, F.} \& \bibinfo{author}{{Vogelsberger}, M.}
\newblock \bibinfo{journal}{\bibinfo{title}{{Effects of simulated cosmological
  magnetic fields on the galaxy population}}}.
\newblock {\emph{\JournalTitle{\mnras}}} \textbf{\bibinfo{volume}{456}},
  \bibinfo{pages}{L69--L73} (\bibinfo{year}{2016}).
\newblock \doiprefix 10.1093/mnrasl/slv176.
\newblock \eprint{1508.06631}.

\bibitem{Ferriere2001}
\bibinfo{author}{{Ferri{\`e}re}, K.~M.}
\newblock \bibinfo{journal}{\bibinfo{title}{{The interstellar environment of
  our galaxy}}}.
\newblock {\emph{\JournalTitle{Reviews of Modern Physics}}}
  \textbf{\bibinfo{volume}{73}}, \bibinfo{pages}{1031--1066}
  (\bibinfo{year}{2001}).
\newblock \doiprefix 10.1103/RevModPhys.73.1031.
\newblock \eprint{astro-ph/0106359}.

\bibitem{Kotera2011}
\bibinfo{author}{{Kotera}, K.} \& \bibinfo{author}{{Olinto}, A.~V.}
\newblock \bibinfo{journal}{\bibinfo{title}{{The Astrophysics of
  Ultrahigh-Energy Cosmic Rays}}}.
\newblock {\emph{\JournalTitle{\araa}}} \textbf{\bibinfo{volume}{49}},
  \bibinfo{pages}{119--153} (\bibinfo{year}{2011}).
\newblock \doiprefix 10.1146/annurev-astro-081710-102620.
\newblock \eprint{1101.4256}.

\bibitem{Donnert2009}
\bibinfo{author}{{Donnert}, J.}, \bibinfo{author}{{Dolag}, K.},
  \bibinfo{author}{{Lesch}, H.} \& \bibinfo{author}{{M{\"u}ller}, E.}
\newblock \bibinfo{journal}{\bibinfo{title}{{Cluster magnetic fields from
  galactic outflows}}}.
\newblock {\emph{\JournalTitle{\mnras}}} \textbf{\bibinfo{volume}{392}},
  \bibinfo{pages}{1008--1021} (\bibinfo{year}{2009}).
\newblock \doiprefix 10.1111/j.1365-2966.2008.14132.x.
\newblock \eprint{0808.0919}.

\bibitem{Marinacci2018}
\bibinfo{author}{{Marinacci}, F.} \emph{et~al.}
\newblock \bibinfo{journal}{\bibinfo{title}{{First results from the
  IllustrisTNG simulations: radio haloes and magnetic fields}}}.
\newblock {\emph{\JournalTitle{\mnras}}} \textbf{\bibinfo{volume}{480}},
  \bibinfo{pages}{5113--5139} (\bibinfo{year}{2018}).
\newblock \doiprefix 10.1093/mnras/sty2206.
\newblock \eprint{1707.03396}.

\bibitem{Pakmor2017}
\bibinfo{author}{{Pakmor}, R.} \emph{et~al.}
\newblock \bibinfo{journal}{\bibinfo{title}{{Magnetic field formation in the
  Milky Way like disc galaxies of the Auriga project}}}.
\newblock {\emph{\JournalTitle{\mnras}}} \textbf{\bibinfo{volume}{469}},
  \bibinfo{pages}{3185--3199} (\bibinfo{year}{2017}).
\newblock \doiprefix 10.1093/mnras/stx1074.
\newblock \eprint{1701.07028}.

\bibitem{Powell1999}
\bibinfo{author}{{Powell}, K.~G.}, \bibinfo{author}{{Roe}, P.~L.},
  \bibinfo{author}{{Linde}, T.~J.}, \bibinfo{author}{{Gombosi}, T.~I.} \&
  \bibinfo{author}{{De Zeeuw}, D.~L.}
\newblock \bibinfo{journal}{\bibinfo{title}{{A Solution-Adaptive Upwind Scheme
  for Ideal Magnetohydrodynamics}}}.
\newblock {\emph{\JournalTitle{Journal of Computational Physics}}}
  \textbf{\bibinfo{volume}{154}}, \bibinfo{pages}{284--309}
  (\bibinfo{year}{1999}).
\newblock \doiprefix 10.1006/jcph.1999.6299.

\bibitem{Dedner2002}
\bibinfo{author}{{Dedner}, A.} \emph{et~al.}
\newblock \bibinfo{journal}{\bibinfo{title}{{Hyperbolic Divergence Cleaning for
  the MHD Equations}}}.
\newblock {\emph{\JournalTitle{Journal of Computational Physics}}}
  \textbf{\bibinfo{volume}{175}}, \bibinfo{pages}{645--673}
  (\bibinfo{year}{2002}).
\newblock \doiprefix 10.1006/jcph.2001.6961.

\bibitem{Evans1988}
\bibinfo{author}{{Evans}, C.~R.} \& \bibinfo{author}{{Hawley}, J.~F.}
\newblock \bibinfo{journal}{\bibinfo{title}{{Simulation of magnetohydrodynamic
  flows - A constrained transport method}}}.
\newblock {\emph{\JournalTitle{\apj}}} \textbf{\bibinfo{volume}{332}},
  \bibinfo{pages}{659--677} (\bibinfo{year}{1988}).
\newblock \doiprefix 10.1086/166684.

\bibitem{Mocz2016}
\bibinfo{author}{{Mocz}, P.} \emph{et~al.}
\newblock \bibinfo{journal}{\bibinfo{title}{{A moving mesh unstaggered
  constrained transport scheme for magnetohydrodynamics}}}.
\newblock {\emph{\JournalTitle{\mnras}}} \textbf{\bibinfo{volume}{463}},
  \bibinfo{pages}{477--488} (\bibinfo{year}{2016}).
\newblock \doiprefix 10.1093/mnras/stw2004.
\newblock \eprint{1606.02310}.

\bibitem{Rosswog2007}
\bibinfo{author}{{Rosswog}, S.} \& \bibinfo{author}{{Price}, D.}
\newblock \bibinfo{journal}{\bibinfo{title}{{MAGMA: a three-dimensional,
  Lagrangian magnetohydrodynamics code for merger applications}}}.
\newblock {\emph{\JournalTitle{\mnras}}} \textbf{\bibinfo{volume}{379}},
  \bibinfo{pages}{915--931} (\bibinfo{year}{2007}).
\newblock \doiprefix 10.1111/j.1365-2966.2007.11984.x.
\newblock \eprint{0705.1441}.

\bibitem{Dolag2009}
\bibinfo{author}{{Dolag}, K.} \& \bibinfo{author}{{Stasyszyn}, F.}
\newblock \bibinfo{journal}{\bibinfo{title}{{An MHD GADGET for cosmological
  simulations}}}.
\newblock {\emph{\JournalTitle{\mnras}}} \textbf{\bibinfo{volume}{398}},
  \bibinfo{pages}{1678--1697} (\bibinfo{year}{2009}).
\newblock \doiprefix 10.1111/j.1365-2966.2009.15181.x.
\newblock \eprint{0807.3553}.

\bibitem{Stone1992}
\bibinfo{author}{{Stone}, J.~M.} \& \bibinfo{author}{{Norman}, M.~L.}
\newblock \bibinfo{journal}{\bibinfo{title}{{ZEUS-2D: A Radiation
  Magnetohydrodynamics Code for Astrophysical Flows in Two Space Dimensions.
  II. The Magnetohydrodynamic Algorithms and Tests}}}.
\newblock {\emph{\JournalTitle{\apjs}}} \textbf{\bibinfo{volume}{80}},
  \bibinfo{pages}{791} (\bibinfo{year}{1992}).
\newblock \doiprefix 10.1086/191681.

\bibitem{Londrillo2004}
\bibinfo{author}{{Londrillo}, P.} \& \bibinfo{author}{{del Zanna}, L.}
\newblock \bibinfo{journal}{\bibinfo{title}{{On the divergence-free condition
  in Godunov-type schemes for ideal magnetohydrodynamics: the upwind
  constrained transport method}}}.
\newblock {\emph{\JournalTitle{Journal of Computational Physics}}}
  \textbf{\bibinfo{volume}{195}}, \bibinfo{pages}{17--48}
  (\bibinfo{year}{2004}).
\newblock \doiprefix 10.1016/j.jcp.2003.09.016.
\newblock \eprint{astro-ph/0310183}.

\bibitem{Fromang2006}
\bibinfo{author}{{Fromang}, S.}, \bibinfo{author}{{Hennebelle}, P.} \&
  \bibinfo{author}{{Teyssier}, R.}
\newblock \bibinfo{journal}{\bibinfo{title}{{A high order Godunov scheme with
  constrained transport and adaptive mesh refinement for astrophysical
  magnetohydrodynamics}}}.
\newblock {\emph{\JournalTitle{\aap}}} \textbf{\bibinfo{volume}{457}},
  \bibinfo{pages}{371--384} (\bibinfo{year}{2006}).
\newblock \doiprefix 10.1051/0004-6361:20065371.
\newblock \eprint{astro-ph/0607230}.

\bibitem{Teyssier2006}
\bibinfo{author}{{Teyssier}, R.}, \bibinfo{author}{{Fromang}, S.} \&
  \bibinfo{author}{{Dormy}, E.}
\newblock \bibinfo{journal}{\bibinfo{title}{{Kinematic dynamos using
  constrained transport with high order Godunov schemes and adaptive mesh
  refinement}}}.
\newblock {\emph{\JournalTitle{Journal of Computational Physics}}}
  \textbf{\bibinfo{volume}{218}}, \bibinfo{pages}{44--67}
  (\bibinfo{year}{2006}).
\newblock \doiprefix 10.1016/j.jcp.2006.01.042.
\newblock \eprint{astro-ph/0601715}.

\bibitem{Balsara2012}
\bibinfo{author}{{Balsara}, D.~S.}
\newblock \bibinfo{journal}{\bibinfo{title}{{A two-dimensional HLLC Riemann
  solver for conservation laws: Application to Euler and magnetohydrodynamic
  flows}}}.
\newblock {\emph{\JournalTitle{Journal of Computational Physics}}}
  \textbf{\bibinfo{volume}{231}}, \bibinfo{pages}{7476--7503}
  (\bibinfo{year}{2012}).
\newblock \doiprefix 10.1016/j.jcp.2011.12.025.
\newblock \eprint{1110.0750}.

\bibitem{Ferriere2003}
\bibinfo{author}{{Ferriere}, K.}
\newblock \bibinfo{journal}{\bibinfo{title}{{Magnetic Fields in Galaxies: Their
  Origin and Their Impact on the Interstellar Medium}}}.
\newblock {\emph{\JournalTitle{Acta Astronomica Sinica}}}
  \textbf{\bibinfo{volume}{44}}, \bibinfo{pages}{115--122}
  (\bibinfo{year}{2003}).

\bibitem{Cox2005}
\bibinfo{author}{{Cox}, D.~P.}
\newblock \bibinfo{journal}{\bibinfo{title}{{The Three-Phase Interstellar
  Medium Revisited}}}.
\newblock {\emph{\JournalTitle{\araa}}} \textbf{\bibinfo{volume}{43}},
  \bibinfo{pages}{337--385} (\bibinfo{year}{2005}).
\newblock \doiprefix 10.1146/annurev.astro.43.072103.150615.

\bibitem{Field1969}
\bibinfo{author}{{Field}, G.~B.}, \bibinfo{author}{{Goldsmith}, D.~W.} \&
  \bibinfo{author}{{Habing}, H.~J.}
\newblock \bibinfo{journal}{\bibinfo{title}{{Cosmic-Ray Heating of the
  Interstellar Gas}}}.
\newblock {\emph{\JournalTitle{\apjl}}} \textbf{\bibinfo{volume}{155}},
  \bibinfo{pages}{L149} (\bibinfo{year}{1969}).
\newblock \doiprefix 10.1086/180324.

\bibitem{Wolfire1995}
\bibinfo{author}{{Wolfire}, M.~G.}, \bibinfo{author}{{Hollenbach}, D.},
  \bibinfo{author}{{McKee}, C.~F.}, \bibinfo{author}{{Tielens}, A.~G.~G.~M.} \&
  \bibinfo{author}{{Bakes}, E.~L.~O.}
\newblock \bibinfo{journal}{\bibinfo{title}{{The neutral atomic phases of the
  interstellar medium}}}.
\newblock {\emph{\JournalTitle{\apj}}} \textbf{\bibinfo{volume}{443}},
  \bibinfo{pages}{152--168} (\bibinfo{year}{1995}).
\newblock \doiprefix 10.1086/175510.

\bibitem{Uhlig2012}
\bibinfo{author}{{Uhlig}, M.} \emph{et~al.}
\newblock \bibinfo{journal}{\bibinfo{title}{{Galactic winds driven by cosmic
  ray streaming}}}.
\newblock {\emph{\JournalTitle{\mnras}}} \textbf{\bibinfo{volume}{423}},
  \bibinfo{pages}{2374--2396} (\bibinfo{year}{2012}).
\newblock \doiprefix 10.1111/j.1365-2966.2012.21045.x.
\newblock \eprint{1203.1038}.

\bibitem{Booth2013}
\bibinfo{author}{{Booth}, C.~M.}, \bibinfo{author}{{Agertz}, O.},
  \bibinfo{author}{{Kravtsov}, A.~V.} \& \bibinfo{author}{{Gnedin}, N.~Y.}
\newblock \bibinfo{journal}{\bibinfo{title}{{Simulations of Disk Galaxies with
  Cosmic Ray Driven Galactic Winds}}}.
\newblock {\emph{\JournalTitle{\apjl}}} \textbf{\bibinfo{volume}{777}},
  \bibinfo{pages}{L16} (\bibinfo{year}{2013}).
\newblock \doiprefix 10.1088/2041-8205/777/1/L16.
\newblock \eprint{1308.4974}.

\bibitem{Hanasz2013}
\bibinfo{author}{{Hanasz}, M.} \emph{et~al.}
\newblock \bibinfo{journal}{\bibinfo{title}{{Cosmic Rays Can Drive Strong
  Outflows from Gas-rich High-redshift Disk Galaxies}}}.
\newblock {\emph{\JournalTitle{\apjl}}} \textbf{\bibinfo{volume}{777}},
  \bibinfo{pages}{L38} (\bibinfo{year}{2013}).
\newblock \doiprefix 10.1088/2041-8205/777/2/L38.
\newblock \eprint{1310.3273}.

\bibitem{Salem2014}
\bibinfo{author}{{Salem}, M.}, \bibinfo{author}{{Bryan}, G.~L.} \&
  \bibinfo{author}{{Hummels}, C.}
\newblock \bibinfo{journal}{\bibinfo{title}{{Cosmological Simulations of Galaxy
  Formation with Cosmic Rays}}}.
\newblock {\emph{\JournalTitle{\apjl}}} \textbf{\bibinfo{volume}{797}},
  \bibinfo{pages}{L18} (\bibinfo{year}{2014}).
\newblock \doiprefix 10.1088/2041-8205/797/2/L18.
\newblock \eprint{1412.0661}.

\bibitem{Pakmor2016}
\bibinfo{author}{{Pakmor}, R.}, \bibinfo{author}{{Pfrommer}, C.},
  \bibinfo{author}{{Simpson}, C.~M.} \& \bibinfo{author}{{Springel}, V.}
\newblock \bibinfo{journal}{\bibinfo{title}{{Galactic Winds Driven by Isotropic
  and Anisotropic Cosmic-Ray Diffusion in Disk Galaxies}}}.
\newblock {\emph{\JournalTitle{\apjl}}} \textbf{\bibinfo{volume}{824}},
  \bibinfo{pages}{L30} (\bibinfo{year}{2016}).
\newblock \doiprefix 10.3847/2041-8205/824/2/L30.
\newblock \eprint{1605.00643}.

\bibitem{Simpson2016}
\bibinfo{author}{{Simpson}, C.~M.} \emph{et~al.}
\newblock \bibinfo{journal}{\bibinfo{title}{{The Role of Cosmic-Ray Pressure in
  Accelerating Galactic Outflows}}}.
\newblock {\emph{\JournalTitle{\apjl}}} \textbf{\bibinfo{volume}{827}},
  \bibinfo{pages}{L29} (\bibinfo{year}{2016}).
\newblock \doiprefix 10.3847/2041-8205/827/2/L29.
\newblock \eprint{1606.02324}.

\bibitem{Ruszkowski2017}
\bibinfo{author}{{Ruszkowski}, M.}, \bibinfo{author}{{Yang}, H.-Y.~K.} \&
  \bibinfo{author}{{Zweibel}, E.}
\newblock \bibinfo{journal}{\bibinfo{title}{{Global Simulations of Galactic
  Winds Including Cosmic-ray Streaming}}}.
\newblock {\emph{\JournalTitle{\apj}}} \textbf{\bibinfo{volume}{834}},
  \bibinfo{pages}{208} (\bibinfo{year}{2017}).
\newblock \doiprefix 10.3847/1538-4357/834/2/208.
\newblock \eprint{1602.04856}.

\bibitem{Farber2018}
\bibinfo{author}{{Farber}, R.}, \bibinfo{author}{{Ruszkowski}, M.},
  \bibinfo{author}{{Yang}, H.-Y.~K.} \& \bibinfo{author}{{Zweibel}, E.~G.}
\newblock \bibinfo{journal}{\bibinfo{title}{{Impact of Cosmic-Ray Transport on
  Galactic Winds}}}.
\newblock {\emph{\JournalTitle{\apj}}} \textbf{\bibinfo{volume}{856}},
  \bibinfo{pages}{112} (\bibinfo{year}{2018}).
\newblock \doiprefix 10.3847/1538-4357/aab26d.
\newblock \eprint{1707.04579}.

\bibitem{Girichidis2018}
\bibinfo{author}{{Girichidis}, P.}, \bibinfo{author}{{Naab}, T.},
  \bibinfo{author}{{Hanasz}, M.} \& \bibinfo{author}{{Walch}, S.}
\newblock \bibinfo{journal}{\bibinfo{title}{{Cooler and smoother - the impact
  of cosmic rays on the phase structure of galactic outflows}}}.
\newblock {\emph{\JournalTitle{\mnras}}} \textbf{\bibinfo{volume}{479}},
  \bibinfo{pages}{3042--3067} (\bibinfo{year}{2018}).
\newblock \doiprefix 10.1093/mnras/sty1653.
\newblock \eprint{1805.09333}.

\bibitem{Jacob2018}
\bibinfo{author}{{Jacob}, S.}, \bibinfo{author}{{Pakmor}, R.},
  \bibinfo{author}{{Simpson}, C.~M.}, \bibinfo{author}{{Springel}, V.} \&
  \bibinfo{author}{{Pfrommer}, C.}
\newblock \bibinfo{journal}{\bibinfo{title}{{The dependence of cosmic
  ray-driven galactic winds on halo mass}}}.
\newblock {\emph{\JournalTitle{\mnras}}} \textbf{\bibinfo{volume}{475}},
  \bibinfo{pages}{570--584} (\bibinfo{year}{2018}).
\newblock \doiprefix 10.1093/mnras/stx3221.
\newblock \eprint{1712.04947}.

\bibitem{Pfrommer2017}
\bibinfo{author}{{Pfrommer}, C.}, \bibinfo{author}{{Pakmor}, R.},
  \bibinfo{author}{{Schaal}, K.}, \bibinfo{author}{{Simpson}, C.~M.} \&
  \bibinfo{author}{{Springel}, V.}
\newblock \bibinfo{journal}{\bibinfo{title}{{Simulating cosmic ray physics on a
  moving mesh}}}.
\newblock {\emph{\JournalTitle{\mnras}}} \textbf{\bibinfo{volume}{465}},
  \bibinfo{pages}{4500--4529} (\bibinfo{year}{2017}).
\newblock \doiprefix 10.1093/mnras/stw2941.
\newblock \eprint{1604.07399}.

\bibitem{Sharma2007}
\bibinfo{author}{{Sharma}, P.} \& \bibinfo{author}{{Hammett}, G.~W.}
\newblock \bibinfo{journal}{\bibinfo{title}{{Preserving monotonicity in
  anisotropic diffusion}}}.
\newblock {\emph{\JournalTitle{Journal of Computational Physics}}}
  \textbf{\bibinfo{volume}{227}}, \bibinfo{pages}{123--142}
  (\bibinfo{year}{2007}).
\newblock \doiprefix 10.1016/j.jcp.2007.07.026.
\newblock \eprint{0707.2616}.

\bibitem{Kannan2016}
\bibinfo{author}{{Kannan}, R.}, \bibinfo{author}{{Springel}, V.},
  \bibinfo{author}{{Pakmor}, R.}, \bibinfo{author}{{Marinacci}, F.} \&
  \bibinfo{author}{{Vogelsberger}, M.}
\newblock \bibinfo{journal}{\bibinfo{title}{{Accurately simulating anisotropic
  thermal conduction on a moving mesh}}}.
\newblock {\emph{\JournalTitle{\mnras}}} \textbf{\bibinfo{volume}{458}},
  \bibinfo{pages}{410--424} (\bibinfo{year}{2016}).
\newblock \doiprefix 10.1093/mnras/stw294.
\newblock \eprint{1512.03053}.

\bibitem{Butsky2018}
\bibinfo{author}{{Butsky}, I.~S.} \& \bibinfo{author}{{Quinn}, T.~R.}
\newblock \bibinfo{journal}{\bibinfo{title}{{The Role of Cosmic-ray Transport
  in Shaping the Simulated Circumgalactic Medium}}}.
\newblock {\emph{\JournalTitle{\apj}}} \textbf{\bibinfo{volume}{868}},
  \bibinfo{pages}{108} (\bibinfo{year}{2018}).
\newblock \doiprefix 10.3847/1538-4357/aaeac2.
\newblock \eprint{1803.06345}.

\bibitem{Sharma2009}
\bibinfo{author}{{Sharma}, P.}, \bibinfo{author}{{Colella}, P.} \&
  \bibinfo{author}{{Martin}, D.}
\newblock \bibinfo{journal}{\bibinfo{title}{Numerical implementation of
  streaming down the gradient: Application to fluid modeling of cosmic rays and
  saturated conduction}}.
\newblock {\emph{\JournalTitle{Siam Journal on Scientific Computing}}}
  \textbf{\bibinfo{volume}{32}}, \bibinfo{pages}{3476--3494}
  (\bibinfo{year}{2009}).
\newblock \doiprefix 10.1137/100792135.

\bibitem{Jiang2018}
\bibinfo{author}{{Jiang}, Y.-F.} \& \bibinfo{author}{{Oh}, S.~P.}
\newblock \bibinfo{journal}{\bibinfo{title}{{A New Numerical Scheme for
  Cosmic-Ray Transport}}}.
\newblock {\emph{\JournalTitle{\apj}}} \textbf{\bibinfo{volume}{854}},
  \bibinfo{pages}{5} (\bibinfo{year}{2018}).
\newblock \doiprefix 10.3847/1538-4357/aaa6ce.
\newblock \eprint{1712.07117}.

\bibitem{Thomas2019}
\bibinfo{author}{{Thomas}, T.} \& \bibinfo{author}{{Pfrommer}, C.}
\newblock \bibinfo{journal}{\bibinfo{title}{{Cosmic-ray hydrodynamics:
  Alfv{\'e}n-wave regulated transport of cosmic rays}}}.
\newblock {\emph{\JournalTitle{\mnras}}} \textbf{\bibinfo{volume}{485}},
  \bibinfo{pages}{2977--3008} (\bibinfo{year}{2019}).
\newblock \doiprefix 10.1093/mnras/stz263.
\newblock \eprint{1805.11092}.

\bibitem{Gnedin2014}
\bibinfo{author}{{Gnedin}, N.~Y.} \& \bibinfo{author}{{Kaurov}, A.~A.}
\newblock \bibinfo{journal}{\bibinfo{title}{{Cosmic Reionization on Computers.
  II. Reionization History and Its Back-reaction on Early Galaxies}}}.
\newblock {\emph{\JournalTitle{\apj}}} \textbf{\bibinfo{volume}{793}},
  \bibinfo{pages}{30} (\bibinfo{year}{2014}).
\newblock \doiprefix 10.1088/0004-637X/793/1/30.
\newblock \eprint{1403.4251}.

\bibitem{Ocvirk2018}
\bibinfo{author}{{Ocvirk}, P.} \emph{et~al.}
\newblock \bibinfo{journal}{\bibinfo{title}{{Cosmic Dawn II (CoDa II): a new
  radiation-hydrodynamics simulation of the self-consistent coupling of galaxy
  formation and reionization}}}.
\newblock {\emph{\JournalTitle{arXiv e-prints}}}  (\bibinfo{year}{2018}).
\newblock \eprint{1811.11192}.

\bibitem{Wise2012}
\bibinfo{author}{{Wise}, J.~H.}, \bibinfo{author}{{Abel}, T.},
  \bibinfo{author}{{Turk}, M.~J.}, \bibinfo{author}{{Norman}, M.~L.} \&
  \bibinfo{author}{{Smith}, B.~D.}
\newblock \bibinfo{journal}{\bibinfo{title}{{The birth of a galaxy - II. The
  role of radiation pressure}}}.
\newblock {\emph{\JournalTitle{\mnras}}} \textbf{\bibinfo{volume}{427}},
  \bibinfo{pages}{311--326} (\bibinfo{year}{2012}).
\newblock \doiprefix 10.1111/j.1365-2966.2012.21809.x.
\newblock \eprint{1206.1043}.

\bibitem{Gnedin2001}
\bibinfo{author}{{Gnedin}, N.~Y.} \& \bibinfo{author}{{Abel}, T.}
\newblock \bibinfo{journal}{\bibinfo{title}{{Multi-dimensional cosmological
  radiative transfer with a Variable Eddington Tensor formalism}}}.
\newblock {\emph{\JournalTitle{\na}}} \textbf{\bibinfo{volume}{6}},
  \bibinfo{pages}{437--455} (\bibinfo{year}{2001}).
\newblock \doiprefix 10.1016/S1384-1076(01)00068-9.
\newblock \eprint{astro-ph/0106278}.

\bibitem{Gnedin2016}
\bibinfo{author}{{Gnedin}, N.~Y.}
\newblock \bibinfo{journal}{\bibinfo{title}{{On the Proper Use of the Reduced
  Speed of Light Approximation}}}.
\newblock {\emph{\JournalTitle{\apj}}} \textbf{\bibinfo{volume}{833}},
  \bibinfo{pages}{66} (\bibinfo{year}{2016}).
\newblock \doiprefix 10.3847/1538-4357/833/1/66.
\newblock \eprint{1607.07869}.

\bibitem{Deparis2019}
\bibinfo{author}{{Deparis}, N.}, \bibinfo{author}{{Aubert}, D.},
  \bibinfo{author}{{Ocvirk}, P.}, \bibinfo{author}{{Chardin}, J.} \&
  \bibinfo{author}{{Lewis}, J.}
\newblock \bibinfo{journal}{\bibinfo{title}{{Impact of the reduced speed of
  light approximation on ionization front velocities in cosmological
  simulations of the epoch of reionization}}}.
\newblock {\emph{\JournalTitle{\aap}}} \textbf{\bibinfo{volume}{622}},
  \bibinfo{pages}{A142} (\bibinfo{year}{2019}).
\newblock \doiprefix 10.1051/0004-6361/201832889.
\newblock \eprint{1803.01634}.

\bibitem{Ocvirk2019}
\bibinfo{author}{{Ocvirk}, P.}, \bibinfo{author}{{Aubert}, D.},
  \bibinfo{author}{{Chardin}, J.}, \bibinfo{author}{{Deparis}, N.} \&
  \bibinfo{author}{{Lewis}, J.}
\newblock \bibinfo{journal}{\bibinfo{title}{{Impact of the reduced speed of
  light approximation on the post-overlap neutral hydrogen fraction in
  numerical simulations of the epoch of reionization}}}.
\newblock {\emph{\JournalTitle{\aap}}} \textbf{\bibinfo{volume}{626}},
  \bibinfo{pages}{A77} (\bibinfo{year}{2019}).
\newblock \doiprefix 10.1051/0004-6361/201832923.
\newblock \eprint{1803.02434}.

\bibitem{Mihalas1984}
\bibinfo{author}{{Mihalas}, D.} \& \bibinfo{author}{{Mihalas}, B.~W.}
\newblock \emph{\bibinfo{title}{{Foundations of radiation hydrodynamics}}}
  (\bibinfo{publisher}{New York, Oxford University Press},
  \bibinfo{year}{1984}).

\bibitem{Abel1999}
\bibinfo{author}{{Abel}, T.}, \bibinfo{author}{{Norman}, M.~L.} \&
  \bibinfo{author}{{Madau}, P.}
\newblock \bibinfo{journal}{\bibinfo{title}{{Photon-conserving Radiative
  Transfer around Point Sources in Multidimensional Numerical Cosmology}}}.
\newblock {\emph{\JournalTitle{\apj}}} \textbf{\bibinfo{volume}{523}},
  \bibinfo{pages}{66--71} (\bibinfo{year}{1999}).
\newblock \doiprefix 10.1086/307739.
\newblock \eprint{astro-ph/9812151}.

\bibitem{Jaura2018}
\bibinfo{author}{{Jaura}, O.}, \bibinfo{author}{{Glover}, S.~C.~O.},
  \bibinfo{author}{{Klessen}, R.~S.} \& \bibinfo{author}{{Paardekooper}, J.-P.}
\newblock \bibinfo{journal}{\bibinfo{title}{{SPRAI: coupling of radiative
  feedback and primordial chemistry in moving mesh hydrodynamics}}}.
\newblock {\emph{\JournalTitle{\mnras}}} \textbf{\bibinfo{volume}{475}},
  \bibinfo{pages}{2822--2834} (\bibinfo{year}{2018}).
\newblock \doiprefix 10.1093/mnras/stx3356.
\newblock \eprint{1711.02542}.

\bibitem{Rijkhorst2006}
\bibinfo{author}{{Rijkhorst}, E.-J.}, \bibinfo{author}{{Plewa}, T.},
  \bibinfo{author}{{Dubey}, A.} \& \bibinfo{author}{{Mellema}, G.}
\newblock \bibinfo{journal}{\bibinfo{title}{{Hybrid characteristics: 3D
  radiative transfer for parallel adaptive mesh refinement hydrodynamics}}}.
\newblock {\emph{\JournalTitle{\aap}}} \textbf{\bibinfo{volume}{452}},
  \bibinfo{pages}{907--920} (\bibinfo{year}{2006}).
\newblock \doiprefix 10.1051/0004-6361:20053401.
\newblock \eprint{astro-ph/0505213}.

\bibitem{Rosen2017}
\bibinfo{author}{{Rosen}, A.~L.}, \bibinfo{author}{{Krumholz}, M.~R.},
  \bibinfo{author}{{Oishi}, J.~S.}, \bibinfo{author}{{Lee}, A.~T.} \&
  \bibinfo{author}{{Klein}, R.~I.}
\newblock \bibinfo{journal}{\bibinfo{title}{{Hybrid Adaptive Ray-Moment Method
  (HARM$^{2}$): A highly parallel method for radiation hydrodynamics on
  adaptive grids}}}.
\newblock {\emph{\JournalTitle{Journal of Computational Physics}}}
  \textbf{\bibinfo{volume}{330}}, \bibinfo{pages}{924--942}
  (\bibinfo{year}{2017}).
\newblock \doiprefix 10.1016/j.jcp.2016.10.048.
\newblock \eprint{1607.01802}.

\bibitem{Whalen2006}
\bibinfo{author}{{Whalen}, D.} \& \bibinfo{author}{{Norman}, M.~L.}
\newblock \bibinfo{journal}{\bibinfo{title}{{A Multistep Algorithm for the
  Radiation Hydrodynamical Transport of Cosmological Ionization Fronts and
  Ionized Flows}}}.
\newblock {\emph{\JournalTitle{\apjs}}} \textbf{\bibinfo{volume}{162}},
  \bibinfo{pages}{281--303} (\bibinfo{year}{2006}).
\newblock \doiprefix 10.1086/499072.
\newblock \eprint{astro-ph/0508214}.

\bibitem{Trac2007}
\bibinfo{author}{{Trac}, H.} \& \bibinfo{author}{{Cen}, R.}
\newblock \bibinfo{journal}{\bibinfo{title}{{Radiative Transfer Simulations of
  Cosmic Reionization. I. Methodology and Initial Results}}}.
\newblock {\emph{\JournalTitle{\apj}}} \textbf{\bibinfo{volume}{671}},
  \bibinfo{pages}{1--13} (\bibinfo{year}{2007}).
\newblock \doiprefix 10.1086/522566.
\newblock \eprint{astro-ph/0612406}.

\bibitem{Pawlik2008}
\bibinfo{author}{{Pawlik}, A.~H.} \& \bibinfo{author}{{Schaye}, J.}
\newblock \bibinfo{journal}{\bibinfo{title}{{TRAPHIC - radiative transfer for
  smoothed particle hydrodynamics simulations}}}.
\newblock {\emph{\JournalTitle{\mnras}}} \textbf{\bibinfo{volume}{389}},
  \bibinfo{pages}{651--677} (\bibinfo{year}{2008}).
\newblock \doiprefix 10.1111/j.1365-2966.2008.13601.x.
\newblock \eprint{0802.1715}.

\bibitem{Petkova2011}
\bibinfo{author}{{Petkova}, M.} \& \bibinfo{author}{{Springel}, V.}
\newblock \bibinfo{journal}{\bibinfo{title}{{A novel approach for accurate
  radiative transfer in cosmological hydrodynamic simulations}}}.
\newblock {\emph{\JournalTitle{\mnras}}} \textbf{\bibinfo{volume}{415}},
  \bibinfo{pages}{3731--3749} (\bibinfo{year}{2011}).
\newblock \doiprefix 10.1111/j.1365-2966.2011.18986.x.
\newblock \eprint{1012.1017}.

\bibitem{Ciardi2001}
\bibinfo{author}{{Ciardi}, B.}, \bibinfo{author}{{Ferrara}, A.},
  \bibinfo{author}{{Marri}, S.} \& \bibinfo{author}{{Raimondo}, G.}
\newblock \bibinfo{journal}{\bibinfo{title}{{Cosmological reionization around
  the first stars: Monte Carlo radiative transfer}}}.
\newblock {\emph{\JournalTitle{\mnras}}} \textbf{\bibinfo{volume}{324}},
  \bibinfo{pages}{381--388} (\bibinfo{year}{2001}).
\newblock \doiprefix 10.1046/j.1365-8711.2001.04316.x.
\newblock \eprint{astro-ph/0005181}.

\bibitem{Oxley2003}
\bibinfo{author}{{Oxley}, S.} \& \bibinfo{author}{{Woolfson}, M.~M.}
\newblock \bibinfo{journal}{\bibinfo{title}{{Smoothed particle hydrodynamics
  with radiation transfer}}}.
\newblock {\emph{\JournalTitle{\mnras}}} \textbf{\bibinfo{volume}{343}},
  \bibinfo{pages}{900--912} (\bibinfo{year}{2003}).
\newblock \doiprefix 10.1046/j.1365-8711.2003.06751.x.

\bibitem{Tasitsiomi2006}
\bibinfo{author}{{Tasitsiomi}, A.}
\newblock \bibinfo{journal}{\bibinfo{title}{{Ly{\ensuremath{\alpha}} Radiative
  Transfer in Cosmological Simulations and Application to a z \textasciitilde=
  8 Ly{\ensuremath{\alpha}} Emitter}}}.
\newblock {\emph{\JournalTitle{\apj}}} \textbf{\bibinfo{volume}{645}},
  \bibinfo{pages}{792--813} (\bibinfo{year}{2006}).
\newblock \doiprefix 10.1086/504460.
\newblock \eprint{astro-ph/0510347}.

\bibitem{Semelin2007}
\bibinfo{author}{{Semelin}, B.}, \bibinfo{author}{{Combes}, F.} \&
  \bibinfo{author}{{Baek}, S.}
\newblock \bibinfo{journal}{\bibinfo{title}{{Lyman-alpha radiative transfer
  during the epoch of reionization: contribution to 21-cm signal
  fluctuations}}}.
\newblock {\emph{\JournalTitle{\aap}}} \textbf{\bibinfo{volume}{474}},
  \bibinfo{pages}{365--374} (\bibinfo{year}{2007}).
\newblock \doiprefix 10.1051/0004-6361:20077965.
\newblock \eprint{0707.2483}.

\bibitem{Dullemond2012}
\bibinfo{author}{{Dullemond}, C.~P.} \emph{et~al.}
\newblock \bibinfo{title}{{RADMC-3D: A multi-purpose radiative transfer tool}}.
\newblock \bibinfo{howpublished}{Astrophysics Source Code Library}
  (\bibinfo{year}{2012}).
\newblock \eprint{1202.015}.

\bibitem{Smith2019}
\bibinfo{author}{{Smith}, A.} \emph{et~al.}
\newblock \bibinfo{journal}{\bibinfo{title}{{The physics of Lyman
  {\ensuremath{\alpha}} escape from high-redshift galaxies}}}.
\newblock {\emph{\JournalTitle{\mnras}}} \textbf{\bibinfo{volume}{484}},
  \bibinfo{pages}{39--59} (\bibinfo{year}{2019}).
\newblock \doiprefix 10.1093/mnras/sty3483.
\newblock \eprint{1810.08185}.

\bibitem{Lucy1999}
\bibinfo{author}{{Lucy}, L.~B.}
\newblock \bibinfo{journal}{\bibinfo{title}{{Improved Monte Carlo techniques
  for the spectral synthesis of supernovae}}}.
\newblock {\emph{\JournalTitle{\aap}}} \textbf{\bibinfo{volume}{345}},
  \bibinfo{pages}{211--220} (\bibinfo{year}{1999}).

\bibitem{Gentile2001}
\bibinfo{author}{{Gentile}, N.~A.}
\newblock \bibinfo{journal}{\bibinfo{title}{{Implicit Monte Carlo
  Diffusion{\textemdash}An Acceleration Method for Monte Carlo Time-Dependent
  Radiative Transfer Simulations}}}.
\newblock {\emph{\JournalTitle{Journal of Computational Physics}}}
  \textbf{\bibinfo{volume}{172}}, \bibinfo{pages}{543--571}
  (\bibinfo{year}{2001}).
\newblock \doiprefix 10.1006/jcph.2001.6836.

\bibitem{Densmore2007}
\bibinfo{author}{{Densmore}, J.~D.}, \bibinfo{author}{{Urbatsch}, T.~J.},
  \bibinfo{author}{{Evans}, T.~M.} \& \bibinfo{author}{{Buksas}, M.~W.}
\newblock \bibinfo{journal}{\bibinfo{title}{{A hybrid transport-diffusion
  method for Monte Carlo radiative-transfer simulations}}}.
\newblock {\emph{\JournalTitle{Journal of Computational Physics}}}
  \textbf{\bibinfo{volume}{222}}, \bibinfo{pages}{485--503}
  (\bibinfo{year}{2007}).
\newblock \doiprefix 10.1016/j.jcp.2006.07.031.

\bibitem{SmithTsang2018}
\bibinfo{author}{{Smith}, A.}, \bibinfo{author}{{Tsang}, B. T.~H.},
  \bibinfo{author}{{Bromm}, V.} \& \bibinfo{author}{{Milosavljevi{\'c}}, M.}
\newblock \bibinfo{journal}{\bibinfo{title}{{Discrete diffusion Lyman
  {\ensuremath{\alpha}} radiative transfer}}}.
\newblock {\emph{\JournalTitle{\mnras}}} \textbf{\bibinfo{volume}{479}},
  \bibinfo{pages}{2065--2078} (\bibinfo{year}{2018}).
\newblock \doiprefix 10.1093/mnras/sty1509.
\newblock \eprint{1709.10187}.

\bibitem{Levermore1984}
\bibinfo{author}{{Levermore}, C.~D.}
\newblock \bibinfo{journal}{\bibinfo{title}{{Relating Eddington factors to flux
  limiters.}}}
\newblock {\emph{\JournalTitle{Journal of Quantitative Spectroscopy and
  Radiative Transfer}}} \textbf{\bibinfo{volume}{31}},
  \bibinfo{pages}{149--160} (\bibinfo{year}{1984}).
\newblock \doiprefix 10.1016/0022-4073(84)90112-2.

\bibitem{Gonzalez2007}
\bibinfo{author}{{Gonz{\'a}lez}, M.}, \bibinfo{author}{{Audit}, E.} \&
  \bibinfo{author}{{Huynh}, P.}
\newblock \bibinfo{journal}{\bibinfo{title}{{HERACLES: a three-dimensional
  radiation hydrodynamics code}}}.
\newblock {\emph{\JournalTitle{\aap}}} \textbf{\bibinfo{volume}{464}},
  \bibinfo{pages}{429--435} (\bibinfo{year}{2007}).
\newblock \doiprefix 10.1051/0004-6361:20065486.

\bibitem{Rosdahl2013}
\bibinfo{author}{{Rosdahl}, J.}, \bibinfo{author}{{Blaizot}, J.},
  \bibinfo{author}{{Aubert}, D.}, \bibinfo{author}{{Stranex}, T.} \&
  \bibinfo{author}{{Teyssier}, R.}
\newblock \bibinfo{journal}{\bibinfo{title}{{RAMSES-RT: radiation hydrodynamics
  in the cosmological context}}}.
\newblock {\emph{\JournalTitle{\mnras}}} \textbf{\bibinfo{volume}{436}},
  \bibinfo{pages}{2188--2231} (\bibinfo{year}{2013}).
\newblock \doiprefix 10.1093/mnras/stt1722.
\newblock \eprint{1304.7126}.

\bibitem{Kannan2018}
\bibinfo{author}{{Kannan}, R.} \emph{et~al.}
\newblock \bibinfo{journal}{\bibinfo{title}{{AREPO-RT: radiation hydrodynamics
  on a moving mesh}}}.
\newblock {\emph{\JournalTitle{\mnras}}} \textbf{\bibinfo{volume}{485}},
  \bibinfo{pages}{117--149} (\bibinfo{year}{2019}).
\newblock \doiprefix 10.1093/mnras/stz287.
\newblock \eprint{1804.01987}.

\bibitem{Krumholz2007}
\bibinfo{author}{{Krumholz}, M.~R.}, \bibinfo{author}{{Klein}, R.~I.},
  \bibinfo{author}{{McKee}, C.~F.} \& \bibinfo{author}{{Bolstad}, J.}
\newblock \bibinfo{journal}{\bibinfo{title}{{Equations and Algorithms for
  Mixed-frame Flux-limited Diffusion Radiation Hydrodynamics}}}.
\newblock {\emph{\JournalTitle{\apj}}} \textbf{\bibinfo{volume}{667}},
  \bibinfo{pages}{626--643} (\bibinfo{year}{2007}).
\newblock \doiprefix 10.1086/520791.
\newblock \eprint{astro-ph/0611003}.

\bibitem{Filantor2009}
\bibinfo{author}{{Finlator}, K.}, \bibinfo{author}{{{\"O}zel}, F.} \&
  \bibinfo{author}{{Dav{\'e}}, R.}
\newblock \bibinfo{journal}{\bibinfo{title}{{A new moment method for continuum
  radiative transfer in cosmological re-ionization}}}.
\newblock {\emph{\JournalTitle{\mnras}}} \textbf{\bibinfo{volume}{393}},
  \bibinfo{pages}{1090--1106} (\bibinfo{year}{2009}).
\newblock \doiprefix 10.1111/j.1365-2966.2008.14190.x.
\newblock \eprint{0808.3578}.

\bibitem{Petkova2009}
\bibinfo{author}{{Petkova}, M.} \& \bibinfo{author}{{Springel}, V.}
\newblock \bibinfo{journal}{\bibinfo{title}{{An implementation of radiative
  transfer in the cosmological simulation code GADGET}}}.
\newblock {\emph{\JournalTitle{\mnras}}} \textbf{\bibinfo{volume}{396}},
  \bibinfo{pages}{1383--1403} (\bibinfo{year}{2009}).
\newblock \doiprefix 10.1111/j.1365-2966.2009.14843.x.
\newblock \eprint{0812.1801}.

\bibitem{Dubroca1999}
\bibinfo{author}{{Dubroca}, B.} \& \bibinfo{author}{{Feugeas}, J.}
\newblock \bibinfo{journal}{\bibinfo{title}{{Etude th{\'e}orique et
  num{\'e}rique d'une hi{\'e}rarchie de mod{\`e}les aux moments pour le
  transfert radiatif}}}.
\newblock {\emph{\JournalTitle{Academie des Sciences Paris Comptes Rendus Serie
  Sciences Mathematiques}}} \textbf{\bibinfo{volume}{329}},
  \bibinfo{pages}{915--920} (\bibinfo{year}{1999}).
\newblock \doiprefix 10.1016/S0764-4442(00)87499-6.

\bibitem{Ripoll2001}
\bibinfo{author}{{Ripoll}, J.-F.}, \bibinfo{author}{{Dubroca}, B.} \&
  \bibinfo{author}{{Duffa}, G.}
\newblock \bibinfo{journal}{\bibinfo{title}{{Modelling radiative mean
  absorption coefficients}}}.
\newblock {\emph{\JournalTitle{Combustion Theory and Modelling}}}
  \textbf{\bibinfo{volume}{5}}, \bibinfo{pages}{261--274}
  (\bibinfo{year}{2001}).
\newblock \doiprefix 10.1088/1364-7830/5/3/301.

\bibitem{McKinnon2016}
\bibinfo{author}{{McKinnon}, R.}, \bibinfo{author}{{Torrey}, P.} \&
  \bibinfo{author}{{Vogelsberger}, M.}
\newblock \bibinfo{journal}{\bibinfo{title}{{Dust formation in Milky Way-like
  galaxies}}}.
\newblock {\emph{\JournalTitle{\mnras}}} \textbf{\bibinfo{volume}{457}},
  \bibinfo{pages}{3775--3800} (\bibinfo{year}{2016}).
\newblock \doiprefix 10.1093/mnras/stw253.
\newblock \eprint{1505.04792}.

\bibitem{Aoyama2017}
\bibinfo{author}{{Aoyama}, S.} \emph{et~al.}
\newblock \bibinfo{journal}{\bibinfo{title}{{Galaxy simulation with dust
  formation and destruction}}}.
\newblock {\emph{\JournalTitle{\mnras}}} \textbf{\bibinfo{volume}{466}},
  \bibinfo{pages}{105--121} (\bibinfo{year}{2017}).
\newblock \doiprefix 10.1093/mnras/stw3061.
\newblock \eprint{1609.07547}.

\bibitem{Hou2017}
\bibinfo{author}{{Hou}, K.-C.}, \bibinfo{author}{{Hirashita}, H.},
  \bibinfo{author}{{Nagamine}, K.}, \bibinfo{author}{{Aoyama}, S.} \&
  \bibinfo{author}{{Shimizu}, I.}
\newblock \bibinfo{journal}{\bibinfo{title}{{Evolution of dust extinction
  curves in galaxy simulation}}}.
\newblock {\emph{\JournalTitle{\mnras}}} \textbf{\bibinfo{volume}{469}},
  \bibinfo{pages}{870--885} (\bibinfo{year}{2017}).
\newblock \doiprefix 10.1093/mnras/stx877.
\newblock \eprint{1704.01769}.

\bibitem{McKinnon2017}
\bibinfo{author}{{McKinnon}, R.}, \bibinfo{author}{{Torrey}, P.},
  \bibinfo{author}{{Vogelsberger}, M.}, \bibinfo{author}{{Hayward}, C.~C.} \&
  \bibinfo{author}{{Marinacci}, F.}
\newblock \bibinfo{journal}{\bibinfo{title}{{Simulating the dust content of
  galaxies: successes and failures}}}.
\newblock {\emph{\JournalTitle{\mnras}}} \textbf{\bibinfo{volume}{468}},
  \bibinfo{pages}{1505--1521} (\bibinfo{year}{2017}).
\newblock \doiprefix 10.1093/mnras/stx467.
\newblock \eprint{1606.02714}.

\bibitem{Aoyama2018}
\bibinfo{author}{{Aoyama}, S.}, \bibinfo{author}{{Hou}, K.-C.},
  \bibinfo{author}{{Hirashita}, H.}, \bibinfo{author}{{Nagamine}, K.} \&
  \bibinfo{author}{{Shimizu}, I.}
\newblock \bibinfo{journal}{\bibinfo{title}{{Cosmological simulation with dust
  formation and destruction}}}.
\newblock {\emph{\JournalTitle{\mnras}}} \textbf{\bibinfo{volume}{478}},
  \bibinfo{pages}{4905--4921} (\bibinfo{year}{2018}).
\newblock \doiprefix 10.1093/mnras/sty1431.
\newblock \eprint{1802.04027}.

\bibitem{McKinnon2018}
\bibinfo{author}{{McKinnon}, R.}, \bibinfo{author}{{Vogelsberger}, M.},
  \bibinfo{author}{{Torrey}, P.}, \bibinfo{author}{{Marinacci}, F.} \&
  \bibinfo{author}{{Kannan}, R.}
\newblock \bibinfo{journal}{\bibinfo{title}{{Simulating galactic dust grain
  evolution on a moving mesh}}}.
\newblock {\emph{\JournalTitle{\mnras}}} \textbf{\bibinfo{volume}{478}},
  \bibinfo{pages}{2851--2886} (\bibinfo{year}{2018}).
\newblock \doiprefix 10.1093/mnras/sty1248.
\newblock \eprint{1805.04521}.

\bibitem{Vogelsberger2018}
\bibinfo{author}{{Vogelsberger}, M.} \emph{et~al.}
\newblock \bibinfo{journal}{\bibinfo{title}{{Dust in and around galaxies: dust
  in cluster environments and its impact on gas cooling}}}.
\newblock {\emph{\JournalTitle{\mnras}}} \textbf{\bibinfo{volume}{487}},
  \bibinfo{pages}{4870--4883} (\bibinfo{year}{2019}).
\newblock \doiprefix 10.1093/mnras/stz1644.
\newblock \eprint{1811.05477}.

\bibitem{Gjergo2018}
\bibinfo{author}{{Gjergo}, E.} \emph{et~al.}
\newblock \bibinfo{journal}{\bibinfo{title}{{Dust evolution in galaxy cluster
  simulations}}}.
\newblock {\emph{\JournalTitle{\mnras}}} \textbf{\bibinfo{volume}{479}},
  \bibinfo{pages}{2588--2606} (\bibinfo{year}{2018}).
\newblock \doiprefix 10.1093/mnras/sty1564.
\newblock \eprint{1804.06855}.

\bibitem{Hou2019}
\bibinfo{author}{{Hou}, K.-C.}, \bibinfo{author}{{Aoyama}, S.},
  \bibinfo{author}{{Hirashita}, H.}, \bibinfo{author}{{Nagamine}, K.} \&
  \bibinfo{author}{{Shimizu}, I.}
\newblock \bibinfo{journal}{\bibinfo{title}{{Dust scaling relations in a
  cosmological simulation}}}.
\newblock {\emph{\JournalTitle{\mnras}}} \textbf{\bibinfo{volume}{485}},
  \bibinfo{pages}{1727--1744} (\bibinfo{year}{2019}).
\newblock \doiprefix 10.1093/mnras/stz121.
\newblock \eprint{1901.02886}.

\bibitem{Ruszkowski2011b}
\bibinfo{author}{{Ruszkowski}, M.}, \bibinfo{author}{{Lee}, D.},
  \bibinfo{author}{{Br{\"u}ggen}, M.}, \bibinfo{author}{{Parrish}, I.} \&
  \bibinfo{author}{{Oh}, S.~P.}
\newblock \bibinfo{journal}{\bibinfo{title}{{Cosmological Magnetohydrodynamic
  Simulations of Cluster Formation with Anisotropic Thermal Conduction}}}.
\newblock {\emph{\JournalTitle{\apj}}} \textbf{\bibinfo{volume}{740}},
  \bibinfo{pages}{81} (\bibinfo{year}{2011}).
\newblock \doiprefix 10.1088/0004-637X/740/2/81.
\newblock \eprint{1010.2277}.

\bibitem{Smith2013}
\bibinfo{author}{{Smith}, B.}, \bibinfo{author}{{O'Shea}, B.~W.},
  \bibinfo{author}{{Voit}, G.~M.}, \bibinfo{author}{{Ventimiglia}, D.} \&
  \bibinfo{author}{{Skillman}, S.~W.}
\newblock \bibinfo{journal}{\bibinfo{title}{{Cosmological Simulations of
  Isotropic Conduction in Galaxy Clusters}}}.
\newblock {\emph{\JournalTitle{\apj}}} \textbf{\bibinfo{volume}{778}},
  \bibinfo{pages}{152} (\bibinfo{year}{2013}).
\newblock \doiprefix 10.1088/0004-637X/778/2/152.
\newblock \eprint{1306.5748}.

\bibitem{Arth2014}
\bibinfo{author}{{Arth}, A.}, \bibinfo{author}{{Dolag}, K.},
  \bibinfo{author}{{Beck}, A.~M.}, \bibinfo{author}{{Petkova}, M.} \&
  \bibinfo{author}{{Lesch}, H.}
\newblock \bibinfo{journal}{\bibinfo{title}{{Anisotropic thermal conduction in
  galaxy clusters with MHD in Gadget}}}.
\newblock {\emph{\JournalTitle{arXiv e-prints}}}  (\bibinfo{year}{2014}).
\newblock \eprint{1412.6533}.

\bibitem{Yang2016}
\bibinfo{author}{{Yang}, H.-Y.~K.} \& \bibinfo{author}{{Reynolds}, C.~S.}
\newblock \bibinfo{journal}{\bibinfo{title}{{Interplay Among Cooling, AGN
  Feedback, and Anisotropic Conduction in the Cool Cores of Galaxy Clusters}}}.
\newblock {\emph{\JournalTitle{\apj}}} \textbf{\bibinfo{volume}{818}},
  \bibinfo{pages}{181} (\bibinfo{year}{2016}).
\newblock \doiprefix 10.3847/0004-637X/818/2/181.
\newblock \eprint{1512.05796}.

\bibitem{Kannan2017}
\bibinfo{author}{{Kannan}, R.} \emph{et~al.}
\newblock \bibinfo{journal}{\bibinfo{title}{{Increasing Black Hole
  Feedback-induced Quenching with Anisotropic Thermal Conduction}}}.
\newblock {\emph{\JournalTitle{\apjl}}} \textbf{\bibinfo{volume}{837}},
  \bibinfo{pages}{L18} (\bibinfo{year}{2017}).
\newblock \doiprefix 10.3847/2041-8213/aa624b.
\newblock \eprint{1612.01522}.

\bibitem{Parrish2012}
\bibinfo{author}{{Parrish}, I.~J.}, \bibinfo{author}{{McCourt}, M.},
  \bibinfo{author}{{Quataert}, E.} \& \bibinfo{author}{{Sharma}, P.}
\newblock \bibinfo{journal}{\bibinfo{title}{{The effects of anisotropic
  viscosity on turbulence and heat transport in the intracluster medium}}}.
\newblock {\emph{\JournalTitle{\mnras}}} \textbf{\bibinfo{volume}{422}},
  \bibinfo{pages}{704--718} (\bibinfo{year}{2012}).
\newblock \doiprefix 10.1111/j.1365-2966.2012.20650.x.
\newblock \eprint{1201.0754}.

\bibitem{Suzuki2013}
\bibinfo{author}{{Suzuki}, K.}, \bibinfo{author}{{Ogawa}, T.},
  \bibinfo{author}{{Matsumoto}, Y.} \& \bibinfo{author}{{Matsumoto}, R.}
\newblock \bibinfo{journal}{\bibinfo{title}{{Magnetohydrodynamic Simulations of
  the Formation of Cold Fronts in Clusters of Galaxies: Effects of Anisotropic
  Viscosity}}}.
\newblock {\emph{\JournalTitle{\apj}}} \textbf{\bibinfo{volume}{768}},
  \bibinfo{pages}{175} (\bibinfo{year}{2013}).
\newblock \doiprefix 10.1088/0004-637X/768/2/175.

\bibitem{ZuHone2015}
\bibinfo{author}{{ZuHone}, J.~A.}, \bibinfo{author}{{Kunz}, M.~W.},
  \bibinfo{author}{{Markevitch}, M.}, \bibinfo{author}{{Stone}, J.~M.} \&
  \bibinfo{author}{{Biffi}, V.}
\newblock \bibinfo{journal}{\bibinfo{title}{{The Effect of Anisotropic
  Viscosity on Cold Fronts in Galaxy Clusters}}}.
\newblock {\emph{\JournalTitle{\apj}}} \textbf{\bibinfo{volume}{798}},
  \bibinfo{pages}{90} (\bibinfo{year}{2015}).
\newblock \doiprefix 10.1088/0004-637X/798/2/90.
\newblock \eprint{1406.4031}.

\bibitem{Su2017}
\bibinfo{author}{{Su}, K.-Y.} \emph{et~al.}
\newblock \bibinfo{journal}{\bibinfo{title}{{Feedback first: the surprisingly
  weak effects of magnetic fields, viscosity, conduction and metal diffusion on
  sub-L* galaxy formation}}}.
\newblock {\emph{\JournalTitle{\mnras}}} \textbf{\bibinfo{volume}{471}},
  \bibinfo{pages}{144--166} (\bibinfo{year}{2017}).
\newblock \doiprefix 10.1093/mnras/stx1463.
\newblock \eprint{1607.05274}.

\bibitem{Balbus2000}
\bibinfo{author}{{Balbus}, S.~A.}
\newblock \bibinfo{journal}{\bibinfo{title}{{Stability, Instability, and
  ``Backward'' Transport in Stratified Fluids}}}.
\newblock {\emph{\JournalTitle{\apj}}} \textbf{\bibinfo{volume}{534}},
  \bibinfo{pages}{420--427} (\bibinfo{year}{2000}).
\newblock \doiprefix 10.1086/308732.
\newblock \eprint{astro-ph/9906315}.

\bibitem{Quataert2008}
\bibinfo{author}{{Quataert}, E.}
\newblock \bibinfo{journal}{\bibinfo{title}{{Buoyancy Instabilities in Weakly
  Magnetized Low-Collisionality Plasmas}}}.
\newblock {\emph{\JournalTitle{\apj}}} \textbf{\bibinfo{volume}{673}},
  \bibinfo{pages}{758--762} (\bibinfo{year}{2008}).
\newblock \doiprefix 10.1086/525248.
\newblock \eprint{0710.5521}.

\bibitem{Barnes2019}
\bibinfo{author}{{Barnes}, D.~J.} \emph{et~al.}
\newblock \bibinfo{journal}{\bibinfo{title}{{Enhancing AGN efficiency and
  cool-core formation with anisotropic thermal conduction}}}.
\newblock {\emph{\JournalTitle{arXiv e-prints}}}  (\bibinfo{year}{2018}).
\newblock \eprint{1805.04109}.

\bibitem{Torrey2015}
\bibinfo{author}{{Torrey}, P.} \emph{et~al.}
\newblock \bibinfo{journal}{\bibinfo{title}{{Synthetic galaxy images and
  spectra from the Illustris simulation}}}.
\newblock {\emph{\JournalTitle{\mnras}}} \textbf{\bibinfo{volume}{447}},
  \bibinfo{pages}{2753--2771} (\bibinfo{year}{2015}).
\newblock \doiprefix 10.1093/mnras/stu2592.
\newblock \eprint{1411.3717}.

\bibitem{Trayford2017}
\bibinfo{author}{{Trayford}, J.~W.} \emph{et~al.}
\newblock \bibinfo{journal}{\bibinfo{title}{{Optical colours and spectral
  indices of z = 0.1 eagle galaxies with the 3D dust radiative transfer code
  skirt}}}.
\newblock {\emph{\JournalTitle{\mnras}}} \textbf{\bibinfo{volume}{470}},
  \bibinfo{pages}{771--799} (\bibinfo{year}{2017}).
\newblock \doiprefix 10.1093/mnras/stx1051.
\newblock \eprint{1705.02331}.

\bibitem{Hernquist1996}
\bibinfo{author}{{Hernquist}, L.}, \bibinfo{author}{{Katz}, N.},
  \bibinfo{author}{{Weinberg}, D.~H.} \& \bibinfo{author}{{Miralda-Escud{\'e}},
  J.}
\newblock \bibinfo{journal}{\bibinfo{title}{{The Lyman-Alpha Forest in the Cold
  Dark Matter Model}}}.
\newblock {\emph{\JournalTitle{\apjl}}} \textbf{\bibinfo{volume}{457}},
  \bibinfo{pages}{L51} (\bibinfo{year}{1996}).
\newblock \doiprefix 10.1086/309899.
\newblock \eprint{astro-ph/9509105}.

\bibitem{Katz1991}
\bibinfo{author}{{Katz}, N.} \& \bibinfo{author}{{Gunn}, J.~E.}
\newblock \bibinfo{journal}{\bibinfo{title}{{Dissipational galaxy formation. I
  - Effects of gasdynamics}}}.
\newblock {\emph{\JournalTitle{\apj}}} \textbf{\bibinfo{volume}{377}},
  \bibinfo{pages}{365--381} (\bibinfo{year}{1991}).
\newblock \doiprefix 10.1086/170367.

\bibitem{Navarro1991}
\bibinfo{author}{{Navarro}, J.~F.} \& \bibinfo{author}{{Benz}, W.}
\newblock \bibinfo{journal}{\bibinfo{title}{{Dynamics of cooling gas in
  galactic dark halos}}}.
\newblock {\emph{\JournalTitle{\apj}}} \textbf{\bibinfo{volume}{380}},
  \bibinfo{pages}{320--329} (\bibinfo{year}{1991}).
\newblock \doiprefix 10.1086/170590.

\bibitem{Katz1992}
\bibinfo{author}{{Katz}, N.}
\newblock \bibinfo{journal}{\bibinfo{title}{{Dissipational galaxy formation. II
  - Effects of star formation}}}.
\newblock {\emph{\JournalTitle{\apj}}} \textbf{\bibinfo{volume}{391}},
  \bibinfo{pages}{502--517} (\bibinfo{year}{1992}).
\newblock \doiprefix 10.1086/171366.

\bibitem{Katz1992b}
\bibinfo{author}{{Katz}, N.}, \bibinfo{author}{{Hernquist}, L.} \&
  \bibinfo{author}{{Weinberg}, D.~H.}
\newblock \bibinfo{journal}{\bibinfo{title}{{Galaxies and gas in a cold dark
  matter universe}}}.
\newblock {\emph{\JournalTitle{\apjl}}} \textbf{\bibinfo{volume}{399}},
  \bibinfo{pages}{L109--L112} (\bibinfo{year}{1992}).
\newblock \doiprefix 10.1086/186619.

\bibitem{Navarro1997b}
\bibinfo{author}{{Navarro}, J.~F.} \& \bibinfo{author}{{Steinmetz}, M.}
\newblock \bibinfo{journal}{\bibinfo{title}{{The Effects of a Photoionizing
  Ultraviolet Background on the Formation of Disk Galaxies}}}.
\newblock {\emph{\JournalTitle{\apj}}} \textbf{\bibinfo{volume}{478}},
  \bibinfo{pages}{13--28} (\bibinfo{year}{1997}).
\newblock \doiprefix 10.1086/303763.
\newblock \eprint{astro-ph/9605043}.

\bibitem{Navarro2000}
\bibinfo{author}{{Navarro}, J.~F.} \& \bibinfo{author}{{Steinmetz}, M.}
\newblock \bibinfo{journal}{\bibinfo{title}{{Dark Halo and Disk Galaxy Scaling
  Laws in Hierarchical Universes}}}.
\newblock {\emph{\JournalTitle{\apj}}} \textbf{\bibinfo{volume}{538}},
  \bibinfo{pages}{477--488} (\bibinfo{year}{2000}).
\newblock \doiprefix 10.1086/309175.
\newblock \eprint{arXiv:astro-ph/0001003}.

\bibitem{Abadi2003}
\bibinfo{author}{{Abadi}, M.~G.}, \bibinfo{author}{{Navarro}, J.~F.},
  \bibinfo{author}{{Steinmetz}, M.} \& \bibinfo{author}{{Eke}, V.~R.}
\newblock \bibinfo{journal}{\bibinfo{title}{{Simulations of Galaxy Formation in
  a {$\Lambda$} Cold Dark Matter Universe. I. Dynamical and Photometric
  Properties of a Simulated Disk Galaxy}}}.
\newblock {\emph{\JournalTitle{\apj}}} \textbf{\bibinfo{volume}{591}},
  \bibinfo{pages}{499--514} (\bibinfo{year}{2003}).
\newblock \doiprefix 10.1086/375512.
\newblock \eprint{astro-ph/0211331}.

\bibitem{Scannapieco2008}
\bibinfo{author}{{Scannapieco}, C.}, \bibinfo{author}{{Tissera}, P.~B.},
  \bibinfo{author}{{White}, S.~D.~M.} \& \bibinfo{author}{{Springel}, V.}
\newblock \bibinfo{journal}{\bibinfo{title}{{Effects of supernova feedback on
  the formation of galaxy discs}}}.
\newblock {\emph{\JournalTitle{MNRAS}}} \textbf{\bibinfo{volume}{389}},
  \bibinfo{pages}{1137--1149} (\bibinfo{year}{2008}).
\newblock \doiprefix 10.1111/j.1365-2966.2008.13678.x.
\newblock \eprint{0804.3795}.

\bibitem{Grand2017}
\bibinfo{author}{{Grand}, R. J.~J.} \emph{et~al.}
\newblock \bibinfo{journal}{\bibinfo{title}{{The Auriga Project: the properties
  and formation mechanisms of disc galaxies across cosmic time}}}.
\newblock {\emph{\JournalTitle{\mnras}}} \textbf{\bibinfo{volume}{467}},
  \bibinfo{pages}{179--207} (\bibinfo{year}{2017}).
\newblock \doiprefix 10.1093/mnras/stx071.

\bibitem{Kaviraj2017}
\bibinfo{author}{{Kaviraj}, S.} \emph{et~al.}
\newblock \bibinfo{journal}{\bibinfo{title}{{The Horizon-AGN simulation:
  evolution of galaxy properties over cosmic time}}}.
\newblock {\emph{\JournalTitle{\mnras}}} \textbf{\bibinfo{volume}{467}},
  \bibinfo{pages}{4739--4752} (\bibinfo{year}{2017}).
\newblock \doiprefix 10.1093/mnras/stx126.
\newblock \eprint{1605.09379}.

\bibitem{Schechter1976}
\bibinfo{author}{{Schechter}, P.}
\newblock \bibinfo{journal}{\bibinfo{title}{{An analytic expression for the
  luminosity function for galaxies.}}}
\newblock {\emph{\JournalTitle{\apj}}} \textbf{\bibinfo{volume}{203}},
  \bibinfo{pages}{297--306} (\bibinfo{year}{1976}).
\newblock \doiprefix 10.1086/154079.

\bibitem{Panter2007}
\bibinfo{author}{{Panter}, B.}, \bibinfo{author}{{Jimenez}, R.},
  \bibinfo{author}{{Heavens}, A.~F.} \& \bibinfo{author}{{Charlot}, S.}
\newblock \bibinfo{journal}{\bibinfo{title}{{The star formation histories of
  galaxies in the Sloan Digital Sky Survey}}}.
\newblock {\emph{\JournalTitle{\mnras}}} \textbf{\bibinfo{volume}{378}},
  \bibinfo{pages}{1550--1564} (\bibinfo{year}{2007}).
\newblock \doiprefix 10.1111/j.1365-2966.2007.11909.x.
\newblock \eprint{astro-ph/0608531}.

\bibitem{Pozzetti2010}
\bibinfo{author}{{Pozzetti}, L.} \emph{et~al.}
\newblock \bibinfo{journal}{\bibinfo{title}{{zCOSMOS - 10k-bright spectroscopic
  sample. The bimodality in the galaxy stellar mass function: exploring its
  evolution with redshift}}}.
\newblock {\emph{\JournalTitle{\aap}}} \textbf{\bibinfo{volume}{523}},
  \bibinfo{pages}{A13} (\bibinfo{year}{2010}).
\newblock \doiprefix 10.1051/0004-6361/200913020.
\newblock \eprint{0907.5416}.

\bibitem{Baldry2012}
\bibinfo{author}{{Baldry}, I.~K.} \emph{et~al.}
\newblock \bibinfo{journal}{\bibinfo{title}{{Galaxy And Mass Assembly (GAMA):
  the galaxy stellar mass function at z $<$ 0.06}}}.
\newblock {\emph{\JournalTitle{\mnras}}} \textbf{\bibinfo{volume}{421}},
  \bibinfo{pages}{621--634} (\bibinfo{year}{2012}).
\newblock \doiprefix 10.1111/j.1365-2966.2012.20340.x.
\newblock \eprint{1111.5707}.

\bibitem{Ilbert2013}
\bibinfo{author}{{Ilbert}, O.} \emph{et~al.}
\newblock \bibinfo{journal}{\bibinfo{title}{{Mass assembly in quiescent and
  star-forming galaxies since z ${\simeq}$ 4 from UltraVISTA}}}.
\newblock {\emph{\JournalTitle{\aap}}} \textbf{\bibinfo{volume}{556}},
  \bibinfo{pages}{A55} (\bibinfo{year}{2013}).
\newblock \doiprefix 10.1051/0004-6361/201321100.
\newblock \eprint{1301.3157}.

\bibitem{Muzzin2013}
\bibinfo{author}{{Muzzin}, A.} \emph{et~al.}
\newblock \bibinfo{journal}{\bibinfo{title}{{The Evolution of the Stellar Mass
  Functions of Star-forming and Quiescent Galaxies to z = 4 from the
  COSMOS/UltraVISTA Survey}}}.
\newblock {\emph{\JournalTitle{\apj}}} \textbf{\bibinfo{volume}{777}},
  \bibinfo{pages}{18} (\bibinfo{year}{2013}).
\newblock \doiprefix 10.1088/0004-637X/777/1/18.
\newblock \eprint{1303.4409}.

\bibitem{Weigel2016}
\bibinfo{author}{{Weigel}, A.~K.}, \bibinfo{author}{{Schawinski}, K.} \&
  \bibinfo{author}{{Bruderer}, C.}
\newblock \bibinfo{journal}{\bibinfo{title}{{Stellar mass functions: methods,
  systematics and results for the local Universe}}}.
\newblock {\emph{\JournalTitle{\mnras}}} \textbf{\bibinfo{volume}{459}},
  \bibinfo{pages}{2150--2187} (\bibinfo{year}{2016}).
\newblock \doiprefix 10.1093/mnras/stw756.
\newblock \eprint{1604.00008}.

\bibitem{Dekel1986}
\bibinfo{author}{{Dekel}, A.} \& \bibinfo{author}{{Silk}, J.}
\newblock \bibinfo{journal}{\bibinfo{title}{{The Origin of Dwarf Galaxies, Cold
  Dark Matter, and Biased Galaxy Formation}}}.
\newblock {\emph{\JournalTitle{\apj}}} \textbf{\bibinfo{volume}{303}},
  \bibinfo{pages}{39} (\bibinfo{year}{1986}).
\newblock \doiprefix 10.1086/164050.

\bibitem{Larson1974}
\bibinfo{author}{{Larson}, R.~B.}
\newblock \bibinfo{journal}{\bibinfo{title}{{Effects of supernovae on the early
  evolution of galaxies}}}.
\newblock {\emph{\JournalTitle{\mnras}}} \textbf{\bibinfo{volume}{169}},
  \bibinfo{pages}{229--246} (\bibinfo{year}{1974}).
\newblock \doiprefix 10.1093/mnras/169.2.229.

\bibitem{White1991}
\bibinfo{author}{{White}, S. D.~M.} \& \bibinfo{author}{{Frenk}, C.~S.}
\newblock \bibinfo{journal}{\bibinfo{title}{{Galaxy Formation through
  Hierarchical Clustering}}}.
\newblock {\emph{\JournalTitle{\apj}}} \textbf{\bibinfo{volume}{379}},
  \bibinfo{pages}{52} (\bibinfo{year}{1991}).
\newblock \doiprefix 10.1086/170483.

\bibitem{Pillepich2018b}
\bibinfo{author}{{Pillepich}, A.} \emph{et~al.}
\newblock \bibinfo{journal}{\bibinfo{title}{{First results from the
  IllustrisTNG simulations: the stellar mass content of groups and clusters of
  galaxies}}}.
\newblock {\emph{\JournalTitle{\mnras}}} \textbf{\bibinfo{volume}{475}},
  \bibinfo{pages}{648--675} (\bibinfo{year}{2018}).
\newblock \doiprefix 10.1093/mnras/stx3112.
\newblock \eprint{1707.03406}.

\bibitem{Werk2014}
\bibinfo{author}{{Werk}, J.~K.} \emph{et~al.}
\newblock \bibinfo{journal}{\bibinfo{title}{{The COS-Halos Survey: Physical
  Conditions and Baryonic Mass in the Low-redshift Circumgalactic Medium}}}.
\newblock {\emph{\JournalTitle{\apj}}} \textbf{\bibinfo{volume}{792}},
  \bibinfo{pages}{8} (\bibinfo{year}{2014}).
\newblock \doiprefix 10.1088/0004-637X/792/1/8.
\newblock \eprint{1403.0947}.

\bibitem{Werk2016}
\bibinfo{author}{{Werk}, J.~K.} \emph{et~al.}
\newblock \bibinfo{journal}{\bibinfo{title}{{The COS-Halos Survey: Origins of
  the Highly Ionized Circumgalactic Medium of Star-Forming Galaxies}}}.
\newblock {\emph{\JournalTitle{\apj}}} \textbf{\bibinfo{volume}{833}},
  \bibinfo{pages}{54} (\bibinfo{year}{2016}).
\newblock \doiprefix 10.3847/1538-4357/833/1/54.
\newblock \eprint{1609.00012}.

\bibitem{Stern2016}
\bibinfo{author}{{Stern}, J.}, \bibinfo{author}{{Hennawi}, J.~F.},
  \bibinfo{author}{{Prochaska}, J.~X.} \& \bibinfo{author}{{Werk}, J.~K.}
\newblock \bibinfo{journal}{\bibinfo{title}{{A Universal Density Structure for
  Circumgalactic Gas}}}.
\newblock {\emph{\JournalTitle{\apj}}} \textbf{\bibinfo{volume}{830}},
  \bibinfo{pages}{87} (\bibinfo{year}{2016}).
\newblock \doiprefix 10.3847/0004-637X/830/2/87.
\newblock \eprint{1604.02168}.

\bibitem{Rubin2018}
\bibinfo{author}{{Rubin}, K.~H.~R.}, \bibinfo{author}{{Diamond-Stanic}, A.~M.},
  \bibinfo{author}{{Coil}, A.~L.}, \bibinfo{author}{{Crighton}, N.~H.~M.} \&
  \bibinfo{author}{{Stewart}, K.~R.}
\newblock \bibinfo{journal}{\bibinfo{title}{{Galaxies Probing Galaxies in
  PRIMUS. II. The Coherence Scale of the Cool Circumgalactic Medium}}}.
\newblock {\emph{\JournalTitle{\apj}}} \textbf{\bibinfo{volume}{868}},
  \bibinfo{pages}{142} (\bibinfo{year}{2018}).
\newblock \doiprefix 10.3847/1538-4357/aad566.
\newblock \eprint{1806.08801}.

\bibitem{Peeples2018}
\bibinfo{author}{{Peeples}, M.~S.} \emph{et~al.}
\newblock \bibinfo{journal}{\bibinfo{title}{{Figuring Out Gas \&amp; Galaxies
  in Enzo (FOGGIE). I. Resolving Simulated Circumgalactic Absorption at 2
  {\ensuremath{\leq}} z {\ensuremath{\leq}} 2.5}}}.
\newblock {\emph{\JournalTitle{\apj}}} \textbf{\bibinfo{volume}{873}},
  \bibinfo{pages}{129} (\bibinfo{year}{2019}).
\newblock \doiprefix 10.3847/1538-4357/ab0654.
\newblock \eprint{1810.06566}.

\bibitem{vandeVoort2019}
\bibinfo{author}{{van de Voort}, F.}, \bibinfo{author}{{Springel}, V.},
  \bibinfo{author}{{Mandelker}, N.}, \bibinfo{author}{{van den Bosch}, F.~C.}
  \& \bibinfo{author}{{Pakmor}, R.}
\newblock \bibinfo{journal}{\bibinfo{title}{{Cosmological simulations of the
  circumgalactic medium with 1 kpc resolution: enhanced H I column
  densities}}}.
\newblock {\emph{\JournalTitle{\mnras}}} \textbf{\bibinfo{volume}{482}},
  \bibinfo{pages}{L85--L89} (\bibinfo{year}{2019}).
\newblock \doiprefix 10.1093/mnrasl/sly190.
\newblock \eprint{1808.04369}.

\bibitem{Suresh2019}
\bibinfo{author}{{Suresh}, J.}, \bibinfo{author}{{Nelson}, D.},
  \bibinfo{author}{{Genel}, S.}, \bibinfo{author}{{Rubin}, K.~H.~R.} \&
  \bibinfo{author}{{Hernquist}, L.}
\newblock \bibinfo{journal}{\bibinfo{title}{{Zooming in on accretion - II. Cold
  circumgalactic gas simulated with a super-Lagrangian refinement scheme}}}.
\newblock {\emph{\JournalTitle{\mnras}}} \textbf{\bibinfo{volume}{483}},
  \bibinfo{pages}{4040--4059} (\bibinfo{year}{2019}).
\newblock \doiprefix 10.1093/mnras/sty3402.
\newblock \eprint{1811.01949}.

\bibitem{Hummels2019}
\bibinfo{author}{{Hummels}, C.~B.} \emph{et~al.}
\newblock \bibinfo{journal}{\bibinfo{title}{{The Impact of Enhanced Halo
  Resolution on the Simulated Circumgalactic Medium}}}.
\newblock {\emph{\JournalTitle{\apj}}} \textbf{\bibinfo{volume}{882}},
  \bibinfo{pages}{156} (\bibinfo{year}{2019}).
\newblock \doiprefix 10.3847/1538-4357/ab378f.
\newblock \eprint{1811.12410}.

\bibitem{Rasia2015}
\bibinfo{author}{{Rasia}, E.} \emph{et~al.}
\newblock \bibinfo{journal}{\bibinfo{title}{{Cool Core Clusters from
  Cosmological Simulations}}}.
\newblock {\emph{\JournalTitle{\apjl}}} \textbf{\bibinfo{volume}{813}},
  \bibinfo{pages}{L17} (\bibinfo{year}{2015}).
\newblock \doiprefix 10.1088/2041-8205/813/1/L17.
\newblock \eprint{1509.04247}.

\bibitem{Planelles2014}
\bibinfo{author}{{Planelles}, S.} \emph{et~al.}
\newblock \bibinfo{journal}{\bibinfo{title}{{On the role of AGN feedback on the
  thermal and chemodynamical properties of the hot intracluster medium}}}.
\newblock {\emph{\JournalTitle{\mnras}}} \textbf{\bibinfo{volume}{438}},
  \bibinfo{pages}{195--216} (\bibinfo{year}{2014}).
\newblock \doiprefix 10.1093/mnras/stt2141.
\newblock \eprint{1311.0818}.

\bibitem{Biffi2016}
\bibinfo{author}{{Biffi}, V.} \emph{et~al.}
\newblock \bibinfo{journal}{\bibinfo{title}{{On the Nature of Hydrostatic
  Equilibrium in Galaxy Clusters}}}.
\newblock {\emph{\JournalTitle{\apj}}} \textbf{\bibinfo{volume}{827}},
  \bibinfo{pages}{112} (\bibinfo{year}{2016}).
\newblock \doiprefix 10.3847/0004-637X/827/2/112.
\newblock \eprint{1606.02293}.

\bibitem{Vogelsberger2018b}
\bibinfo{author}{{Vogelsberger}, M.} \emph{et~al.}
\newblock \bibinfo{journal}{\bibinfo{title}{{The uniformity and time-invariance
  of the intra-cluster metal distribution in galaxy clusters from the
  IllustrisTNG simulations}}}.
\newblock {\emph{\JournalTitle{\mnras}}} \textbf{\bibinfo{volume}{474}},
  \bibinfo{pages}{2073--2093} (\bibinfo{year}{2018}).
\newblock \doiprefix 10.1093/mnras/stx2955.
\newblock \eprint{1707.05318}.

\bibitem{Barnes2018b}
\bibinfo{author}{{Barnes}, D.~J.} \emph{et~al.}
\newblock \bibinfo{journal}{\bibinfo{title}{{A census of cool-core galaxy
  clusters in IllustrisTNG}}}.
\newblock {\emph{\JournalTitle{\mnras}}} \textbf{\bibinfo{volume}{481}},
  \bibinfo{pages}{1809--1831} (\bibinfo{year}{2018}).
\newblock \doiprefix 10.1093/mnras/sty2078.
\newblock \eprint{1710.08420}.

\bibitem{Meneux2008}
\bibinfo{author}{{Meneux}, B.} \emph{et~al.}
\newblock \bibinfo{journal}{\bibinfo{title}{{The VIMOS-VLT Deep Survey (VVDS).
  The dependence of clustering on galaxy stellar mass at z \~{} 1}}}.
\newblock {\emph{\JournalTitle{\aap}}} \textbf{\bibinfo{volume}{478}},
  \bibinfo{pages}{299--310} (\bibinfo{year}{2008}).
\newblock \doiprefix 10.1051/0004-6361:20078182.
\newblock \eprint{0706.4371}.

\bibitem{Foucaud2010}
\bibinfo{author}{{Foucaud}, S.} \emph{et~al.}
\newblock \bibinfo{journal}{\bibinfo{title}{{Clustering properties of galaxies
  selected in stellar mass: breaking down the link between luminous and dark
  matter in massive galaxies from z = 0 to z = 2}}}.
\newblock {\emph{\JournalTitle{\mnras}}} \textbf{\bibinfo{volume}{406}},
  \bibinfo{pages}{147--164} (\bibinfo{year}{2010}).
\newblock \doiprefix 10.1111/j.1365-2966.2010.16682.x.
\newblock \eprint{1003.2755}.

\bibitem{Wake2011}
\bibinfo{author}{{Wake}, D.~A.} \emph{et~al.}
\newblock \bibinfo{journal}{\bibinfo{title}{{Galaxy Clustering in the NEWFIRM
  Medium Band Survey: The Relationship Between Stellar Mass and Dark Matter
  Halo Mass at 1 $<$ z $<$ 2}}}.
\newblock {\emph{\JournalTitle{\apj}}} \textbf{\bibinfo{volume}{728}},
  \bibinfo{pages}{46} (\bibinfo{year}{2011}).
\newblock \doiprefix 10.1088/0004-637X/728/1/46.
\newblock \eprint{1012.1317}.

\bibitem{Artale2017}
\bibinfo{author}{{Artale}, M.~C.} \emph{et~al.}
\newblock \bibinfo{journal}{\bibinfo{title}{{Small-scale galaxy clustering in
  the eagle simulation}}}.
\newblock {\emph{\JournalTitle{\mnras}}} \textbf{\bibinfo{volume}{470}},
  \bibinfo{pages}{1771--1787} (\bibinfo{year}{2017}).
\newblock \doiprefix 10.1093/mnras/stx1263.
\newblock \eprint{1611.05064}.

\bibitem{Marulli2013}
\bibinfo{author}{{Marulli}, F.} \emph{et~al.}
\newblock \bibinfo{journal}{\bibinfo{title}{{The VIMOS Public Extragalactic
  Redshift Survey (VIPERS) . Luminosity and stellar mass dependence of galaxy
  clustering at 0.5 $<$ z $<$ 1.1}}}.
\newblock {\emph{\JournalTitle{\aap}}} \textbf{\bibinfo{volume}{557}},
  \bibinfo{pages}{A17} (\bibinfo{year}{2013}).
\newblock \doiprefix 10.1051/0004-6361/201321476.
\newblock \eprint{1303.2633}.

\bibitem{Shen2003}
\bibinfo{author}{{Shen}, S.} \emph{et~al.}
\newblock \bibinfo{journal}{\bibinfo{title}{{The size distribution of galaxies
  in the Sloan Digital Sky Survey}}}.
\newblock {\emph{\JournalTitle{\mnras}}} \textbf{\bibinfo{volume}{343}},
  \bibinfo{pages}{978--994} (\bibinfo{year}{2003}).
\newblock \doiprefix 10.1046/j.1365-8711.2003.06740.x.
\newblock \eprint{astro-ph/0301527}.

\bibitem{Kormendy2013}
\bibinfo{author}{{Kormendy}, J.} \& \bibinfo{author}{{Ho}, L.~C.}
\newblock \bibinfo{journal}{\bibinfo{title}{{Coevolution (Or Not) of
  Supermassive Black Holes and Host Galaxies}}}.
\newblock {\emph{\JournalTitle{\araa}}} \textbf{\bibinfo{volume}{51}},
  \bibinfo{pages}{511--653} (\bibinfo{year}{2013}).
\newblock \doiprefix 10.1146/annurev-astro-082708-101811.
\newblock \eprint{1304.7762}.

\bibitem{Tremonti2004}
\bibinfo{author}{{Tremonti}, C.~A.} \emph{et~al.}
\newblock \bibinfo{journal}{\bibinfo{title}{{The Origin of the Mass-Metallicity
  Relation: Insights from 53,000 Star-forming Galaxies in the Sloan Digital Sky
  Survey}}}.
\newblock {\emph{\JournalTitle{\apj}}} \textbf{\bibinfo{volume}{613}},
  \bibinfo{pages}{898--913} (\bibinfo{year}{2004}).
\newblock \doiprefix 10.1086/423264.
\newblock \eprint{astro-ph/0405537}.

\bibitem{Dave2017}
\bibinfo{author}{{Dav{\'e}}, R.}, \bibinfo{author}{{Rafieferantsoa}, M.~H.},
  \bibinfo{author}{{Thompson}, R.~J.} \& \bibinfo{author}{{Hopkins}, P.~F.}
\newblock \bibinfo{journal}{\bibinfo{title}{{MUFASA: Galaxy star formation,
  gas, and metal properties across cosmic time}}}.
\newblock {\emph{\JournalTitle{\mnras}}} \textbf{\bibinfo{volume}{467}},
  \bibinfo{pages}{115--132} (\bibinfo{year}{2017}).
\newblock \doiprefix 10.1093/mnras/stx108.
\newblock \eprint{1610.01626}.

\bibitem{DeRossi2017}
\bibinfo{author}{{De Rossi}, M.~E.}, \bibinfo{author}{{Bower}, R.~G.},
  \bibinfo{author}{{Font}, A.~S.}, \bibinfo{author}{{Schaye}, J.} \&
  \bibinfo{author}{{Theuns}, T.}
\newblock \bibinfo{journal}{\bibinfo{title}{{Galaxy metallicity scaling
  relations in the EAGLE simulations}}}.
\newblock {\emph{\JournalTitle{\mnras}}} \textbf{\bibinfo{volume}{472}},
  \bibinfo{pages}{3354--3377} (\bibinfo{year}{2017}).
\newblock \doiprefix 10.1093/mnras/stx2158.
\newblock \eprint{1704.00006}.

\bibitem{Torrey2017}
\bibinfo{author}{{Torrey}, P.} \emph{et~al.}
\newblock \bibinfo{journal}{\bibinfo{title}{{The evolution of the
  mass-metallicity relation and its scatter in IllustrisTNG}}}.
\newblock {\emph{\JournalTitle{\mnras}}} \textbf{\bibinfo{volume}{484}},
  \bibinfo{pages}{5587--5607} (\bibinfo{year}{2019}).
\newblock \doiprefix 10.1093/mnras/stz243.
\newblock \eprint{1711.05261}.

\bibitem{Torrey2018}
\bibinfo{author}{{Torrey}, P.} \emph{et~al.}
\newblock \bibinfo{journal}{\bibinfo{title}{{Similar star formation rate and
  metallicity variability time-scales drive the fundamental metallicity
  relation}}}.
\newblock {\emph{\JournalTitle{\mnras}}} \textbf{\bibinfo{volume}{477}},
  \bibinfo{pages}{L16--L20} (\bibinfo{year}{2018}).
\newblock \doiprefix 10.1093/mnrasl/sly031.
\newblock \eprint{1711.11039}.

\bibitem{Trayford2015}
\bibinfo{author}{{Trayford}, J.~W.} \emph{et~al.}
\newblock \bibinfo{journal}{\bibinfo{title}{{Colours and luminosities of z =
  0.1 galaxies in the EAGLE simulation}}}.
\newblock {\emph{\JournalTitle{\mnras}}} \textbf{\bibinfo{volume}{452}},
  \bibinfo{pages}{2879--2896} (\bibinfo{year}{2015}).
\newblock \doiprefix 10.1093/mnras/stv1461.
\newblock \eprint{1504.04374}.

\bibitem{Nelson2018}
\bibinfo{author}{{Nelson}, D.} \emph{et~al.}
\newblock \bibinfo{journal}{\bibinfo{title}{{First results from the
  IllustrisTNG simulations: the galaxy colour bimodality}}}.
\newblock {\emph{\JournalTitle{\mnras}}} \textbf{\bibinfo{volume}{475}},
  \bibinfo{pages}{624--647} (\bibinfo{year}{2018}).
\newblock \doiprefix 10.1093/mnras/stx3040.
\newblock \eprint{1707.03395}.

\bibitem{Kauffmann1993}
\bibinfo{author}{{Kauffmann}, G.}, \bibinfo{author}{{White}, S.~D.~M.} \&
  \bibinfo{author}{{Guiderdoni}, B.}
\newblock \bibinfo{journal}{\bibinfo{title}{{The Formation and Evolution of
  Galaxies Within Merging Dark Matter Haloes}}}.
\newblock {\emph{\JournalTitle{\mnras}}} \textbf{\bibinfo{volume}{264}},
  \bibinfo{pages}{201} (\bibinfo{year}{1993}).
\newblock \doiprefix 10.1093/mnras/264.1.201.

\bibitem{Somerville1999}
\bibinfo{author}{{Somerville}, R.~S.} \& \bibinfo{author}{{Primack}, J.~R.}
\newblock \bibinfo{journal}{\bibinfo{title}{{Semi-analytic modelling of galaxy
  formation: the local Universe}}}.
\newblock {\emph{\JournalTitle{\mnras}}} \textbf{\bibinfo{volume}{310}},
  \bibinfo{pages}{1087--1110} (\bibinfo{year}{1999}).
\newblock \doiprefix 10.1046/j.1365-8711.1999.03032.x.
\newblock \eprint{astro-ph/9802268}.

\bibitem{Bower2006}
\bibinfo{author}{{Bower}, R.~G.} \emph{et~al.}
\newblock \bibinfo{journal}{\bibinfo{title}{{Breaking the hierarchy of galaxy
  formation}}}.
\newblock {\emph{\JournalTitle{\mnras}}} \textbf{\bibinfo{volume}{370}},
  \bibinfo{pages}{645--655} (\bibinfo{year}{2006}).
\newblock \doiprefix 10.1111/j.1365-2966.2006.10519.x.
\newblock \eprint{astro-ph/0511338}.

\bibitem{Croton2006}
\bibinfo{author}{{Croton}, D.~J.} \emph{et~al.}
\newblock \bibinfo{journal}{\bibinfo{title}{{The many lives of active galactic
  nuclei: cooling flows, black holes and the luminosities and colours of
  galaxies}}}.
\newblock {\emph{\JournalTitle{\mnras}}} \textbf{\bibinfo{volume}{365}},
  \bibinfo{pages}{11--28} (\bibinfo{year}{2006}).
\newblock \doiprefix 10.1111/j.1365-2966.2005.09675.x.
\newblock \eprint{astro-ph/0508046}.

\bibitem{Guo2011}
\bibinfo{author}{{Guo}, Q.} \emph{et~al.}
\newblock \bibinfo{journal}{\bibinfo{title}{{From dwarf spheroidals to cD
  galaxies: simulating the galaxy population in a {$\Lambda$}CDM cosmology}}}.
\newblock {\emph{\JournalTitle{\mnras}}} \textbf{\bibinfo{volume}{413}},
  \bibinfo{pages}{101--131} (\bibinfo{year}{2011}).
\newblock \doiprefix 10.1111/j.1365-2966.2010.18114.x.
\newblock \eprint{1006.0106}.

\bibitem{Fall1980}
\bibinfo{author}{{Fall}, S.~M.} \& \bibinfo{author}{{Efstathiou}, G.}
\newblock \bibinfo{journal}{\bibinfo{title}{{Formation and rotation of disc
  galaxies with haloes}}}.
\newblock {\emph{\JournalTitle{\mnras}}} \textbf{\bibinfo{volume}{193}},
  \bibinfo{pages}{189--206} (\bibinfo{year}{1980}).

\bibitem{Mo1998}
\bibinfo{author}{{Mo}, H.~J.}, \bibinfo{author}{{Mao}, S.} \&
  \bibinfo{author}{{White}, S.~D.~M.}
\newblock \bibinfo{journal}{\bibinfo{title}{{The formation of galactic
  discs}}}.
\newblock {\emph{\JournalTitle{\mnras}}} \textbf{\bibinfo{volume}{295}},
  \bibinfo{pages}{319--336} (\bibinfo{year}{1998}).
\newblock \doiprefix 10.1046/j.1365-8711.1998.01227.x.
\newblock \eprint{astro-ph/9707093}.

\bibitem{Scannapieco2009}
\bibinfo{author}{{Scannapieco}, C.}, \bibinfo{author}{{White}, S.~D.~M.},
  \bibinfo{author}{{Springel}, V.} \& \bibinfo{author}{{Tissera}, P.~B.}
\newblock \bibinfo{journal}{\bibinfo{title}{{The formation and survival of
  discs in a {$\Lambda$}CDM universe}}}.
\newblock {\emph{\JournalTitle{\mnras}}} \textbf{\bibinfo{volume}{396}},
  \bibinfo{pages}{696--708} (\bibinfo{year}{2009}).
\newblock \doiprefix 10.1111/j.1365-2966.2009.14764.x.
\newblock \eprint{0812.0976}.

\bibitem{Balogh2001}
\bibinfo{author}{{Balogh}, M.~L.}, \bibinfo{author}{{Pearce}, F.~R.},
  \bibinfo{author}{{Bower}, R.~G.} \& \bibinfo{author}{{Kay}, S.~T.}
\newblock \bibinfo{journal}{\bibinfo{title}{{Revisiting the cosmic cooling
  crisis}}}.
\newblock {\emph{\JournalTitle{\mnras}}} \textbf{\bibinfo{volume}{326}},
  \bibinfo{pages}{1228--1234} (\bibinfo{year}{2001}).
\newblock \doiprefix 10.1111/j.1365-8711.2001.04667.x.
\newblock \eprint{arXiv:astro-ph/0104041}.

\bibitem{Brook2011}
\bibinfo{author}{{Brook}, C.~B.} \emph{et~al.}
\newblock \bibinfo{journal}{\bibinfo{title}{{Hierarchical formation of
  bulgeless galaxies: why outflows have low angular momentum}}}.
\newblock {\emph{\JournalTitle{\mnras}}} \textbf{\bibinfo{volume}{415}},
  \bibinfo{pages}{1051--1060} (\bibinfo{year}{2011}).
\newblock \doiprefix 10.1111/j.1365-2966.2011.18545.x.
\newblock \eprint{1010.1004}.

\bibitem{Okamoto2005}
\bibinfo{author}{{Okamoto}, T.}, \bibinfo{author}{{Eke}, V.~R.},
  \bibinfo{author}{{Frenk}, C.~S.} \& \bibinfo{author}{{Jenkins}, A.}
\newblock \bibinfo{journal}{\bibinfo{title}{{Effects of feedback on the
  morphology of galaxy discs}}}.
\newblock {\emph{\JournalTitle{\mnras}}} \textbf{\bibinfo{volume}{363}},
  \bibinfo{pages}{1299--1314} (\bibinfo{year}{2005}).
\newblock \doiprefix 10.1111/j.1365-2966.2005.09525.x.
\newblock \eprint{astro-ph/0503676}.

\bibitem{Brooks2011}
\bibinfo{author}{{Brooks}, A.~M.} \emph{et~al.}
\newblock \bibinfo{journal}{\bibinfo{title}{{Interpreting the Evolution of the
  Size-Luminosity Relation for Disk Galaxies from Redshift 1 to the Present}}}.
\newblock {\emph{\JournalTitle{\apj}}} \textbf{\bibinfo{volume}{728}},
  \bibinfo{pages}{51} (\bibinfo{year}{2011}).
\newblock \doiprefix 10.1088/0004-637X/728/1/51.
\newblock \eprint{1011.0432}.

\bibitem{Guedes2011}
\bibinfo{author}{{Guedes}, J.}, \bibinfo{author}{{Callegari}, S.},
  \bibinfo{author}{{Madau}, P.} \& \bibinfo{author}{{Mayer}, L.}
\newblock \bibinfo{journal}{\bibinfo{title}{{Forming Realistic Late-type
  Spirals in a {$\Lambda$}CDM Universe: The Eris Simulation}}}.
\newblock {\emph{\JournalTitle{\apj}}} \textbf{\bibinfo{volume}{742}},
  \bibinfo{pages}{76} (\bibinfo{year}{2011}).
\newblock \doiprefix 10.1088/0004-637X/742/2/76.
\newblock \eprint{1103.6030}.

\bibitem{Aumer2013b}
\bibinfo{author}{{Aumer}, M.}, \bibinfo{author}{{White}, S.~D.~M.},
  \bibinfo{author}{{Naab}, T.} \& \bibinfo{author}{{Scannapieco}, C.}
\newblock \bibinfo{journal}{\bibinfo{title}{{Towards a more realistic
  population of bright spiral galaxies in cosmological simulations}}}.
\newblock {\emph{\JournalTitle{\mnras}}} \textbf{\bibinfo{volume}{434}},
  \bibinfo{pages}{3142--3164} (\bibinfo{year}{2013}).
\newblock \doiprefix 10.1093/mnras/stt1230.
\newblock \eprint{1304.1559}.

\bibitem{Marinacci2014}
\bibinfo{author}{{Marinacci}, F.}, \bibinfo{author}{{Pakmor}, R.} \&
  \bibinfo{author}{{Springel}, V.}
\newblock \bibinfo{journal}{\bibinfo{title}{{The formation of disc galaxies in
  high-resolution moving-mesh cosmological simulations}}}.
\newblock {\emph{\JournalTitle{\mnras}}} \textbf{\bibinfo{volume}{437}},
  \bibinfo{pages}{1750--1775} (\bibinfo{year}{2014}).
\newblock \doiprefix 10.1093/mnras/stt2003.
\newblock \eprint{1305.5360}.

\bibitem{Wetzel2016}
\bibinfo{author}{{Wetzel}, A.~R.} \emph{et~al.}
\newblock \bibinfo{journal}{\bibinfo{title}{{Reconciling Dwarf Galaxies with
  {$\Lambda$}CDM Cosmology: Simulating a Realistic Population of Satellites
  around a Milky Way-mass Galaxy}}}.
\newblock {\emph{\JournalTitle{\apjl}}} \textbf{\bibinfo{volume}{827}},
  \bibinfo{pages}{L23} (\bibinfo{year}{2016}).
\newblock \doiprefix 10.3847/2041-8205/827/2/L23.
\newblock \eprint{1602.05957}.

\bibitem{Pakmor2014}
\bibinfo{author}{{Pakmor}, R.}, \bibinfo{author}{{Marinacci}, F.} \&
  \bibinfo{author}{{Springel}, V.}
\newblock \bibinfo{journal}{\bibinfo{title}{{Magnetic Fields in Cosmological
  Simulations of Disk Galaxies}}}.
\newblock {\emph{\JournalTitle{\apj}}} \textbf{\bibinfo{volume}{783}},
  \bibinfo{pages}{L20} (\bibinfo{year}{2014}).
\newblock \doiprefix 10.1088/2041-8205/783/1/L20.
\newblock \eprint{1312.2620}.

\bibitem{Beck2015}
\bibinfo{author}{{Beck}, R.}
\newblock \bibinfo{journal}{\bibinfo{title}{{Magnetic fields in spiral
  galaxies}}}.
\newblock {\emph{\JournalTitle{Astronomy \& Astropysics Review}}}
  \textbf{\bibinfo{volume}{24}}, \bibinfo{pages}{4} (\bibinfo{year}{2015}).
\newblock \doiprefix 10.1007/s00159-015-0084-4.
\newblock \eprint{1509.04522}.

\bibitem{Rieder2016}
\bibinfo{author}{{Rieder}, M.} \& \bibinfo{author}{{Teyssier}, R.}
\newblock \bibinfo{journal}{\bibinfo{title}{{A small-scale dynamo in
  feedback-dominated galaxies as the origin of cosmic magnetic fields - I. The
  kinematic phase}}}.
\newblock {\emph{\JournalTitle{\mnras}}} \textbf{\bibinfo{volume}{457}},
  \bibinfo{pages}{1722--1738} (\bibinfo{year}{2016}).
\newblock \doiprefix 10.1093/mnras/stv2985.
\newblock \eprint{1506.00849}.

\bibitem{Rieder2017}
\bibinfo{author}{{Rieder}, M.} \& \bibinfo{author}{{Teyssier}, R.}
\newblock \bibinfo{journal}{\bibinfo{title}{{A small-scale dynamo in
  feedback-dominated galaxies - II. The saturation phase and the final magnetic
  configuration}}}.
\newblock {\emph{\JournalTitle{\mnras}}} \textbf{\bibinfo{volume}{471}},
  \bibinfo{pages}{2674--2686} (\bibinfo{year}{2017}).
\newblock \doiprefix 10.1093/mnras/stx1670.
\newblock \eprint{1704.05845}.

\bibitem{Rieder2017b}
\bibinfo{author}{{Rieder}, M.} \& \bibinfo{author}{{Teyssier}, R.}
\newblock \bibinfo{journal}{\bibinfo{title}{{A small-scale dynamo in
  feedback-dominated galaxies - III. Cosmological simulations}}}.
\newblock {\emph{\JournalTitle{\mnras}}} \textbf{\bibinfo{volume}{472}},
  \bibinfo{pages}{4368--4373} (\bibinfo{year}{2017}).
\newblock \doiprefix 10.1093/mnras/stx2276.
\newblock \eprint{1708.01486}.

\bibitem{Naab2014}
\bibinfo{author}{{Naab}, T.} \emph{et~al.}
\newblock \bibinfo{journal}{\bibinfo{title}{{The ATLAS$^{3D}$ project - XXV.
  Two-dimensional kinematic analysis of simulated galaxies and the cosmological
  origin of fast and slow rotators}}}.
\newblock {\emph{\JournalTitle{\mnras}}} \textbf{\bibinfo{volume}{444}},
  \bibinfo{pages}{3357--3387} (\bibinfo{year}{2014}).
\newblock \doiprefix 10.1093/mnras/stt1919.
\newblock \eprint{1311.0284}.

\bibitem{Kobayashi2005}
\bibinfo{author}{{Kobayashi}, C.}
\newblock \bibinfo{journal}{\bibinfo{title}{{GRAPE-SPH chemodynamical
  simulation of elliptical galaxies - II. Scaling relations and the fundamental
  plane}}}.
\newblock {\emph{\JournalTitle{\mnras}}} \textbf{\bibinfo{volume}{361}},
  \bibinfo{pages}{1216--1226} (\bibinfo{year}{2005}).
\newblock \doiprefix 10.1111/j.1365-2966.2005.09248.x.
\newblock \eprint{astro-ph/0506094}.

\bibitem{Feldmann2011}
\bibinfo{author}{{Feldmann}, R.}, \bibinfo{author}{{Carollo}, C.~M.} \&
  \bibinfo{author}{{Mayer}, L.}
\newblock \bibinfo{journal}{\bibinfo{title}{{The Hubble Sequence in Groups: The
  Birth of the Early-type Galaxies}}}.
\newblock {\emph{\JournalTitle{\apj}}} \textbf{\bibinfo{volume}{736}},
  \bibinfo{pages}{88} (\bibinfo{year}{2011}).
\newblock \doiprefix 10.1088/0004-637X/736/2/88.
\newblock \eprint{1008.3386}.

\bibitem{Kobayashi2004}
\bibinfo{author}{{Kobayashi}, C.}
\newblock \bibinfo{journal}{\bibinfo{title}{{GRAPE-SPH chemodynamical
  simulation of elliptical galaxies - I. Evolution of metallicity gradients}}}.
\newblock {\emph{\JournalTitle{\mnras}}} \textbf{\bibinfo{volume}{347}},
  \bibinfo{pages}{740--758} (\bibinfo{year}{2004}).
\newblock \doiprefix 10.1111/j.1365-2966.2004.07258.x.
\newblock \eprint{astro-ph/0310160}.

\bibitem{Oser2010}
\bibinfo{author}{{Oser}, L.}, \bibinfo{author}{{Ostriker}, J.~P.},
  \bibinfo{author}{{Naab}, T.}, \bibinfo{author}{{Johansson}, P.~H.} \&
  \bibinfo{author}{{Burkert}, A.}
\newblock \bibinfo{journal}{\bibinfo{title}{{The Two Phases of Galaxy
  Formation}}}.
\newblock {\emph{\JournalTitle{\apj}}} \textbf{\bibinfo{volume}{725}},
  \bibinfo{pages}{2312--2323} (\bibinfo{year}{2010}).
\newblock \doiprefix 10.1088/0004-637X/725/2/2312.
\newblock \eprint{1010.1381}.

\bibitem{Huang2013}
\bibinfo{author}{{Huang}, S.}, \bibinfo{author}{{Ho}, L.~C.},
  \bibinfo{author}{{Peng}, C.~Y.}, \bibinfo{author}{{Li}, Z.-Y.} \&
  \bibinfo{author}{{Barth}, A.~J.}
\newblock \bibinfo{journal}{\bibinfo{title}{{Fossil Evidence for the Two-phase
  Formation of Elliptical Galaxies}}}.
\newblock {\emph{\JournalTitle{\apjl}}} \textbf{\bibinfo{volume}{768}},
  \bibinfo{pages}{L28} (\bibinfo{year}{2013}).
\newblock \doiprefix 10.1088/2041-8205/768/2/L28.
\newblock \eprint{1304.2299}.

\bibitem{Rodriguez-Gomez2016}
\bibinfo{author}{{Rodriguez-Gomez}, V.} \emph{et~al.}
\newblock \bibinfo{journal}{\bibinfo{title}{{The stellar mass assembly of
  galaxies in the Illustris simulation: growth by mergers and the spatial
  distribution of accreted stars}}}.
\newblock {\emph{\JournalTitle{\mnras}}} \textbf{\bibinfo{volume}{458}},
  \bibinfo{pages}{2371--2390} (\bibinfo{year}{2016}).
\newblock \doiprefix 10.1093/mnras/stw456.
\newblock \eprint{1511.08804}.

\bibitem{Clauwens2018}
\bibinfo{author}{{Clauwens}, B.}, \bibinfo{author}{{Schaye}, J.},
  \bibinfo{author}{{Franx}, M.} \& \bibinfo{author}{{Bower}, R.~G.}
\newblock \bibinfo{journal}{\bibinfo{title}{{The three phases of galaxy
  formation}}}.
\newblock {\emph{\JournalTitle{\mnras}}} \textbf{\bibinfo{volume}{478}},
  \bibinfo{pages}{3994--4009} (\bibinfo{year}{2018}).
\newblock \doiprefix 10.1093/mnras/sty1229.
\newblock \eprint{1711.00030}.

\bibitem{Lagos2018}
\bibinfo{author}{{Lagos}, C.~d.~P.} \emph{et~al.}
\newblock \bibinfo{journal}{\bibinfo{title}{{The connection between mass,
  environment, and slow rotation in simulated galaxies}}}.
\newblock {\emph{\JournalTitle{\mnras}}} \textbf{\bibinfo{volume}{476}},
  \bibinfo{pages}{4327--4345} (\bibinfo{year}{2018}).
\newblock \doiprefix 10.1093/mnras/sty489.
\newblock \eprint{1712.01398}.

\bibitem{Schulze2018}
\bibinfo{author}{{Schulze}, F.} \emph{et~al.}
\newblock \bibinfo{journal}{\bibinfo{title}{{Kinematics of simulated galaxies -
  I. Connecting dynamical and morphological properties of early-type galaxies
  at different redshifts}}}.
\newblock {\emph{\JournalTitle{\mnras}}} \textbf{\bibinfo{volume}{480}},
  \bibinfo{pages}{4636--4658} (\bibinfo{year}{2018}).
\newblock \doiprefix 10.1093/mnras/sty2090.
\newblock \eprint{1802.01583}.

\bibitem{Millennium}
\bibinfo{author}{{Springel}, V.} \emph{et~al.}
\newblock \bibinfo{journal}{\bibinfo{title}{{Simulations of the formation,
  evolution and clustering of galaxies and quasars}}}.
\newblock {\emph{\JournalTitle{\nat}}} \textbf{\bibinfo{volume}{435}},
  \bibinfo{pages}{629--636} (\bibinfo{year}{2005b}).
\newblock \doiprefix 10.1038/nature03597.
\newblock \eprint{arXiv:astro-ph/0504097}.

\bibitem{MilleniumII}
\bibinfo{author}{{Boylan-Kolchin}, M.}, \bibinfo{author}{{Springel}, V.},
  \bibinfo{author}{{White}, S.~D.~M.}, \bibinfo{author}{{Jenkins}, A.} \&
  \bibinfo{author}{{Lemson}, G.}
\newblock \bibinfo{journal}{\bibinfo{title}{{Resolving cosmic structure
  formation with the Millennium-II Simulation}}}.
\newblock {\emph{\JournalTitle{\mnras}}} \textbf{\bibinfo{volume}{398}},
  \bibinfo{pages}{1150--1164} (\bibinfo{year}{2009}).
\newblock \doiprefix 10.1111/j.1365-2966.2009.15191.x.
\newblock \eprint{0903.3041}.

\bibitem{Teyssier2009}
\bibinfo{author}{{Teyssier}, R.} \emph{et~al.}
\newblock \bibinfo{journal}{\bibinfo{title}{{Full-sky weak-lensing simulation
  with 70 billion particles}}}.
\newblock {\emph{\JournalTitle{\aap}}} \textbf{\bibinfo{volume}{497}},
  \bibinfo{pages}{335--341} (\bibinfo{year}{2009}).
\newblock \doiprefix 10.1051/0004-6361/200810657.
\newblock \eprint{0807.3651}.

\bibitem{Klypin2011}
\bibinfo{author}{{Klypin}, A.~A.}, \bibinfo{author}{{Trujillo-Gomez}, S.} \&
  \bibinfo{author}{{Primack}, J.}
\newblock \bibinfo{journal}{\bibinfo{title}{{Dark Matter Halos in the Standard
  Cosmological Model: Results from the Bolshoi Simulation}}}.
\newblock {\emph{\JournalTitle{\apj}}} \textbf{\bibinfo{volume}{740}},
  \bibinfo{pages}{102} (\bibinfo{year}{2011}).
\newblock \doiprefix 10.1088/0004-637X/740/2/102.
\newblock \eprint{1002.3660}.

\bibitem{Alimi2012}
\bibinfo{author}{{Alimi}, J.-M.} \emph{et~al.}
\newblock \bibinfo{journal}{\bibinfo{title}{{DEUS Full Observable LCDM Universe
  Simulation: the numerical challenge}}}.
\newblock {\emph{\JournalTitle{arXiv e-prints}}}
  \bibinfo{pages}{arXiv:1206.2838} (\bibinfo{year}{2012}).
\newblock \eprint{1206.2838}.

\bibitem{Prada2012}
\bibinfo{author}{{Prada}, F.}, \bibinfo{author}{{Klypin}, A.~A.},
  \bibinfo{author}{{Cuesta}, A.~J.}, \bibinfo{author}{{Betancort-Rijo}, J.~E.}
  \& \bibinfo{author}{{Primack}, J.}
\newblock \bibinfo{journal}{\bibinfo{title}{{Halo concentrations in the
  standard {$\Lambda$} cold dark matter cosmology}}}.
\newblock {\emph{\JournalTitle{\mnras}}} \textbf{\bibinfo{volume}{423}},
  \bibinfo{pages}{3018--3030} (\bibinfo{year}{2012}).
\newblock \doiprefix 10.1111/j.1365-2966.2012.21007.x.
\newblock \eprint{1104.5130}.

\bibitem{Ishiyama2015}
\bibinfo{author}{{Ishiyama}, T.} \emph{et~al.}
\newblock \bibinfo{journal}{\bibinfo{title}{{The {$\nu$}$^{2}$GC simulations:
  Quantifying the dark side of the universe in the Planck cosmology}}}.
\newblock {\emph{\JournalTitle{\pasj}}} \textbf{\bibinfo{volume}{67}},
  \bibinfo{pages}{61} (\bibinfo{year}{2015}).
\newblock \doiprefix 10.1093/pasj/psv021.
\newblock \eprint{1412.2860}.

\bibitem{Diemand2008}
\bibinfo{author}{{Diemand}, J.} \emph{et~al.}
\newblock \bibinfo{journal}{\bibinfo{title}{{Clumps and streams in the local
  dark matter distribution}}}.
\newblock {\emph{\JournalTitle{\nat}}} \textbf{\bibinfo{volume}{454}},
  \bibinfo{pages}{735--738} (\bibinfo{year}{2008}).
\newblock \doiprefix 10.1038/nature07153.
\newblock \eprint{0805.1244}.

\bibitem{Stadel2009}
\bibinfo{author}{{Stadel}, J.} \emph{et~al.}
\newblock \bibinfo{journal}{\bibinfo{title}{{Quantifying the heart of darkness
  with GHALO - a multibillion particle simulation of a galactic halo}}}.
\newblock {\emph{\JournalTitle{\mnras}}} \textbf{\bibinfo{volume}{398}},
  \bibinfo{pages}{L21--L25} (\bibinfo{year}{2009}).
\newblock \doiprefix 10.1111/j.1745-3933.2009.00699.x.
\newblock \eprint{0808.2981}.

\bibitem{Libeskind2010}
\bibinfo{author}{{Libeskind}, N.~I.} \emph{et~al.}
\newblock \bibinfo{journal}{\bibinfo{title}{{Constrained simulations of the
  Local Group: on the radial distribution of substructures}}}.
\newblock {\emph{\JournalTitle{\mnras}}} \textbf{\bibinfo{volume}{401}},
  \bibinfo{pages}{1889--1897} (\bibinfo{year}{2010}).
\newblock \doiprefix 10.1111/j.1365-2966.2009.15766.x.
\newblock \eprint{0909.4423}.

\bibitem{Garrison-Kimmel2014b}
\bibinfo{author}{{Garrison-Kimmel}, S.}, \bibinfo{author}{{Boylan-Kolchin},
  M.}, \bibinfo{author}{{Bullock}, J.~S.} \& \bibinfo{author}{{Lee}, K.}
\newblock \bibinfo{journal}{\bibinfo{title}{{ELVIS: Exploring the Local Volume
  in Simulations}}}.
\newblock {\emph{\JournalTitle{\mnras}}} \textbf{\bibinfo{volume}{438}},
  \bibinfo{pages}{2578--2596} (\bibinfo{year}{2014}).
\newblock \doiprefix 10.1093/mnras/stt2377.
\newblock \eprint{1310.6746}.

\bibitem{Hellwing2016}
\bibinfo{author}{{Hellwing}, W.~A.} \emph{et~al.}
\newblock \bibinfo{journal}{\bibinfo{title}{{The Copernicus Complexio: a
  high-resolution view of the small-scale Universe}}}.
\newblock {\emph{\JournalTitle{\mnras}}} \textbf{\bibinfo{volume}{457}},
  \bibinfo{pages}{3492--3509} (\bibinfo{year}{2016}).
\newblock \doiprefix 10.1093/mnras/stw214.
\newblock \eprint{1505.06436}.

\bibitem{Dubois2014}
\bibinfo{author}{{Dubois}, Y.} \emph{et~al.}
\newblock \bibinfo{journal}{\bibinfo{title}{{Dancing in the dark: galactic
  properties trace spin swings along the cosmic web}}}.
\newblock {\emph{\JournalTitle{\mnras}}} \textbf{\bibinfo{volume}{444}},
  \bibinfo{pages}{1453--1468} (\bibinfo{year}{2014}).
\newblock \doiprefix 10.1093/mnras/stu1227.
\newblock \eprint{1402.1165}.

\bibitem{Khandai2015}
\bibinfo{author}{{Khandai}, N.} \emph{et~al.}
\newblock \bibinfo{journal}{\bibinfo{title}{{The MassiveBlack-II simulation:
  the evolution of haloes and galaxies to z $\sim$ 0}}}.
\newblock {\emph{\JournalTitle{\mnras}}} \textbf{\bibinfo{volume}{450}},
  \bibinfo{pages}{1349--1374} (\bibinfo{year}{2015}).
\newblock \doiprefix 10.1093/mnras/stv627.
\newblock \eprint{1402.0888}.

\bibitem{Feng2016}
\bibinfo{author}{{Feng}, Y.} \emph{et~al.}
\newblock \bibinfo{journal}{\bibinfo{title}{{The BlueTides simulation: first
  galaxies and reionization}}}.
\newblock {\emph{\JournalTitle{\mnras}}} \textbf{\bibinfo{volume}{455}},
  \bibinfo{pages}{2778--2791} (\bibinfo{year}{2016}).
\newblock \doiprefix 10.1093/mnras/stv2484.
\newblock \eprint{1504.06619}.

\bibitem{Dave2016}
\bibinfo{author}{{Dav{\'e}}, R.}, \bibinfo{author}{{Thompson}, R.} \&
  \bibinfo{author}{{Hopkins}, P.~F.}
\newblock \bibinfo{journal}{\bibinfo{title}{{MUFASA: galaxy formation
  simulations with meshless hydrodynamics}}}.
\newblock {\emph{\JournalTitle{\mnras}}} \textbf{\bibinfo{volume}{462}},
  \bibinfo{pages}{3265--3284} (\bibinfo{year}{2016}).
\newblock \doiprefix 10.1093/mnras/stw1862.
\newblock \eprint{1604.01418}.

\bibitem{McCarthy2017}
\bibinfo{author}{{McCarthy}, I.~G.}, \bibinfo{author}{{Schaye}, J.},
  \bibinfo{author}{{Bird}, S.} \& \bibinfo{author}{{Le Brun}, A.~M.~C.}
\newblock \bibinfo{journal}{\bibinfo{title}{{The BAHAMAS project: calibrated
  hydrodynamical simulations for large-scale structure cosmology}}}.
\newblock {\emph{\JournalTitle{\mnras}}} \textbf{\bibinfo{volume}{465}},
  \bibinfo{pages}{2936--2965} (\bibinfo{year}{2017}).
\newblock \doiprefix 10.1093/mnras/stw2792.
\newblock \eprint{1603.02702}.

\bibitem{Tremmel2017}
\bibinfo{author}{{Tremmel}, M.} \emph{et~al.}
\newblock \bibinfo{journal}{\bibinfo{title}{{The Romulus cosmological
  simulations: a physical approach to the formation, dynamics and accretion
  models of SMBHs}}}.
\newblock {\emph{\JournalTitle{\mnras}}} \textbf{\bibinfo{volume}{470}},
  \bibinfo{pages}{1121--1139} (\bibinfo{year}{2017}).
\newblock \doiprefix 10.1093/mnras/stx1160.
\newblock \eprint{1607.02151}.

\bibitem{Ceverino2014}
\bibinfo{author}{{Ceverino}, D.} \emph{et~al.}
\newblock \bibinfo{journal}{\bibinfo{title}{{Radiative feedback and the low
  efficiency of galaxy formation in low-mass haloes at high redshift}}}.
\newblock {\emph{\JournalTitle{\mnras}}} \textbf{\bibinfo{volume}{442}},
  \bibinfo{pages}{1545--1559} (\bibinfo{year}{2014}).
\newblock \doiprefix 10.1093/mnras/stu956.
\newblock \eprint{1307.0943}.

\bibitem{Sawala2016}
\bibinfo{author}{{Sawala}, T.} \emph{et~al.}
\newblock \bibinfo{journal}{\bibinfo{title}{{The APOSTLE simulations: solutions
  to the Local Group's cosmic puzzles}}}.
\newblock {\emph{\JournalTitle{\mnras}}} \textbf{\bibinfo{volume}{457}},
  \bibinfo{pages}{1931--1943} (\bibinfo{year}{2016}).
\newblock \doiprefix 10.1093/mnras/stw145.
\newblock \eprint{1511.01098}.

\bibitem{Barnes2017b}
\bibinfo{author}{{Barnes}, D.~J.} \emph{et~al.}
\newblock \bibinfo{journal}{\bibinfo{title}{{The redshift evolution of massive
  galaxy clusters in the MACSIS simulations}}}.
\newblock {\emph{\JournalTitle{\mnras}}} \textbf{\bibinfo{volume}{465}},
  \bibinfo{pages}{213--233} (\bibinfo{year}{2017}).
\newblock \doiprefix 10.1093/mnras/stw2722.
\newblock \eprint{1607.04569}.

\bibitem{Cui2018}
\bibinfo{author}{{Cui}, W.} \emph{et~al.}
\newblock \bibinfo{journal}{\bibinfo{title}{{The Three Hundred project: a large
  catalogue of theoretically modelled galaxy clusters for cosmological and
  astrophysical applications}}}.
\newblock {\emph{\JournalTitle{\mnras}}} \textbf{\bibinfo{volume}{480}},
  \bibinfo{pages}{2898--2915} (\bibinfo{year}{2018}).
\newblock \doiprefix 10.1093/mnras/sty2111.
\newblock \eprint{1809.04622}.

\bibitem{Henden2018}
\bibinfo{author}{{Henden}, N.~A.}, \bibinfo{author}{{Puchwein}, E.},
  \bibinfo{author}{{Shen}, S.} \& \bibinfo{author}{{Sijacki}, D.}
\newblock \bibinfo{journal}{\bibinfo{title}{{The FABLE simulations: a feedback
  model for galaxies, groups, and clusters}}}.
\newblock {\emph{\JournalTitle{\mnras}}} \textbf{\bibinfo{volume}{479}},
  \bibinfo{pages}{5385--5412} (\bibinfo{year}{2018}).
\newblock \doiprefix 10.1093/mnras/sty1780.
\newblock \eprint{1804.05064}.

\bibitem{Tremmel2019}
\bibinfo{author}{{Tremmel}, M.} \emph{et~al.}
\newblock \bibinfo{journal}{\bibinfo{title}{{Introducing ROMULUSC: a
  cosmological simulation of a galaxy cluster with an unprecedented
  resolution}}}.
\newblock {\emph{\JournalTitle{\mnras}}} \textbf{\bibinfo{volume}{483}},
  \bibinfo{pages}{3336--3362} (\bibinfo{year}{2019}).
\newblock \doiprefix 10.1093/mnras/sty3336.
\newblock \eprint{1806.01282}.

\bibitem{Bullock2017}
\bibinfo{author}{{Bullock}, J.~S.} \& \bibinfo{author}{{Boylan-Kolchin}, M.}
\newblock \bibinfo{journal}{\bibinfo{title}{{Small-Scale Challenges to the
  {$\Lambda$}CDM Paradigm}}}.
\newblock {\emph{\JournalTitle{\araa}}} \textbf{\bibinfo{volume}{55}},
  \bibinfo{pages}{343--387} (\bibinfo{year}{2017}).
\newblock \doiprefix 10.1146/annurev-astro-091916-055313.
\newblock \eprint{1707.04256}.

\bibitem{Klypin1999}
\bibinfo{author}{{Klypin}, A.}, \bibinfo{author}{{Kravtsov}, A.~V.},
  \bibinfo{author}{{Valenzuela}, O.} \& \bibinfo{author}{{Prada}, F.}
\newblock \bibinfo{journal}{\bibinfo{title}{{Where Are the Missing Galactic
  Satellites?}}}
\newblock {\emph{\JournalTitle{\apj}}} \textbf{\bibinfo{volume}{522}},
  \bibinfo{pages}{82--92} (\bibinfo{year}{1999}).
\newblock \doiprefix 10.1086/307643.
\newblock \eprint{astro-ph/9901240}.

\bibitem{Zavala2009}
\bibinfo{author}{{Zavala}, J.} \emph{et~al.}
\newblock \bibinfo{journal}{\bibinfo{title}{{The Velocity Function in the Local
  Environment from {$\Lambda$}CDM and {$\Lambda$}WDM Constrained
  Simulations}}}.
\newblock {\emph{\JournalTitle{\apj}}} \textbf{\bibinfo{volume}{700}},
  \bibinfo{pages}{1779--1793} (\bibinfo{year}{2009}).
\newblock \doiprefix 10.1088/0004-637X/700/2/1779.
\newblock \eprint{0906.0585}.

\bibitem{Papastergis2011}
\bibinfo{author}{{Papastergis}, E.}, \bibinfo{author}{{Martin}, A.~M.},
  \bibinfo{author}{{Giovanelli}, R.} \& \bibinfo{author}{{Haynes}, M.~P.}
\newblock \bibinfo{journal}{\bibinfo{title}{{The Velocity Width Function of
  Galaxies from the 40\% ALFALFA Survey: Shedding Light on the Cold Dark Matter
  Overabundance Problem}}}.
\newblock {\emph{\JournalTitle{\apj}}} \textbf{\bibinfo{volume}{739}},
  \bibinfo{pages}{38} (\bibinfo{year}{2011}).
\newblock \doiprefix 10.1088/0004-637X/739/1/38.
\newblock \eprint{1106.0710}.

\bibitem{Klypin2015}
\bibinfo{author}{{Klypin}, A.}, \bibinfo{author}{{Karachentsev}, I.},
  \bibinfo{author}{{Makarov}, D.} \& \bibinfo{author}{{Nasonova}, O.}
\newblock \bibinfo{journal}{\bibinfo{title}{{Abundance of field galaxies}}}.
\newblock {\emph{\JournalTitle{\mnras}}} \textbf{\bibinfo{volume}{454}},
  \bibinfo{pages}{1798--1810} (\bibinfo{year}{2015}).
\newblock \doiprefix 10.1093/mnras/stv2040.
\newblock \eprint{1405.4523}.

\bibitem{deBlok1997}
\bibinfo{author}{{de Blok}, W.~J.~G.} \& \bibinfo{author}{{McGaugh}, S.~S.}
\newblock \bibinfo{journal}{\bibinfo{title}{{The dark and visible matter
  content of low surface brightness disc galaxies}}}.
\newblock {\emph{\JournalTitle{\mnras}}} \textbf{\bibinfo{volume}{290}},
  \bibinfo{pages}{533--552} (\bibinfo{year}{1997}).
\newblock \doiprefix 10.1093/mnras/290.3.533.
\newblock \eprint{astro-ph/9704274}.

\bibitem{Walker2011}
\bibinfo{author}{{Walker}, M.~G.} \& \bibinfo{author}{{Pe{\~n}arrubia}, J.}
\newblock \bibinfo{journal}{\bibinfo{title}{{A Method for Measuring (Slopes of)
  the Mass Profiles of Dwarf Spheroidal Galaxies}}}.
\newblock {\emph{\JournalTitle{\apj}}} \textbf{\bibinfo{volume}{742}},
  \bibinfo{pages}{20} (\bibinfo{year}{2011}).
\newblock \doiprefix 10.1088/0004-637X/742/1/20.
\newblock \eprint{1108.2404}.

\bibitem{BoylanKolchin2011}
\bibinfo{author}{{Boylan-Kolchin}, M.}, \bibinfo{author}{{Bullock}, J.~S.} \&
  \bibinfo{author}{{Kaplinghat}, M.}
\newblock \bibinfo{journal}{\bibinfo{title}{{Too big to fail? The puzzling
  darkness of massive Milky Way subhaloes}}}.
\newblock {\emph{\JournalTitle{\mnras}}} \textbf{\bibinfo{volume}{415}},
  \bibinfo{pages}{L40--L44} (\bibinfo{year}{2011}).
\newblock \doiprefix 10.1111/j.1745-3933.2011.01074.x.
\newblock \eprint{1103.0007}.

\bibitem{Papastergis2015}
\bibinfo{author}{{Papastergis}, E.}, \bibinfo{author}{{Giovanelli}, R.},
  \bibinfo{author}{{Haynes}, M.~P.} \& \bibinfo{author}{{Shankar}, F.}
\newblock \bibinfo{journal}{\bibinfo{title}{{Is there a ``too big to fail''
  problem in the field?}}}
\newblock {\emph{\JournalTitle{\aap}}} \textbf{\bibinfo{volume}{574}},
  \bibinfo{pages}{A113} (\bibinfo{year}{2015}).
\newblock \doiprefix 10.1051/0004-6361/201424909.
\newblock \eprint{1407.4665}.

\bibitem{Oman2015}
\bibinfo{author}{{Oman}, K.~A.} \emph{et~al.}
\newblock \bibinfo{journal}{\bibinfo{title}{{The unexpected diversity of dwarf
  galaxy rotation curves}}}.
\newblock {\emph{\JournalTitle{\mnras}}} \textbf{\bibinfo{volume}{452}},
  \bibinfo{pages}{3650--3665} (\bibinfo{year}{2015}).
\newblock \doiprefix 10.1093/mnras/stv1504.
\newblock \eprint{1504.01437}.

\bibitem{Navarro1996b}
\bibinfo{author}{{Navarro}, J.~F.}, \bibinfo{author}{{Eke}, V.~R.} \&
  \bibinfo{author}{{Frenk}, C.~S.}
\newblock \bibinfo{journal}{\bibinfo{title}{{The cores of dwarf galaxy
  haloes}}}.
\newblock {\emph{\JournalTitle{\mnras}}} \textbf{\bibinfo{volume}{283}},
  \bibinfo{pages}{L72--L78} (\bibinfo{year}{1996}).
\newblock \doiprefix 10.1093/mnras/283.3.L72.
\newblock \eprint{astro-ph/9610187}.

\bibitem{Governato2012}
\bibinfo{author}{{Governato}, F.} \emph{et~al.}
\newblock \bibinfo{journal}{\bibinfo{title}{{Cuspy no more: how outflows affect
  the central dark matter and baryon distribution in {$\Lambda$} cold dark
  matter galaxies}}}.
\newblock {\emph{\JournalTitle{\mnras}}} \textbf{\bibinfo{volume}{422}},
  \bibinfo{pages}{1231--1240} (\bibinfo{year}{2012}).
\newblock \doiprefix 10.1111/j.1365-2966.2012.20696.x.
\newblock \eprint{1202.0554}.

\bibitem{Onorbe2015}
\bibinfo{author}{{O{\~n}orbe}, J.} \emph{et~al.}
\newblock \bibinfo{journal}{\bibinfo{title}{{Forged in FIRE: cusps, cores and
  baryons in low-mass dwarf galaxies}}}.
\newblock {\emph{\JournalTitle{\mnras}}} \textbf{\bibinfo{volume}{454}},
  \bibinfo{pages}{2092--2106} (\bibinfo{year}{2015}).
\newblock \doiprefix 10.1093/mnras/stv2072.
\newblock \eprint{1502.02036}.

\bibitem{Chan2015}
\bibinfo{author}{{Chan}, T.~K.} \emph{et~al.}
\newblock \bibinfo{journal}{\bibinfo{title}{{The impact of baryonic physics on
  the structure of dark matter haloes: the view from the FIRE cosmological
  simulations}}}.
\newblock {\emph{\JournalTitle{\mnras}}} \textbf{\bibinfo{volume}{454}},
  \bibinfo{pages}{2981--3001} (\bibinfo{year}{2015}).
\newblock \doiprefix 10.1093/mnras/stv2165.
\newblock \eprint{1507.02282}.

\bibitem{Read2016}
\bibinfo{author}{{Read}, J.~I.}, \bibinfo{author}{{Agertz}, O.} \&
  \bibinfo{author}{{Collins}, M.~L.~M.}
\newblock \bibinfo{journal}{\bibinfo{title}{{Dark matter cores all the way
  down}}}.
\newblock {\emph{\JournalTitle{\mnras}}} \textbf{\bibinfo{volume}{459}},
  \bibinfo{pages}{2573--2590} (\bibinfo{year}{2016}).
\newblock \doiprefix 10.1093/mnras/stw713.
\newblock \eprint{1508.04143}.

\bibitem{Green2004}
\bibinfo{author}{{Green}, A.~M.}, \bibinfo{author}{{Hofmann}, S.} \&
  \bibinfo{author}{{Schwarz}, D.~J.}
\newblock \bibinfo{journal}{\bibinfo{title}{{The power spectrum of SUSY-CDM on
  subgalactic scales}}}.
\newblock {\emph{\JournalTitle{\mnras}}} \textbf{\bibinfo{volume}{353}},
  \bibinfo{pages}{L23--L27} (\bibinfo{year}{2004}).
\newblock \doiprefix 10.1111/j.1365-2966.2004.08232.x.
\newblock \eprint{astro-ph/0309621}.

\bibitem{Bode2001}
\bibinfo{author}{{Bode}, P.}, \bibinfo{author}{{Ostriker}, J.~P.} \&
  \bibinfo{author}{{Turok}, N.}
\newblock \bibinfo{journal}{\bibinfo{title}{{Halo Formation in Warm Dark Matter
  Models}}}.
\newblock {\emph{\JournalTitle{\apj}}} \textbf{\bibinfo{volume}{556}},
  \bibinfo{pages}{93--107} (\bibinfo{year}{2001}).
\newblock \doiprefix 10.1086/321541.
\newblock \eprint{astro-ph/0010389}.

\bibitem{Carlson1992}
\bibinfo{author}{{Carlson}, E.~D.}, \bibinfo{author}{{Machacek}, M.~E.} \&
  \bibinfo{author}{{Hall}, L.~J.}
\newblock \bibinfo{journal}{\bibinfo{title}{{Self-interacting dark matter}}}.
\newblock {\emph{\JournalTitle{\apj}}} \textbf{\bibinfo{volume}{398}},
  \bibinfo{pages}{43--52} (\bibinfo{year}{1992}).
\newblock \doiprefix 10.1086/171833.

\bibitem{Spergel2000}
\bibinfo{author}{{Spergel}, D.~N.} \& \bibinfo{author}{{Steinhardt}, P.~J.}
\newblock \bibinfo{journal}{\bibinfo{title}{{Observational Evidence for
  Self-Interacting Cold Dark Matter}}}.
\newblock {\emph{\JournalTitle{Physical Review Letters}}}
  \textbf{\bibinfo{volume}{84}}, \bibinfo{pages}{3760--3763}
  (\bibinfo{year}{2000}).
\newblock \doiprefix 10.1103/PhysRevLett.84.3760.
\newblock \eprint{astro-ph/9909386}.

\bibitem{vogelsberger2012}
\bibinfo{author}{{Vogelsberger}, M.}, \bibinfo{author}{{Sijacki}, D.},
  \bibinfo{author}{{Kere{\v s}}, D.}, \bibinfo{author}{{Springel}, V.} \&
  \bibinfo{author}{{Hernquist}, L.}
\newblock \bibinfo{journal}{\bibinfo{title}{{Moving mesh cosmology: numerical
  techniques and global statistics}}}.
\newblock {\emph{\JournalTitle{\mnras}}} \textbf{\bibinfo{volume}{425}},
  \bibinfo{pages}{3024--3057} (\bibinfo{year}{2012}).
\newblock \doiprefix 10.1111/j.1365-2966.2012.21590.x.
\newblock \eprint{1109.1281}.

\bibitem{CyrRacine2016}
\bibinfo{author}{{Cyr-Racine}, F.-Y.} \emph{et~al.}
\newblock \bibinfo{journal}{\bibinfo{title}{{ETHOS -- an effective theory of
  structure formation: From dark particle physics to the matter distribution of
  the Universe}}}.
\newblock {\emph{\JournalTitle{Phys. Rev. D}}} \textbf{\bibinfo{volume}{93}},
  \bibinfo{pages}{123527} (\bibinfo{year}{2016}).
\newblock \doiprefix 10.1103/PhysRevD.93.123527.
\newblock \eprint{1512.05344}.

\bibitem{Vogelsberger2016}
\bibinfo{author}{{Vogelsberger}, M.} \emph{et~al.}
\newblock \bibinfo{journal}{\bibinfo{title}{{ETHOS - an effective theory of
  structure formation: dark matter physics as a possible explanation of the
  small-scale CDM problems}}}.
\newblock {\emph{\JournalTitle{\mnras}}} \textbf{\bibinfo{volume}{460}},
  \bibinfo{pages}{1399--1416} (\bibinfo{year}{2016}).
\newblock \doiprefix 10.1093/mnras/stw1076.
\newblock \eprint{1512.05349}.

\bibitem{Vogelsberger2012b}
\bibinfo{author}{{Vogelsberger}, M.}, \bibinfo{author}{{Zavala}, J.} \&
  \bibinfo{author}{{Loeb}, A.}
\newblock \bibinfo{journal}{\bibinfo{title}{{Subhaloes in self-interacting
  galactic dark matter haloes}}}.
\newblock {\emph{\JournalTitle{\mnras}}} \textbf{\bibinfo{volume}{423}},
  \bibinfo{pages}{3740--3752} (\bibinfo{year}{2012}).
\newblock \doiprefix 10.1111/j.1365-2966.2012.21182.x.
\newblock \eprint{1201.5892}.

\bibitem{Peter2013}
\bibinfo{author}{{Peter}, A.~H.~G.}, \bibinfo{author}{{Rocha}, M.},
  \bibinfo{author}{{Bullock}, J.~S.} \& \bibinfo{author}{{Kaplinghat}, M.}
\newblock \bibinfo{journal}{\bibinfo{title}{{Cosmological simulations with
  self-interacting dark matter - II. Halo shapes versus observations}}}.
\newblock {\emph{\JournalTitle{\mnras}}} \textbf{\bibinfo{volume}{430}},
  \bibinfo{pages}{105--120} (\bibinfo{year}{2013}).
\newblock \doiprefix 10.1093/mnras/sts535.
\newblock \eprint{1208.3026}.

\bibitem{Fry2015}
\bibinfo{author}{{Fry}, A.~B.} \emph{et~al.}
\newblock \bibinfo{journal}{\bibinfo{title}{{All about baryons: revisiting SIDM
  predictions at small halo masses}}}.
\newblock {\emph{\JournalTitle{\mnras}}} \textbf{\bibinfo{volume}{452}},
  \bibinfo{pages}{1468--1479} (\bibinfo{year}{2015}).
\newblock \doiprefix 10.1093/mnras/stv1330.
\newblock \eprint{1501.00497}.

\bibitem{Elbert2015}
\bibinfo{author}{{Elbert}, O.~D.} \emph{et~al.}
\newblock \bibinfo{journal}{\bibinfo{title}{{Core formation in dwarf haloes
  with self-interacting dark matter: no fine-tuning necessary}}}.
\newblock {\emph{\JournalTitle{\mnras}}} \textbf{\bibinfo{volume}{453}},
  \bibinfo{pages}{29--37} (\bibinfo{year}{2015}).
\newblock \doiprefix 10.1093/mnras/stv1470.
\newblock \eprint{1412.1477}.

\bibitem{Hui2017}
\bibinfo{author}{{Hui}, L.}, \bibinfo{author}{{Ostriker}, J.~P.},
  \bibinfo{author}{{Tremaine}, S.} \& \bibinfo{author}{{Witten}, E.}
\newblock \bibinfo{journal}{\bibinfo{title}{{Ultralight scalars as cosmological
  dark matter}}}.
\newblock {\emph{\JournalTitle{Phys. Rev. D}}} \textbf{\bibinfo{volume}{95}},
  \bibinfo{pages}{043541} (\bibinfo{year}{2017}).
\newblock \doiprefix 10.1103/PhysRevD.95.043541.
\newblock \eprint{1610.08297}.

\bibitem{Lee1996}
\bibinfo{author}{{Lee}, J.-W.} \& \bibinfo{author}{{Koh}, I.-G.}
\newblock \bibinfo{journal}{\bibinfo{title}{{Galactic halos as boson stars}}}.
\newblock {\emph{\JournalTitle{Phys. Rev. D}}} \textbf{\bibinfo{volume}{53}},
  \bibinfo{pages}{2236--2239} (\bibinfo{year}{1996}).
\newblock \doiprefix 10.1103/PhysRevD.53.2236.
\newblock \eprint{hep-ph/9507385}.

\bibitem{Hu2000}
\bibinfo{author}{{Hu}, W.}, \bibinfo{author}{{Barkana}, R.} \&
  \bibinfo{author}{{Gruzinov}, A.}
\newblock \bibinfo{journal}{\bibinfo{title}{{Fuzzy Cold Dark Matter: The Wave
  Properties of Ultralight Particles}}}.
\newblock {\emph{\JournalTitle{Physical Review Letters}}}
  \textbf{\bibinfo{volume}{85}}, \bibinfo{pages}{1158--1161}
  (\bibinfo{year}{2000}).
\newblock \doiprefix 10.1103/PhysRevLett.85.1158.
\newblock \eprint{astro-ph/0003365}.

\bibitem{Peebles2000}
\bibinfo{author}{{Peebles}, P.~J.~E.}
\newblock \bibinfo{journal}{\bibinfo{title}{{Fluid Dark Matter}}}.
\newblock {\emph{\JournalTitle{\apjl}}} \textbf{\bibinfo{volume}{534}},
  \bibinfo{pages}{L127--L129} (\bibinfo{year}{2000}).
\newblock \doiprefix 10.1086/312677.
\newblock \eprint{astro-ph/0002495}.

\bibitem{Chavanis2011}
\bibinfo{author}{{Chavanis}, P.-H.}
\newblock \bibinfo{journal}{\bibinfo{title}{{Mass-radius relation of Newtonian
  self-gravitating Bose-Einstein condensates with short-range interactions. I.
  Analytical results}}}.
\newblock {\emph{\JournalTitle{Phys. Rev. D}}} \textbf{\bibinfo{volume}{84}},
  \bibinfo{pages}{043531} (\bibinfo{year}{2011}).
\newblock \doiprefix 10.1103/PhysRevD.84.043531.
\newblock \eprint{1103.2050}.

\bibitem{Suarez2014}
\bibinfo{author}{{Su{\'a}rez}, A.}, \bibinfo{author}{{Robles}, V.~H.} \&
  \bibinfo{author}{{Matos}, T.}
\newblock \bibinfo{title}{{A Review on the Scalar Field/Bose-Einstein
  Condensate Dark Matter Model}}.
\newblock In \bibinfo{editor}{{Moreno Gonz{\'a}lez}, C.},
  \bibinfo{editor}{{Madriz Aguilar}, J.~E.} \& \bibinfo{editor}{{Reyes
  Barrera}, L.~M.} (eds.) \emph{\bibinfo{booktitle}{Accelerated Cosmic
  Expansion}}, vol.~\bibinfo{volume}{38} of \emph{\bibinfo{series}{Astrophysics
  and Space Science Proceedings}}, \bibinfo{pages}{107} (\bibinfo{year}{2014}).
\newblock \doiprefix {10.1007/978-3-319-02063-1\_9}.
\newblock \eprint{1302.0903}.

\bibitem{Matos2009}
\bibinfo{author}{{Matos}, T.}, \bibinfo{author}{{V{\'a}zquez-Gonz{\'a}lez}, A.}
  \& \bibinfo{author}{{Maga{\~n}a}, J.}
\newblock \bibinfo{journal}{\bibinfo{title}{{phi squared as dark matter}}}.
\newblock {\emph{\JournalTitle{\mnras}}} \textbf{\bibinfo{volume}{393}},
  \bibinfo{pages}{1359--1369} (\bibinfo{year}{2009}).
\newblock \doiprefix 10.1111/j.1365-2966.2008.13957.x.
\newblock \eprint{0806.0683}.

\bibitem{Lundgren2010}
\bibinfo{author}{{Lundgren}, A.~P.}, \bibinfo{author}{{Bondarescu}, M.},
  \bibinfo{author}{{Bondarescu}, R.} \& \bibinfo{author}{{Balakrishna}, J.}
\newblock \bibinfo{journal}{\bibinfo{title}{{Lukewarm Dark Matter: Bose
  Condensation of Ultralight Particles}}}.
\newblock {\emph{\JournalTitle{\apjl}}} \textbf{\bibinfo{volume}{715}},
  \bibinfo{pages}{L35--L39} (\bibinfo{year}{2010}).
\newblock \doiprefix 10.1088/2041-8205/715/1/L35.
\newblock \eprint{1001.0051}.

\bibitem{Robles2013}
\bibinfo{author}{{Robles}, V.~H.} \& \bibinfo{author}{{Matos}, T.}
\newblock \bibinfo{journal}{\bibinfo{title}{{Exact Solution to Finite
  Temperature SFDM: Natural Cores without Feedback}}}.
\newblock {\emph{\JournalTitle{\apj}}} \textbf{\bibinfo{volume}{763}},
  \bibinfo{pages}{19} (\bibinfo{year}{2013}).
\newblock \doiprefix 10.1088/0004-637X/763/1/19.
\newblock \eprint{1207.5858}.

\bibitem{Seidel1990}
\bibinfo{author}{{Seidel}, E.} \& \bibinfo{author}{{Suen}, W.-M.}
\newblock \bibinfo{journal}{\bibinfo{title}{{Dynamical evolution of boson
  stars: Perturbing the ground state}}}.
\newblock {\emph{\JournalTitle{Phys. Rev. D}}} \textbf{\bibinfo{volume}{42}},
  \bibinfo{pages}{384--403} (\bibinfo{year}{1990}).
\newblock \doiprefix 10.1103/PhysRevD.42.384.

\bibitem{Sin1994}
\bibinfo{author}{{Sin}, S.-J.}
\newblock \bibinfo{journal}{\bibinfo{title}{{Late-time phase transition and the
  galactic halo as a Bose liquid}}}.
\newblock {\emph{\JournalTitle{Phys. Rev. D}}} \textbf{\bibinfo{volume}{50}},
  \bibinfo{pages}{3650--3654} (\bibinfo{year}{1994}).
\newblock \doiprefix 10.1103/PhysRevD.50.3650.
\newblock \eprint{hep-ph/9205208}.

\bibitem{Mocz2015}
\bibinfo{author}{{Mocz}, P.} \& \bibinfo{author}{{Succi}, S.}
\newblock \bibinfo{journal}{\bibinfo{title}{{Numerical solution of the
  nonlinear Schr{\"o}dinger equation using smoothed-particle hydrodynamics}}}.
\newblock {\emph{\JournalTitle{Phys. Rev. D}}} \textbf{\bibinfo{volume}{91}},
  \bibinfo{pages}{053304} (\bibinfo{year}{2015}).
\newblock \doiprefix 10.1103/PhysRevE.91.053304.
\newblock \eprint{1503.03869}.

\bibitem{Nori2018}
\bibinfo{author}{{Nori}, M.} \& \bibinfo{author}{{Baldi}, M.}
\newblock \bibinfo{journal}{\bibinfo{title}{{AX-GADGET: a new code for
  cosmological simulations of Fuzzy Dark Matter and Axion models}}}.
\newblock {\emph{\JournalTitle{\mnras}}} \textbf{\bibinfo{volume}{478}},
  \bibinfo{pages}{3935--3951} (\bibinfo{year}{2018}).
\newblock \doiprefix 10.1093/mnras/sty1224.
\newblock \eprint{1801.08144}.

\bibitem{Copeland2006}
\bibinfo{author}{{Copeland}, E.~J.}, \bibinfo{author}{{Sami}, M.} \&
  \bibinfo{author}{{Tsujikawa}, S.}
\newblock \bibinfo{journal}{\bibinfo{title}{{Dynamics of Dark Energy}}}.
\newblock {\emph{\JournalTitle{International Journal of Modern Physics D}}}
  \textbf{\bibinfo{volume}{15}}, \bibinfo{pages}{1753--1935}
  (\bibinfo{year}{2006}).
\newblock \doiprefix 10.1142/S021827180600942X.
\newblock \eprint{hep-th/0603057}.

\bibitem{Baldi2012}
\bibinfo{author}{{Baldi}, M.}
\newblock \bibinfo{journal}{\bibinfo{title}{{Dark Energy simulations}}}.
\newblock {\emph{\JournalTitle{Physics of the Dark Universe}}}
  \textbf{\bibinfo{volume}{1}}, \bibinfo{pages}{162--193}
  (\bibinfo{year}{2012}).
\newblock \doiprefix 10.1016/j.dark.2012.10.004.
\newblock \eprint{1210.6650}.

\bibitem{Linder2003}
\bibinfo{author}{{Linder}, E.~V.} \& \bibinfo{author}{{Jenkins}, A.}
\newblock \bibinfo{journal}{\bibinfo{title}{{Cosmic structure growth and dark
  energy}}}.
\newblock {\emph{\JournalTitle{\mnras}}} \textbf{\bibinfo{volume}{346}},
  \bibinfo{pages}{573--583} (\bibinfo{year}{2003}).
\newblock \doiprefix 10.1046/j.1365-2966.2003.07112.x.
\newblock \eprint{astro-ph/0305286}.

\bibitem{Grossi2009}
\bibinfo{author}{{Grossi}, M.} \& \bibinfo{author}{{Springel}, V.}
\newblock \bibinfo{journal}{\bibinfo{title}{{The impact of early dark energy on
  non-linear structure formation}}}.
\newblock {\emph{\JournalTitle{\mnras}}} \textbf{\bibinfo{volume}{394}},
  \bibinfo{pages}{1559--1574} (\bibinfo{year}{2009}).
\newblock \doiprefix 10.1111/j.1365-2966.2009.14432.x.
\newblock \eprint{0809.3404}.

\bibitem{Penzo2014}
\bibinfo{author}{{Penzo}, C.}, \bibinfo{author}{{Macci{\`o}}, A.~V.},
  \bibinfo{author}{{Casarini}, L.}, \bibinfo{author}{{Stinson}, G.~S.} \&
  \bibinfo{author}{{Wadsley}, J.}
\newblock \bibinfo{journal}{\bibinfo{title}{{Dark MaGICC: the effect of dark
  energy on disc galaxy formation. Cosmology does matter}}}.
\newblock {\emph{\JournalTitle{\mnras}}} \textbf{\bibinfo{volume}{442}},
  \bibinfo{pages}{176--186} (\bibinfo{year}{2014}).
\newblock \doiprefix 10.1093/mnras/stu857.
\newblock \eprint{1401.3338}.

\bibitem{Sefusatti2011}
\bibinfo{author}{{Sefusatti}, E.} \& \bibinfo{author}{{Vernizzi}, F.}
\newblock \bibinfo{journal}{\bibinfo{title}{{Cosmological structure formation
  with clustering quintessence}}}.
\newblock {\emph{\JournalTitle{Journal of Cosmology and Astro-Particle
  Physics}}} \textbf{\bibinfo{volume}{2011}}, \bibinfo{pages}{047}
  (\bibinfo{year}{2011}).
\newblock \doiprefix 10.1088/1475-7516/2011/03/047.
\newblock \eprint{1101.1026}.

\bibitem{Amendola2001}
\bibinfo{author}{{Amendola}, L.} \& \bibinfo{author}{{Tocchini-Valentini}, D.}
\newblock \bibinfo{journal}{\bibinfo{title}{{Stationary dark energy: The
  present universe as a global attractor}}}.
\newblock {\emph{\JournalTitle{Phys. Rev. D}}} \textbf{\bibinfo{volume}{64}},
  \bibinfo{pages}{043509} (\bibinfo{year}{2001}).
\newblock \doiprefix 10.1103/PhysRevD.64.043509.
\newblock \eprint{astro-ph/0011243}.

\bibitem{Maccio2004}
\bibinfo{author}{{Macci{\`o}}, A.~V.}, \bibinfo{author}{{Quercellini}, C.},
  \bibinfo{author}{{Mainini}, R.}, \bibinfo{author}{{Amendola}, L.} \&
  \bibinfo{author}{{Bonometto}, S.~A.}
\newblock \bibinfo{journal}{\bibinfo{title}{{Coupled dark energy: Parameter
  constraints from N-body simulations}}}.
\newblock {\emph{\JournalTitle{Phys. Rev. D}}} \textbf{\bibinfo{volume}{69}},
  \bibinfo{pages}{123516} (\bibinfo{year}{2004}).
\newblock \doiprefix 10.1103/PhysRevD.69.123516.
\newblock \eprint{astro-ph/0309671}.

\bibitem{Baldi2010}
\bibinfo{author}{{Baldi}, M.}, \bibinfo{author}{{Pettorino}, V.},
  \bibinfo{author}{{Robbers}, G.} \& \bibinfo{author}{{Springel}, V.}
\newblock \bibinfo{journal}{\bibinfo{title}{{Hydrodynamical N-body simulations
  of coupled dark energy cosmologies}}}.
\newblock {\emph{\JournalTitle{\mnras}}} \textbf{\bibinfo{volume}{403}},
  \bibinfo{pages}{1684--1702} (\bibinfo{year}{2010}).
\newblock \doiprefix 10.1111/j.1365-2966.2009.15987.x.
\newblock \eprint{0812.3901}.

\bibitem{Li2011b}
\bibinfo{author}{{Li}, B.} \& \bibinfo{author}{{Barrow}, J.~D.}
\newblock \bibinfo{journal}{\bibinfo{title}{{N-body simulations for coupled
  scalar-field cosmology}}}.
\newblock {\emph{\JournalTitle{Phys. Rev. D}}} \textbf{\bibinfo{volume}{83}},
  \bibinfo{pages}{024007} (\bibinfo{year}{2011}).
\newblock \doiprefix 10.1103/PhysRevD.83.024007.
\newblock \eprint{1005.4231}.

\bibitem{Li2011}
\bibinfo{author}{{Li}, B.} \& \bibinfo{author}{{Barrow}, J.~D.}
\newblock \bibinfo{journal}{\bibinfo{title}{{On the effects of coupled scalar
  fields on structure formation}}}.
\newblock {\emph{\JournalTitle{\mnras}}} \textbf{\bibinfo{volume}{413}},
  \bibinfo{pages}{262--270} (\bibinfo{year}{2011}).
\newblock \doiprefix 10.1111/j.1365-2966.2010.18130.x.
\newblock \eprint{1010.3748}.

\bibitem{Milgrom1983}
\bibinfo{author}{{Milgrom}, M.}
\newblock \bibinfo{journal}{\bibinfo{title}{{A modification of the Newtonian
  dynamics as a possible alternative to the hidden mass hypothesis}}}.
\newblock {\emph{\JournalTitle{\apj}}} \textbf{\bibinfo{volume}{270}},
  \bibinfo{pages}{365--370} (\bibinfo{year}{1983}).
\newblock \doiprefix 10.1086/161130.

\bibitem{Famaey2012}
\bibinfo{author}{{Famaey}, B.} \& \bibinfo{author}{{McGaugh}, S.~S.}
\newblock \bibinfo{journal}{\bibinfo{title}{{Modified Newtonian Dynamics
  (MOND): Observational Phenomenology and Relativistic Extensions}}}.
\newblock {\emph{\JournalTitle{Living Reviews in Relativity}}}
  \textbf{\bibinfo{volume}{15}}, \bibinfo{pages}{10} (\bibinfo{year}{2012}).
\newblock \doiprefix 10.12942/lrr-2012-10.
\newblock \eprint{1112.3960}.

\bibitem{Bekenstein2004}
\bibinfo{author}{{Bekenstein}, J.~D.}
\newblock \bibinfo{journal}{\bibinfo{title}{{Relativistic gravitation theory
  for the modified Newtonian dynamics paradigm}}}.
\newblock {\emph{\JournalTitle{Phys. Rev. D}}} \textbf{\bibinfo{volume}{70}},
  \bibinfo{pages}{083509} (\bibinfo{year}{2004}).
\newblock \doiprefix 10.1103/PhysRevD.70.083509.
\newblock \eprint{astro-ph/0403694}.

\bibitem{Skordis2008}
\bibinfo{author}{{Skordis}, C.}
\newblock \bibinfo{journal}{\bibinfo{title}{{Generalizing tensor-vector-scalar
  cosmology}}}.
\newblock {\emph{\JournalTitle{Phys. Rev. D}}} \textbf{\bibinfo{volume}{77}},
  \bibinfo{pages}{123502} (\bibinfo{year}{2008}).
\newblock \doiprefix 10.1103/PhysRevD.77.123502.
\newblock \eprint{0801.1985}.

\bibitem{McGaugh2016}
\bibinfo{author}{{McGaugh}, S.~S.}, \bibinfo{author}{{Lelli}, F.} \&
  \bibinfo{author}{{Schombert}, J.~M.}
\newblock \bibinfo{journal}{\bibinfo{title}{{Radial Acceleration Relation in
  Rotationally Supported Galaxies}}}.
\newblock {\emph{\JournalTitle{\prl}}} \textbf{\bibinfo{volume}{117}},
  \bibinfo{pages}{201101} (\bibinfo{year}{2016}).
\newblock \doiprefix 10.1103/PhysRevLett.117.201101.
\newblock \eprint{1609.05917}.

\bibitem{Lelli2017}
\bibinfo{author}{{Lelli}, F.}, \bibinfo{author}{{McGaugh}, S.~S.},
  \bibinfo{author}{{Schombert}, J.~M.} \& \bibinfo{author}{{Pawlowski}, M.~S.}
\newblock \bibinfo{journal}{\bibinfo{title}{{One Law to Rule Them All: The
  Radial Acceleration Relation of Galaxies}}}.
\newblock {\emph{\JournalTitle{\apj}}} \textbf{\bibinfo{volume}{836}},
  \bibinfo{pages}{152} (\bibinfo{year}{2017}).
\newblock \doiprefix 10.3847/1538-4357/836/2/152.
\newblock \eprint{1610.08981}.

\bibitem{Keller2017}
\bibinfo{author}{{Keller}, B.~W.} \& \bibinfo{author}{{Wadsley}, J.~W.}
\newblock \bibinfo{journal}{\bibinfo{title}{{{\ensuremath{\Lambda}}CDM is
  Consistent with SPARC Radial Acceleration Relation}}}.
\newblock {\emph{\JournalTitle{\apj}}} \textbf{\bibinfo{volume}{835}},
  \bibinfo{pages}{L17} (\bibinfo{year}{2017}).
\newblock \doiprefix 10.3847/2041-8213/835/1/L17.
\newblock \eprint{1610.06183}.

\bibitem{Ludlow2017}
\bibinfo{author}{{Ludlow}, A.~D.} \emph{et~al.}
\newblock \bibinfo{journal}{\bibinfo{title}{{Mass-Discrepancy Acceleration
  Relation: A Natural Outcome of Galaxy Formation in Cold Dark Matter Halos}}}.
\newblock {\emph{\JournalTitle{\prl}}} \textbf{\bibinfo{volume}{118}},
  \bibinfo{pages}{161103} (\bibinfo{year}{2017}).
\newblock \doiprefix 10.1103/PhysRevLett.118.161103.
\newblock \eprint{1610.07663}.

\bibitem{Dutton2019}
\bibinfo{author}{{Dutton}, A.~A.}, \bibinfo{author}{{Macci{\`o}}, A.~V.},
  \bibinfo{author}{{Obreja}, A.} \& \bibinfo{author}{{Buck}, T.}
\newblock \bibinfo{journal}{\bibinfo{title}{{NIHAO - XVIII. Origin of the MOND
  phenomenology of galactic rotation curves in a {\ensuremath{\Lambda}}CDM
  universe}}}.
\newblock {\emph{\JournalTitle{\mnras}}} \textbf{\bibinfo{volume}{485}},
  \bibinfo{pages}{1886--1899} (\bibinfo{year}{2019}).
\newblock \doiprefix 10.1093/mnras/stz531.
\newblock \eprint{1902.06751}.

\bibitem{Llinares2018}
\bibinfo{author}{{Llinares}, C.}
\newblock \bibinfo{journal}{\bibinfo{title}{{Simulation techniques for modified
  gravity}}}.
\newblock {\emph{\JournalTitle{International Journal of Modern Physics D}}}
  \textbf{\bibinfo{volume}{27}}, \bibinfo{pages}{1848003}
  (\bibinfo{year}{2018}).
\newblock \doiprefix 10.1142/S0218271818480036.

\bibitem{Li2012}
\bibinfo{author}{{Li}, B.}, \bibinfo{author}{{Zhao}, G.-B.},
  \bibinfo{author}{{Teyssier}, R.} \& \bibinfo{author}{{Koyama}, K.}
\newblock \bibinfo{journal}{\bibinfo{title}{{ECOSMOG: an Efficient COde for
  Simulating MOdified Gravity}}}.
\newblock {\emph{\JournalTitle{\jcap}}} \textbf{\bibinfo{volume}{1}},
  \bibinfo{pages}{051} (\bibinfo{year}{2012}).
\newblock \doiprefix 10.1088/1475-7516/2012/01/051.
\newblock \eprint{1110.1379}.

\bibitem{Puchwein2013}
\bibinfo{author}{{Puchwein}, E.}, \bibinfo{author}{{Baldi}, M.} \&
  \bibinfo{author}{{Springel}, V.}
\newblock \bibinfo{journal}{\bibinfo{title}{{Modified-Gravity-GADGET: a new
  code for cosmological hydrodynamical simulations of modified gravity
  models}}}.
\newblock {\emph{\JournalTitle{\mnras}}} \textbf{\bibinfo{volume}{436}},
  \bibinfo{pages}{348--360} (\bibinfo{year}{2013}).
\newblock \doiprefix 10.1093/mnras/stt1575.
\newblock \eprint{1305.2418}.

\bibitem{Llinares2014}
\bibinfo{author}{{Llinares}, C.}, \bibinfo{author}{{Mota}, D.~F.} \&
  \bibinfo{author}{{Winther}, H.~A.}
\newblock \bibinfo{journal}{\bibinfo{title}{{ISIS: a new N-body cosmological
  code with scalar fields based on RAMSES. Code presentation and application to
  the shapes of clusters}}}.
\newblock {\emph{\JournalTitle{\aap}}} \textbf{\bibinfo{volume}{562}},
  \bibinfo{pages}{A78} (\bibinfo{year}{2014}).
\newblock \doiprefix 10.1051/0004-6361/201322412.
\newblock \eprint{1307.6748}.

\bibitem{Brax2012}
\bibinfo{author}{{Brax}, P.}, \bibinfo{author}{{Davis}, A.-C.},
  \bibinfo{author}{{Li}, B.}, \bibinfo{author}{{Winther}, H.~A.} \&
  \bibinfo{author}{{Zhao}, G.-B.}
\newblock \bibinfo{journal}{\bibinfo{title}{{Systematic simulations of modified
  gravity: symmetron and dilaton models}}}.
\newblock {\emph{\JournalTitle{Journal of Cosmology and Astro-Particle
  Physics}}} \textbf{\bibinfo{volume}{2012}}, \bibinfo{pages}{002}
  (\bibinfo{year}{2012}).
\newblock \doiprefix 10.1088/1475-7516/2012/10/002.
\newblock \eprint{1206.3568}.

\bibitem{Barreira2013}
\bibinfo{author}{{Barreira}, A.}, \bibinfo{author}{{Li}, B.},
  \bibinfo{author}{{Hellwing}, W.~A.}, \bibinfo{author}{{Baugh}, C.~M.} \&
  \bibinfo{author}{{Pascoli}, S.}
\newblock \bibinfo{journal}{\bibinfo{title}{{Nonlinear structure formation in
  the cubic Galileon gravity model}}}.
\newblock {\emph{\JournalTitle{Journal of Cosmology and Astro-Particle
  Physics}}} \textbf{\bibinfo{volume}{2013}}, \bibinfo{pages}{027}
  (\bibinfo{year}{2013}).
\newblock \doiprefix 10.1088/1475-7516/2013/10/027.
\newblock \eprint{1306.3219}.

\bibitem{Winther2015}
\bibinfo{author}{{Winther}, H.~A.} \emph{et~al.}
\newblock \bibinfo{journal}{\bibinfo{title}{{Modified gravity N-body code
  comparison project}}}.
\newblock {\emph{\JournalTitle{\mnras}}} \textbf{\bibinfo{volume}{454}},
  \bibinfo{pages}{4208--4234} (\bibinfo{year}{2015}).
\newblock \doiprefix 10.1093/mnras/stv2253.
\newblock \eprint{1506.06384}.

\bibitem{Fontanot2013}
\bibinfo{author}{{Fontanot}, F.}, \bibinfo{author}{{Puchwein}, E.},
  \bibinfo{author}{{Springel}, V.} \& \bibinfo{author}{{Bianchi}, D.}
\newblock \bibinfo{journal}{\bibinfo{title}{{Semi-analytic galaxy formation in
  f(R)-gravity cosmologies}}}.
\newblock {\emph{\JournalTitle{\mnras}}} \textbf{\bibinfo{volume}{436}},
  \bibinfo{pages}{2672--2679} (\bibinfo{year}{2013}).
\newblock \doiprefix 10.1093/mnras/stt1763.
\newblock \eprint{1307.5065}.

\bibitem{Jennings2012}
\bibinfo{author}{{Jennings}, E.}, \bibinfo{author}{{Baugh}, C.~M.},
  \bibinfo{author}{{Li}, B.}, \bibinfo{author}{{Zhao}, G.-B.} \&
  \bibinfo{author}{{Koyama}, K.}
\newblock \bibinfo{journal}{\bibinfo{title}{{Redshift-space distortions in f(R)
  gravity}}}.
\newblock {\emph{\JournalTitle{\mnras}}} \textbf{\bibinfo{volume}{425}},
  \bibinfo{pages}{2128--2143} (\bibinfo{year}{2012}).
\newblock \doiprefix 10.1111/j.1365-2966.2012.21567.x.
\newblock \eprint{1205.2698}.

\bibitem{Naik2018}
\bibinfo{author}{{Naik}, A.~P.}, \bibinfo{author}{{Puchwein}, E.},
  \bibinfo{author}{{Davis}, A.-C.} \& \bibinfo{author}{{Arnold}, C.}
\newblock \bibinfo{journal}{\bibinfo{title}{{Imprints of Chameleon f(R) gravity
  on Galaxy rotation curves}}}.
\newblock {\emph{\JournalTitle{\mnras}}} \textbf{\bibinfo{volume}{480}},
  \bibinfo{pages}{5211--5225} (\bibinfo{year}{2018}).
\newblock \doiprefix 10.1093/mnras/sty2199.
\newblock \eprint{1805.12221}.

\bibitem{Lombriser2015}
\bibinfo{author}{{Lombriser}, L.} \& \bibinfo{author}{{Pe{\~n}arrubia}, J.}
\newblock \bibinfo{journal}{\bibinfo{title}{{How chameleons core dwarfs with
  cusps}}}.
\newblock {\emph{\JournalTitle{\prd}}} \textbf{\bibinfo{volume}{91}},
  \bibinfo{pages}{084022} (\bibinfo{year}{2015}).
\newblock \doiprefix 10.1103/PhysRevD.91.084022.
\newblock \eprint{1407.7862}.

\bibitem{Arnold2019}
\bibinfo{author}{{Arnold}, C.}, \bibinfo{author}{{Leo}, M.} \&
  \bibinfo{author}{{Li}, B.}
\newblock \bibinfo{journal}{\bibinfo{title}{{Realistic simulations of galaxy
  formation in f(R) modified gravity}}}.
\newblock {\emph{\JournalTitle{Nature Astronomy}}} \bibinfo{pages}{386}
  (\bibinfo{year}{2019}).
\newblock \doiprefix 10.1038/s41550-019-0823-y.
\newblock \eprint{1907.02977}.

\bibitem{Davis2012}
\bibinfo{author}{{Davis}, A.-C.}, \bibinfo{author}{{Lim}, E.~A.},
  \bibinfo{author}{{Sakstein}, J.} \& \bibinfo{author}{{Shaw}, D.~J.}
\newblock \bibinfo{journal}{\bibinfo{title}{{Modified gravity makes galaxies
  brighter}}}.
\newblock {\emph{\JournalTitle{\prd}}} \textbf{\bibinfo{volume}{85}},
  \bibinfo{pages}{123006} (\bibinfo{year}{2012}).
\newblock \doiprefix 10.1103/PhysRevD.85.123006.
\newblock \eprint{1102.5278}.

\end{thebibliography}

\end{multicols}
\end{document}